\documentclass[namedreferences,hyperref,optionalrh,solaromanenum]{spr-sola}

\usepackage{graphicx}                    
\usepackage{color}   
\usepackage{gensymb}
\usepackage[version=4]{mhchem}
\usepackage[nolist]{acronym}
\usepackage{bm}

\renewcommand{\textbf}[1]{{\textrm{#1}}}

\newcommand{\ki}{K\,{\sc i}}
\newcommand{\sii}{Si\,{\sc i}}
\newcommand{\fei}{Fe\,{\sc i}}
\newcommand{\caii}{Ca\,{\sc ii}}
\newcommand{\mgi}{Mg\,{\sc i}}

\chardef\us=`\_

\newcommand{\sunrise}{{\sc Sunrise}}
\newcommand{\sunriset}{{\sc Sunrise iii}}

\newcommand{\arcsec}{$^{\prime\prime}$}

\begin{acronym}
\acro{scip}[SCIP]{Sunrise Chromospheric Infrared spectroPolarimeter}
\acro{pmu}[PMU]{polarization modulation unit}
\acro{smm}[SMM]{scan mirror mechanism}
\acro{cte}[CTE]{coefficient of thermal expansion}
\acro{islid}[{\sc ISLiD}]{Image Stabilization and Light Distribution unit}
\acro{susi}[SUSI]{Sunrise UV Spectropolarimeter and Imager}
\acro{cws}[CWS]{Correlating Wavefront Sensor}
\acro{pbs}[PBS]{polarization beam splitter}
\acro{pfi}[PFI]{Post Focus Instrumentation}
\acro{fov}[FoV]{field-of-view}
\acro{rms}[rms]{root-mean-square}
\acro{olr}[OLR]{outgoing longwave radiation}
\acro{leo}[LEO]{low Earth orbit}
\acro{sza}[SZA]{solar zenith angle}
\acro{sli}[SLI]{single-layer insulation}
\acro{mli}[MLI]{multi-layer insulation}
\acro{eps}[EPS]{expanded polystyrene}
\acro{pcb}[PCB]{printed circuit board}
\acro{dpu}[DPU]{digital processing unit}
\acro{amhd}[AMHD]{analog, mechanisms, and heaters driver board}
\acro{pcm}[PCM]{power converter module}
\acro{syc}[SyC]{system controller}
\acro{fg}[FG]{frame grabber} 
\acro{ssd}[SSD]{solid state drive}
\acro{sd}[SD]{secure digital}
\acro{fpga}[FPGA]{Field Programmable Gate Array}
\acro{lvds}[LVDS]{low-voltage differential signaling}
\acro{spi}[SPI]{serial peripheral interface}
\acro{adc}[ADC]{analog-to-digital converter}
\acro{cots}[COTS]{commercial-off-the-shelf}
\acro{csw}[CSW]{control software}
\acro{fgsw}[FGSW]{frame grabber software}
\acro{ics}[ICS]{instrument control system}
\acro{udp}[UDP]{user defined program}
\acro{deco}[DECO]{defined command}
\acro{gpio}[GPIO]{general purpose input/output}
\acro{cxp}[CxP]{CoaXPress} 
\acro{bpp}[bpp]{bits-per-pixel}
\acro{tvt}[TVT]{thermal-vacuum test}
\acro{dma}[DMA]{Direct Memory Access}
\acro{atc}[ATC]{Advanced Technology Center}
\acro{gui}[GUI]{Graphical User Interface}
\acro{mtf}[MTF]{Modulation Transfer Function}
\acro{hk}[HK]{housekeeping}
\acro{aiv}[AIV]{Assembly, Integration, and Verification}
\acro{ddr}[DDR]{Double Data Rate}
\acro{pcie}[PCIe]{Peripheral Component Interconnect Express}
\acro{ip}[IP]{Intellectual Property}
\acro{axi}[AXI]{Advanced eXtensible Interface}
\acro{ptfe}[PTFE]{polytetrafluoroethylene}
\acro{mhd}[MHD]{magneto-hydrodynamics}

\end{acronym}

\begin{document}

\begin{article}
\begin{opening}

\title{The Sunrise Chromospheric Infrared Spectro-Polarimeter SCIP: an instrument for SUNRISE III}

\author[addressref={aff1,aff2,aff3},email={yukio.katsukawa@nao.ac.jp}]{\inits{Y.}\fnm{Y.}~\lnm{Katsukawa}\orcid{0000-0002-5054-8782}}
\author[addressref={aff4,aff15}]{\inits{J.~C.}\fnm{J.~C.}~\lnm{del~Toro~Iniesta}\orcid{0000-0002-3387-026X}}
\author[addressref=aff5]{\inits{S.S.}\fnm{S.~K.}~\lnm{Solanki}\orcid{0000-0002-3418-8449}}
\author[addressref=aff1]{\inits{M.K.}\fnm{M.}~\lnm{Kubo}\orcid{0000-0001-5616-2808}}
\author[addressref=aff1]{\inits{M.K.}\fnm{H.}~\lnm{Hara}\orcid{0000-0001-5686-3081}}
\author[addressref=aff6]{\inits{T.S.}\fnm{T.}~\lnm{Shimizu}\orcid{0000-0003-4764-6856}}
\author[addressref={aff5,aff19}]{\inits{T.O.}\fnm{T.}~\lnm{Oba}\orcid{0000-0002-7044-6281}}
\author[addressref=aff1]{\inits{Y.K.}\fnm{Y.}~\lnm{Kawabata}\orcid{0000-0001-7452-0656}}
\author[addressref=aff1]{\inits{T.T.}\fnm{T.}~\lnm{Tsuzuki}\orcid{0000-0002-8342-8314}}
\author[addressref=aff1]{\inits{F.U.}\fnm{F.}~\lnm{Uraguchi}\orcid{0009-0005-9709-8431}}
\author[addressref=aff1]{\inits{K.S.}\fnm{K.}~\lnm{Shinoda}}
\author[addressref=aff1]{\inits{T.T.}\fnm{T.}~\lnm{Tamura}}
\author[addressref=aff1]{\inits{Y.S.}\fnm{Y.}~\lnm{Suematsu}\orcid{0000-0003-4452-858X}}
\author[addressref={aff11,aff1}]{\inits{T.M.}\fnm{T.}~\lnm{Matsumoto}\orcid{0000-0002-1043-9944}}
\author[addressref=aff12]{\inits{R.I.}\fnm{R.~T.}~\lnm{Ishikawa}\orcid{0000-0002-4669-5376}}
\author[addressref={aff2,aff1}]{\inits{Y.N.}\fnm{Y.}~\lnm{Naito}\orcid{0000-0001-6793-8528}}
\author[addressref=aff8]{\inits{K.I.}\fnm{K.}~\lnm{Ichimoto}\orcid{0000-0001-8689-3564}}
\author[addressref=aff8]{\inits{S.N.}\fnm{S.}~\lnm{Nagata}\orcid{0000-0003-0005-4757}}
\author[addressref=aff10]{\inits{T.A.}\fnm{T.}~\lnm{Anan}\orcid{0000-0001-6824-1108}}
\author[addressref={aff4,aff15}]{\inits{D.O.}\fnm{D.}~\lnm{Orozco~Su\'arez}\orcid{0000-0001-8829-1938}}
\author[addressref={aff20,aff15}]{\inits{E.}\fnm{E.}~\lnm{Sanchis Kilders}\orcid{0000-0002-4208-3575}}
\author[addressref={aff4,aff15}]{\inits{M.B.}\fnm{M.}~\lnm{Balaguer~Jim\'enez}\orcid{0000-0003-4738-7727}}
\author[addressref={aff4,aff15}]{\inits{A.C.}\fnm{A.C.}~\lnm{L\'opez~Jim\'enez}\orcid{0000-0002-6297-0681}}
\author[addressref={aff9,aff18}]{\inits{C.Q.}\fnm{C.}~\lnm{Quintero~Noda}\orcid{0000-0001-5518-8782}}
\author[addressref={aff4,aff15}]{\inits{D.A.}\fnm{D.}~\lnm{\'Alvarez Garc\'{\i}a}\orcid{0000-0002-8169-8476}}
\author[addressref={aff4,aff15}]{\inits{J.L.}\fnm{J.L.}~\lnm{Ramos M\'as}\orcid{0000-0002-8445-2631}}
\author[addressref={aff4,aff15}]{\inits{J.P.}\fnm{J.P.}~\lnm{Cobos Carrascosa}\orcid{0000-0002-5847-7181}}
\author[addressref={aff4,aff15}]{\inits{P.}\fnm{P.}~\lnm{Labrousse}}
\author[addressref={aff4,aff15}]{\inits{B.A.}\fnm{B.}~\lnm{Aparicio del Moral}\orcid{0000-0003-2817-8719}}
\author[addressref={aff4,aff15}]{\inits{A.S.}\fnm{A.}~\lnm{S\'anchez G\'omez}\orcid{0009-0008-7320-5716}}
\author[addressref={aff9,aff15}]{\inits{D.H.}\fnm{D.}~\lnm{Hern\'andez Exp\'osito}}
\author[addressref={aff4,aff15}]{\inits{E.B.}\fnm{E.}~\lnm{Bail\'on Mart\'inez}\orcid{0009-0004-3976-2528}}
\author[addressref={aff4,aff15}]{\inits{J.M.}\fnm{J.M.}~\lnm{Morales Fern\'andez}\orcid{0000-0002-5773-0368}}
\author[addressref={aff4,aff15}]{\inits{A.M.}\fnm{A.}~\lnm{Moreno Mantas}\orcid{0009-0002-3396-3359}}
\author[addressref={aff4,aff15}]{\inits{A.T.}\fnm{A.}~\lnm{Tobaruela}\orcid{0009-0009-4178-4554}}
\author[addressref={aff4,aff15}]{\inits{I.B.}\fnm{I.}~\lnm{Bustamante}}
\author[addressref={aff4,aff15}]{\inits{F.J.}\fnm{F.J.}~\lnm{Bail\'en}\orcid{0000-0002-7318-3536}}
\author[addressref={aff20,aff15}]{\inits{J.}\fnm{J.}~\lnm{Blanco Rodr\'{\i}guez}\orcid{0000-0002-2055-441X}}
\author[addressref={aff20,aff15}]{\inits{J.L.}\fnm{J.L.}~\lnm{Gasent Blesa}\orcid{0000-0002-1225-4177}}
\author[addressref={aff20,aff15}]{\inits{P.}\fnm{P.}~\lnm{Rodr\'{\i}guez Mart\'{\i}nez}}
\author[addressref={aff20,aff15}]{\inits{A.}\fnm{A.}~\lnm{Ferreres}\orcid{0000-0003-1500-1359}}
\author[addressref={aff20,aff15}]{\inits{D.}\fnm{D.}~\lnm{Gilabert Palmer}}
\author[addressref={aff7,aff15}]{\inits{J.P.}\fnm{J.}~\lnm{Piqueras Carre\~no}\orcid{0000-0002-3787-9640}}
\author[addressref={aff7,aff15}]{\inits{I.P.}\fnm{I.}~\lnm{P\'{e}rez~Grande}\orcid{0000-0002-7145-2835}}
\author[addressref={aff7,aff15}]{\inits{I.}\fnm{I.}~\lnm{Torralbo}\orcid{0000-0001-9272-6439}}
\author[addressref={aff17,aff15}]{\inits{A.A.}\fnm{A.}~\lnm{\'Alvarez-Herrero}\orcid{0000-0001-9228-3412}}
\author[addressref=aff5]{\fnm{A.}~\lnm{Korpi-Lagg}\orcid{0000-0003-1459-7074}}
\author[addressref=aff5]{\inits{A.G.}\fnm{A.}~\lnm{Gandorfer}\orcid{0000-0002-9972-9840}}
\author[addressref=aff13]{\inits{T.B.}\fnm{T.}~\lnm{Berkefeld}}
\author[addressref=aff14]{\inits{P.B.}\fnm{P.}~\lnm{Bernasconi}\orcid{0000-0002-0787-8954}}
\author[addressref=aff5]{\inits{A.F.}\fnm{A.}~\lnm{Feller}\orcid{0009-0009-4425-599X}}
\author[addressref=aff5]{\inits{T.R.}\fnm{T.L.}~\lnm{Riethm\"uller}\orcid{0000-0001-6317-4380}}
\author[addressref=aff5]{\inits{S.}\fnm{H.N.}~\lnm{Smitha}\orcid{0000-0003-3490-6532}}
\author[addressref=aff9]{\inits{V.M.}\fnm{V.}~\lnm{Mart\'inez~Pillet}\orcid{0000-0001-7764-6895}}
\author[addressref=aff5]{\inits{B.G.}\fnm{B.}~\lnm{Grauf}}
\author[addressref=aff13]{\inits{A.B.}\fnm{A.}~\lnm{Bell}}
\author[addressref=aff14]{\inits{M.C.}\fnm{M.}~\lnm{Carpenter}}

\address[id=aff1]{National Astronomical Observatory of Japan, 2-21-1 Osawa, Mitaka, Tokyo 181-8588, Japan}
\address[id=aff2]{The Graduate University for Advanced Studies, SOKENDAI}
\address[id=aff3]{Department of Astronomy, Graduate School of Science, The University of Tokyo, 7-3-1 Hongo, Bunkyo-ku, Tokyo 113-0033, Japan}
\address[id=aff4]{Instituto de Astrof\'isica de Andaluc\'ia (IAA-CSIC), Glorieta de la Astronom\'ia s/n, 18008 Granada, Spain}
\address[id=aff15]{Spanish Space Solar Physics Consortium, S$^3$PC}
\address[id=aff5]{Max-Planck-Institut f\"ur Sonnensystemforschung, Justus-von-Liebig-Weg 3, 37077 G\"ottingen, Germany}
\address[id=aff6]{Institute of Space and Astronautical Science, Japan Aerospace Exploration Agency, 3-1-1 Yoshinodai, Chuo-ku, Sagamihara, Kanagawa 252-5210, Japan}
\address[id=aff19]{Advanced Research Center for Space Science and Technology, Institute of Science and Engineering, Kanazawa University, Kakuma-machi, Kanazawa, Ishikawa 920-1192, Japan}
\address[id=aff11]{Institute for Space-Earth Environmental Research, Nagoya University, Furocho, Chikusa-ku, Nagoya, Aichi 464-8601, Japan}
\address[id=aff12]{National Institute of Fusion Science, 322-6 Oroshi-cho, Toki, Gihu 509-5292, Japan}
\address[id=aff8]{Kwasan and Hida Observatory, Kyoto Univ., Kurabashira, Kamitakara, Takayama, Gifu 506-1314, Japan}
\address[id=aff10]{National Solar Observatory, 22 Ohi'a Ku, Makawao, HI, 96768, USA}
\address[id=aff20]{Universitat de Valencia, Avda. de la Universitat s/n, E-46100 Burjassot, Spain}
\address[id=aff9]{Instituto de Astrof\'isica de Canarias, 38205 La Laguna, Tenerife, Spain}
\address[id=aff18]{Departamento de Astrof\'isica, Univ. de La Laguna, La Laguna, Tenerife, E-38200, Spain}
\address[id=aff17]{Instituto Nacional de T\'ecnica Aeroespacial (INTA), Ctra. de Ajalvir, km. 4, E-28850 Torrej\'on de Ardoz, Spain}
\address[id=aff7]{Universidad Polit\'ecnica de Madrid, IDR/UPM, Plaza Cardenal Cisneros 3, 28040 Madrid, Spain}
\address[id=aff13]{Institut f\"ur Sonnenphysik, Sch\"oneckstr. 6, 79104 Freiburg, Germany}
\address[id=aff14]{Johns Hopkins Applied Physics Laboratory, 11100 Johns Hopkins Road, Laurel, Maryland, USA}

\runningauthor{Katsukawa et al.}
\runningtitle{\textit{Solar Physics} \sunriset\ SCIP}

\begin{abstract}
The \sunrise\ balloon-borne solar observatory is equipped with a one-meter aperture optical telescope, offering a unique platform for uninterrupted seeing-free observations across ultraviolet, visible, and infrared wavelengths from altitudes higher than 33 km. For the third flight of the upgraded \sunrise\ observatory conducted in 2024, now called \sunriset, a new spectro-polarimeter called the \ac{scip} was developed for observing near-infrared wavelength ranges around 770~nm and 850~nm. These wavelength ranges contain many spectral lines, including two of the \caii\, infrared triplet, \ki\, D1 and D2 lines, and multiple \fei\, lines, that are sensitive to solar magnetic fields and velocities in the photosphere and chromosphere. SCIP consists of a grating spectrograph in which polarimetric measurements are conducted using a rotating waveplate as a modulator and polarizing beam splitters placed in front of the cameras. The spatial and spectral resolutions are 0.21" and $1\times10^5$, respectively, and a polarimetric sensitivity of 0.03\% (1$\sigma$) of the continuum intensity is achieved with a 10~s integration time per a resolution element. To achieve high-precision detection of small polarization signals, we carefully designed the optical and mechanical systems, polarization components, control electronics, and onboard data processing. Together with the other post-focus instrumentation developed for \sunriset, the \sunrise\ Ultraviolet Spectropolarimeter and Imager (SUSI) and the visible imaging spectro-polarimeter Tunable Magnetograph (TuMag), SCIP provides novel observations that help elucidate energy transfer and time-dependent phenomena across the solar photosphere and chromosphere.
\end{abstract}
\end{opening}

\section{Introduction}
     \label{S-Introduction} 

The Sun provides a unique site to study how magnetic fields drive dynamic phenomena in astrophysical plasmas at a high spatial and temporal resolution. In addition, to understand the physical mechanisms driving the dynamics, it is critically important to obtain physical quantities, such as the temperature, velocity, and magnetic field, in a remote object using spectro-polarimetric measurements. Space-based spectro-polarimeters have been realized and provide valuable data for investigating the behavior of magnetic fields on the solar surface, i.e. the photosphere. 
These include the Michelson Doppler Imager \citep[MDI:][]{1995SoPh..162..129S} on SOHO, the Helioseismic and Magnetic Imager \citep[HMI:][]{2012SoPh..275..207S} on SDO, and the Polarimetric and Helioseismic Imager \citep[PHI:][]{2020A&A...642A..11S} on Solar Orbiter, all of which provided imaging spectropolarimetry using a narrow-band tunable filter. The {\it Hinode} Solar Optical Telescope \citep[SOT:][]{2008SoPh..249..167T} carried both a narrow-band filter imager (NFI) and a spectro-polarimeter \citep[SP:][]{2013SoPh..283..579L}. The NFI provided polarimetric measurements at several photospheric and chromospheric lines with a Lyot filter \citep{2008SoPh..249..233I}. The SP obtained spectro-polarimetric data of the \fei\ lines at 630.25~nm (Land\'e factor $g=2.5$) and 630.15~nm (effective Land\'e factor $g_{\rm eff}=1.67$) with an echelle spectrograph and allowed us to map the magnetic fields in active and quiet regions with a \textbf{spatial} resolution of 0.32" owing to the diffraction-limited optical performance of a 50~cm aperture telescope and the stable observation conditions \citep{2008A&A...484L..17D}. Spectro-polarimetric observations of even larger apertures have been achieved with balloon-borne telescopes \citep[e.g. Flare Genesis Experiment,][]{2000SPIE.4014..214B}. Among them, the \sunrise\ balloon-borne solar telescope \citep{2010ApJ...723L.127S,2011SoPh..268....1B, 2011SoPh..268...35G,2011SoPh..268..103B} carries a 1~m aperture optical telescope and allows us to conduct seeing-free visible observations under stable conditions as well as UV observations with a high spatial resolution of 0.1" at wavelengths shorter than 400~nm, which cannot be well observed using ground-based telescopes due to atmospheric absorption. The Imaging Magnetograph eXperiment (IMaX) \citep{2011SoPh..268...57M} made spectro-polarimetric observations at the \fei\ 525~nm line (Land\'e factor $g=3$) sensitive to the solar photosphere using a narrow-band tunable filter, and provided maps of surface magnetic fields with a spatial resolution of 0.15". 

In the past two flights of \textsc{Sunrise} in 2009 and 2013, spectro-polarimetric observations were conducted only with IMaX in the photosphere \citep[e.g.][]{2010ApJ...723L.164L, 2010ApJ...723L.149D}, whereas imaging observations were used to observe the overlying atmosphere, i.e., the chromosphere \citep[e.g.][]{2013ApJ...776L..13R,2017ApJS..229....9J}. There is a strong demand to expand the capability of spectro-polarimetry to conduct detailed and quantitative investigations of the chromosphere to understand the mechanisms driving the dynamic phenomena, such as jets and \textbf{\ac{mhd}} waves \citep{2018ASSL..449..201K, 2021SoPh..296...70R, 2022ARA&A..60..415T}. Because magnetic fields in the chromosphere are generally weak and spectral lines emanating from the chromosphere are broad compared with photospheric lines, we need a high polarimetric sensitivity to obtain measurable polarization signals at a spectral line sensitive to chromospheric magnetic fields \citep{2017SSRv..210...37L,2018ApJ...866...89C}. The 1~m aperture \textsc{Sunrise} telescope is suitable for collecting the number of photons required for precision polarimetry and to resolve the fine-scale dynamics spatially and temporally. The third successful flight, i.e., \textsc{Sunrise III}, was conducted in 2024 by upgrading the focal plane instruments \citep{2025SoPh..300...75K} and gondola \citep{2025SoPh..300..112B}. IMaX has been replaced by the Tunable Magnetograph (TuMag), which is able to switch the spectral lines to allow observing the mid-chromospheric line \mgi\ 517~nm \citep{2018MNRAS.481.5675Q} in addition to one of the photospheric lines \fei\ 525.02~nm and 525.06~nm \citep{2025SoPh..300..148D}. Observations within the UV wavelength range were conducted using the \textsc{Sunrise} Filter Imager (SuFI) during previous flights \citep{2011SoPh..268...35G}. It has been replaced with the \textsc{Sunrise} UV Spectropolarimeter and Imager (SUSI) in \sunriset\ to make spectro-polarimetric observations within the wavelength range of 309 -- 417~nm \citep{2025SoPh..300...65F}. The Correlating Wavefront Sensor (CWS), as flown in the previous flights, is an instrument for tip-tilt mirror compensation of image motions, for closed-loop autofocus, and for measuring low-order aberrations for better image stability and quality \citep{2026SoPh..301...57B}. 

In addition to the above three instruments, a new instrument called the Sunrise Chromospheric Infrared spectroPolarimeter (SCIP) has been newly developed to observe near-infrared spectral lines sensitive to solar magnetic fields in both the photosphere and the chromosphere. The SCIP instrument \textbf{was} designed and built through international collaboration among Japanese institutes including the Japan Aerospace Exploration Agency (JAXA), the Spanish Space Solar Physics Consortium (S$^3$PC) led by the Instituto de Astrof\'isica de Andaluc\'ia (IAA-CSIC), and the German Max Planck Institute for Solar System Research (MPS) with the leadership of the National Astronomical Observatory of Japan (NAOJ).

\section{Concepts of SCIP}
\label{S-concepts}

\subsection{Science targets and requirements}
\begin{figure}[t]
\centering
\includegraphics[width=\linewidth]{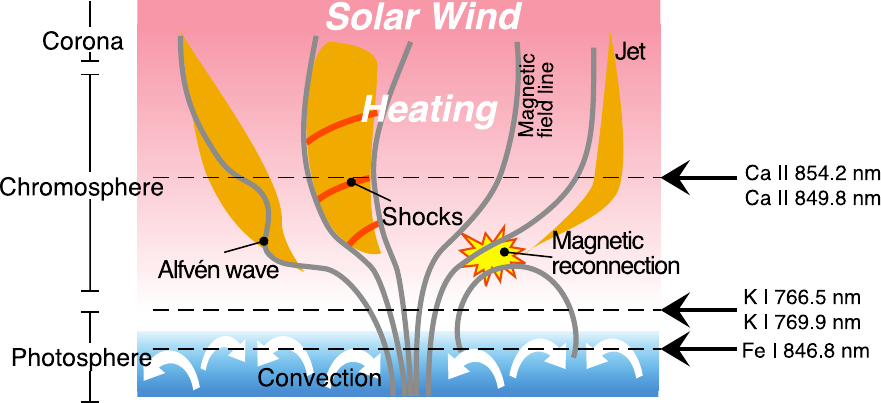}
\caption{Science targets using SCIP. Spectro-polarimetric measurements of multiple spectral lines from the photosphere to the chromosphere allow us to obtain three-dimensional structures of magnetic and velocity fields and reveal the nature of the jets and \ac{mhd} waves responsible for heating the atmosphere and acceleration of the solar wind. Modified from \cite{2020SPIE11447E..0YK}.}
\label{fig:science}
\end{figure}

\begin{figure}[t]
\centering
\includegraphics[width=0.7\linewidth]{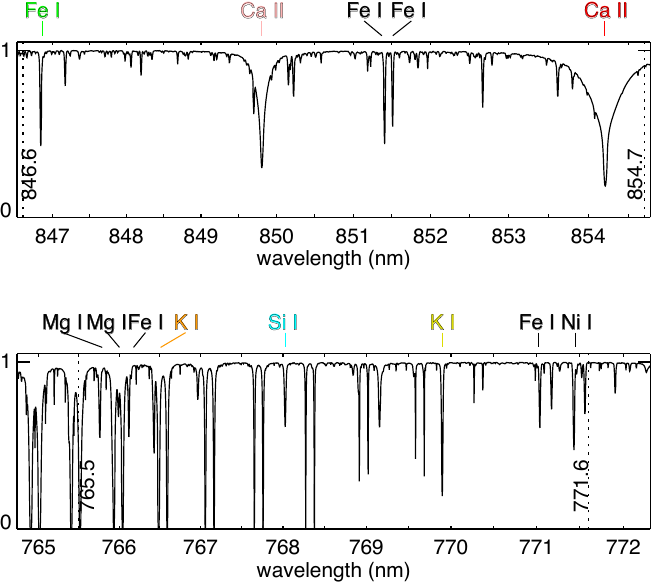}
\caption{Atlas spectra of the two SCIP spectral windows, (top) 850~nm and (bottom) 770~nm windows, and key spectral lines in the wavelength ranges. This solar spectral atlas shown here is compiled using data from the NSO FTS flux atlas \citep{1984sfat.book.....K}.}
\label{fig:atlas}
\end{figure}

\begin{figure}
\centering
\includegraphics[width=\linewidth]{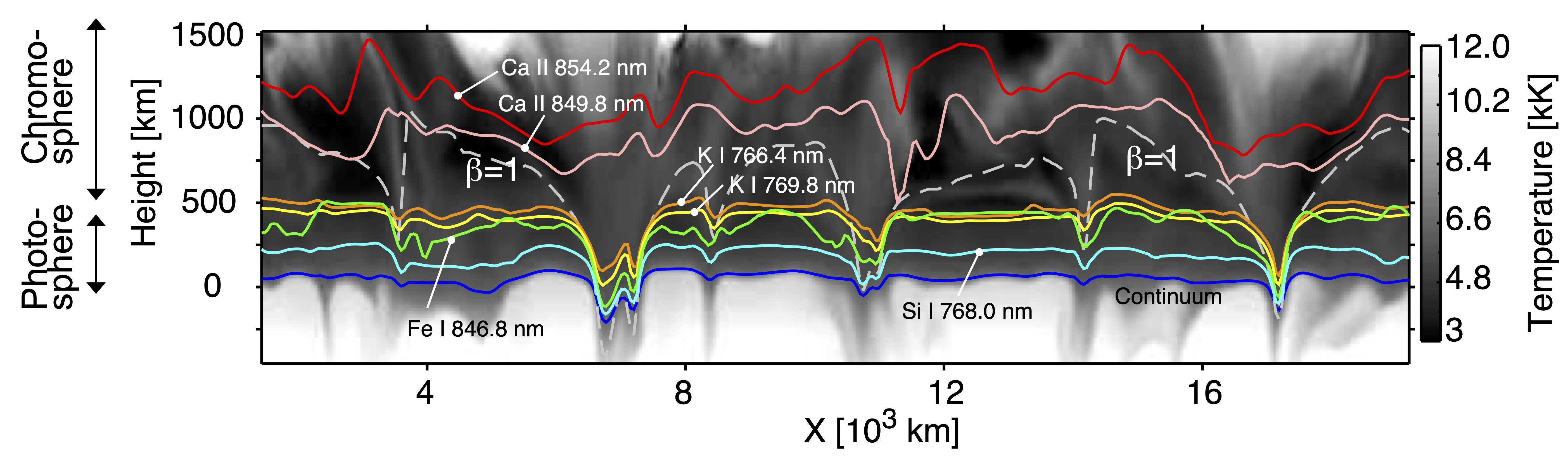}
\caption{Formation heights of key spectral lines observed with SCIP, including \caii\ 854.2~nm and 849.8~nm, \ki\ 766.4~nm and 769.9~nm, \fei\ 846.8~nm, and \sii\ 768.0~nm. The figure \textbf{was} made using a 3D \ac{mhd} simulation and non-LTE radiative transfer calculation \citep{2017MNRAS.470.1453Q}. The gray scale represents temperatures within the solar atmosphere. Modified from \cite{2020SPIE11447E..0YK}.}
\label{fig:height}
\end{figure}

The science targets of \ac{scip} are to (1) determine the 3D magnetic structure from the photosphere to the chromosphere, (2) trace \ac{mhd} waves from the photosphere to the chromosphere, and (3) reveal the mechanisms driving chromospheric jets, by measuring the height- and time-dependent velocities and magnetic fields in the solar atmosphere (Figure \ref{fig:science}). To achieve them, SCIP is designed to observe two spectral regions centered at 850 and 770 nm in the near-infrared wavelength range (Figure \ref{fig:atlas}). The two wavelength ranges contain many spectral lines that produce measurable polarization signals through the Zeeman effect in the presence of magnetic fields. In particular, the \caii\ lines at 849.8~nm (effective Land\'e factor $g_{\rm eff}=1.07$) and 854.2 nm ($g_{\rm eff}=1.10$) are sensitive to magnetic fields in the chromosphere \citep{2017MNRAS.464.4534Q}. The \ki\ lines at 766.4~nm ($g_{\rm eff}=1.16$) and 769.9~nm ($g_{\rm eff}=1.33$) are sensitive to the atmosphere at the interface between the photosphere and the chromosphere \citep{2017MNRAS.470.1453Q}. There are multiple spectral lines sensitive to the photosphere, such as the \fei\ 846.8 nm line (Land\'e factor $g=2.5$), within the wavelength range. By combining these spectral lines, we can seamlessly cover the photosphere and the chromosphere and obtain 3D magnetic and velocity structures in the solar atmosphere, as shown in Figure \ref{fig:height}. It has been demonstrated that spectro-polarimetric observations of these lines provide a good diagnostic tool to understand the mechanism\textbf{s} driving dynamical processes, such as jets, \ac{mhd} waves \citep{2017MNRAS.472..727Q,2019MNRAS.486.4203Q,2023MNRAS.523..974M}, and magnetic reconnection \citep{2024ApJ...960...26K}. It is worth noting that the 770~nm band is heavily affected by the \ce{O2} band in the Earth's atmosphere. In particular, the \ki\ 766.4~nm line is completely \textbf{absorbed by an \ce{O2} line, so neither its intensity nor polarization profiles can be measured with} a ground-based telescope. At the balloon altitude, the absorption becomes much weaker, enabling a reliable measurement of the polarization profile of the \ki\ 766.4~nm line for the first time \citep{2018A&A...610A..79Q}.

In addition to the spectral coverage, a high spatial resolution is also required to resolve the dynamic phenomena occurring in the atmosphere, where we set a goal to achieve a spatial resolution of 0.21" (corresponding to about 150~km on the solar surface), which is the diffraction-limit of the 1~m aperture telescope at a wavelength of 850~nm. A polarimetric sensitivity of 0.03\% (1$\sigma$) of the continuum intensity is necessary to measure weak polarization signals in the chromospheric lines, such as \caii\ 854.2~nm and 849.8~nm. By combining the 1~m aperture telescope with optimized optical elements and \textbf{high} polarization modulation efficiencies, we aim to achieve the polarimetric sensitivity within a resolution element smaller than 0.2" in less than 10~s of integration time. This is critical for obtaining the temporal evolution of magnetic fields that drive dynamics with lifetime as short as 10~s \citep[e.g.][]{2009SSRv..144..317W}. To sample multiple wavelength points across narrow photospheric lines and distinguish the polarization (Stokes) spectral profiles, a spectral resolution higher than $1\times10^5$ is required. The \ac{fov} is limited by the upstream optics but is required to be larger than 50"$\times$50". Seeing-free observations achieved using the stratospheric balloon-borne telescope allow us to map magnetic structures and their temporal evolution in the solar atmosphere with good spatial and spectral resolution, along with the polarimetric sensitivity described above. The key requirements are summarized below.

\begin{itemize}
\item Wavelength coverage to include spectral lines, allowing seamless coverage of the photosphere and the chromosphere (Figure \ref{fig:height}).
\begin{itemize}
\item Two wavelength bands including \caii\ 850~nm and \ki\ 770~nm lines (Figure \ref{fig:atlas})
\end{itemize}
\item Spatial and temporal resolution high enough to resolve fine-scale dynamics in the chromosphere.
\begin{itemize}
\item Spatial resolution of 0.21" (diffraction limit resolution at 850~nm)
\item Stokes IQUV (normal mode): temporal resolution shorter than 10 sec at each slit position
\item Stokes I only (rapid mode): temporal resolution shorter than 45 sec for scanning a full \ac{fov}
\end{itemize}
\item Spectral resolution and polarization sensitivity sufficiently high to get polarization signals at the photospheric and chromospheric spectral lines
\begin{itemize}
\item Spectral resolution and sampling \textbf{larger than} $\lambda/\Delta\lambda=1\times10^5$ and $2\times10^5$, respectively
\item Polarization sensitivity of 0.03\% (1$\sigma$) of the continuum intensity
\end{itemize}
\item Field-of-view large enough to cover a super-granular scale
\begin{itemize}
\item 58"$\times$58"
\end{itemize}
\end{itemize}

\subsection{SCIP instrument overview}
\begin{figure}[t]
\centering
\includegraphics[width=\linewidth]{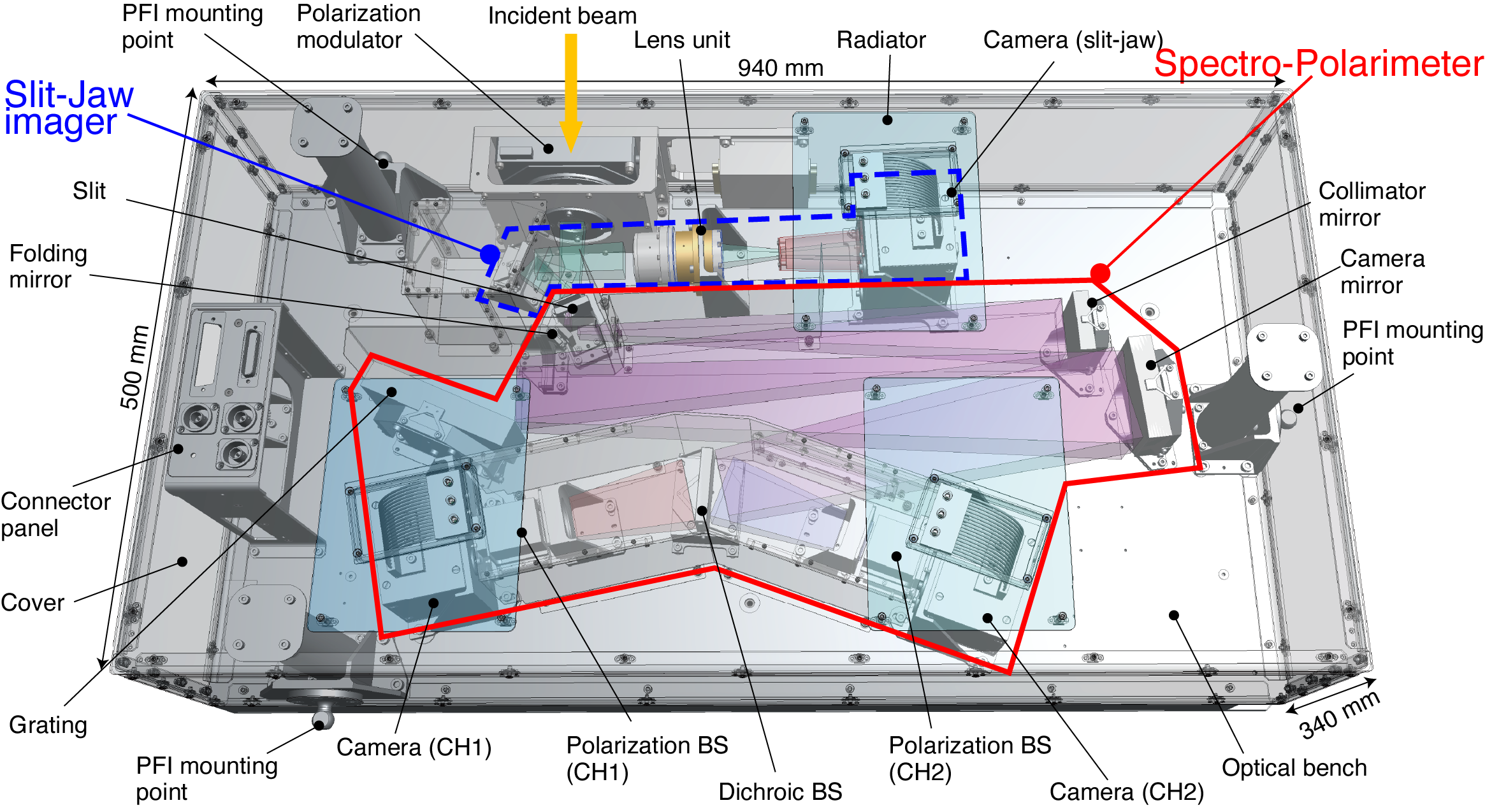}
\caption{The opto-mechanical layout of the SCIP O-unit. The cover and the lighttrap structures are shown as transparent for clarity. Modified from \cite{uraguchi2023jaxarr}.}
\label{fig:cad}
\end{figure}

SCIP consists of two major units: the Optics unit (O-unit) and the Electronics unit (E-unit)  \citep{2020SPIE11447E..0YK}. The O-unit contains optical components and cameras and is installed in the \ac{pfi} structure (see Section \ref{S-ounit} for details). The E-unit contains control electronics and is installed in the gondola E-rack (see Section \ref{S-eunit} for details). The spectro-polarimeter (SP) of the SCIP O-unit is a long-slit spectrograph that uses an echelle grating to observe two wavelength ranges around 850~nm and 770~nm (Figure \ref{fig:cad}). A slit-jaw (SJ) optics system is used to observe a 2D image around the slit and consists of a reflective filter and a doublet lens unit. The reflective filter transmits unnecessary passband light to a light trap behind the filter. Details of the optical and mechanical designs are described in Sections \ref{SS-opt} and \ref{SS-optmecha}, respectively.  We use three cameras (850~nm, 770~nm, and SJ)  employing a back-side illuminated CMOS sensor, with electronics newly developed for \sunriset\ (see Section \ref{SS-cameras} for details).

The polarimetric measurement is carried out by the combination of a rotating waveplate located at the optical entrance of the O-unit and the \acp{pbs} located in front of the cameras (Section \ref{SS-polari}). The rotating waveplate creates a temporally modulated solar image on the slit. Because there are multiple folding mirrors in the telescope and \ac{islid} feeding the light into the O-unit, we need not only an instrument-level, i.e., SCIP only, polarization calibration but also end-to-end polarization calibration combined with the telescope and \ac{islid} after we mount the SCIP O-unit onto the \ac{pfi} structure. Details of the polarization calibration are described in Section \ref{SS-polcal}. 

\begin{figure}[t]
\centering
\includegraphics[width=0.9\linewidth]{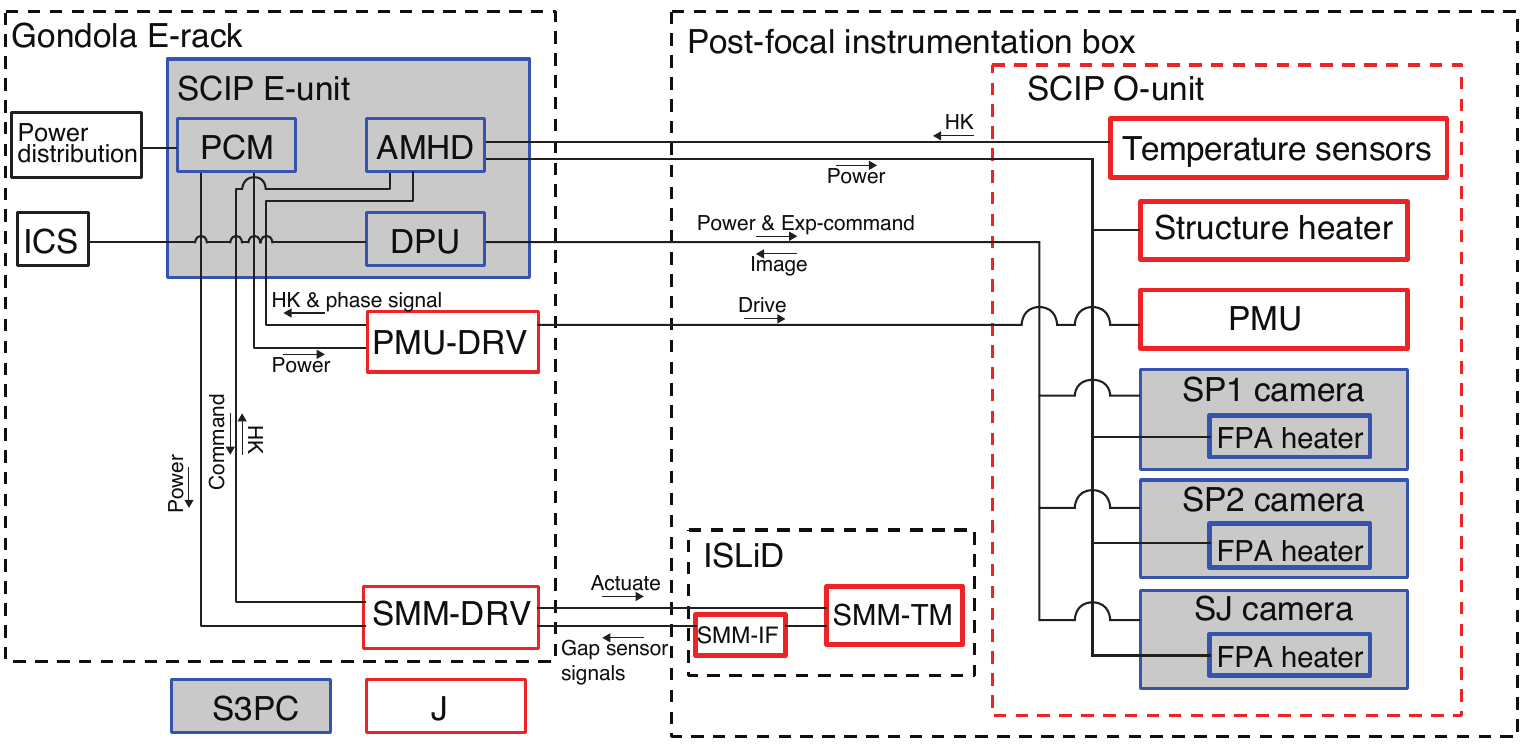}
\caption{Block diagram of the \acs{scip} electronics for observation control. The control electronics of the \acs{scip} E-unit are located on the gondola E-rack. The driver electronics boxes for the \acs{pmu} and \acs{smm} are also located on the E-rack and controlled by the E-unit. Exposures of the three cameras are triggered by phase signals generated by the PMU, and the E-unit sends an exposure command to the cameras. The E-unit receives images taken by the three cameras and applies onboard image processing. The E-unit also sends a command to the \acs{smm} to move the scan mirror. Structure and camera heaters are placed inside the O-unit to stabilize the temperature environment of the O-unit. The red boxes indicate components developed by the Japanese team, whereas the blue boxes indicate components developed by the S$^3$PC. Modified from \cite{2020SPIE11447E..0YK}.}
\label{fig:diagram}
\end{figure}

Figure \ref{fig:diagram} shows a block diagram of the \ac{scip} O-unit and E-unit with associated electronics components. For accurate polarization measurements, it is important to achieve a good synchronization between the exposures by the cameras and the rotating phase of the waveplate by a \ac{pmu} \citep[][see Section \ref{SS-polari} for details]{2020SPIE11447E..A3K}. The E-unit, along with onboard software and firmware, achieves synchronization and control of the O-unit in the following way:
\begin {itemize}
\item The \ac{pmu} sends phase signals to the E-unit. Exposures of the three cameras are triggered by the phase signals, and the E-unit sends an exposure command to the cameras. 
\item The E-unit receives images taken by the three cameras and applies onboard image processing, including polarization demodulation, in which accumulation and subtraction of the images are achieved depending on the rotating phase of the waveplate (See Section \ref{S-software} and Appendix \ref{S-firmware-implementation}).
\item The E-unit synchronizes the \ac{smm}, which scans perpendicular to the slit, with the PMU rotation and camera exposures to prevent image motion during exposure (see Section \ref{SS-smm} for details on \ac{smm}).
\item The E-unit controls the heaters and collects temperature data from sensors distributed in the O-unit.
\end{itemize}

\section{SCIP O-unit}
\label{S-ounit}

The SCIP O-unit contains the optical system for spectro-polarimetric observations using the slit spectrograph and for imaging observations using the slit-jaw optical system. This section describes the optical, mechanical, and thermal designs of the O-unit and its accompanying electrical components: the PMU, SMM, and cameras.

\subsection{Optical design}
\label{SS-opt}
\subsubsection{SP optics}
\begin{figure}[t]
\centering
\includegraphics[width=0.8\linewidth]{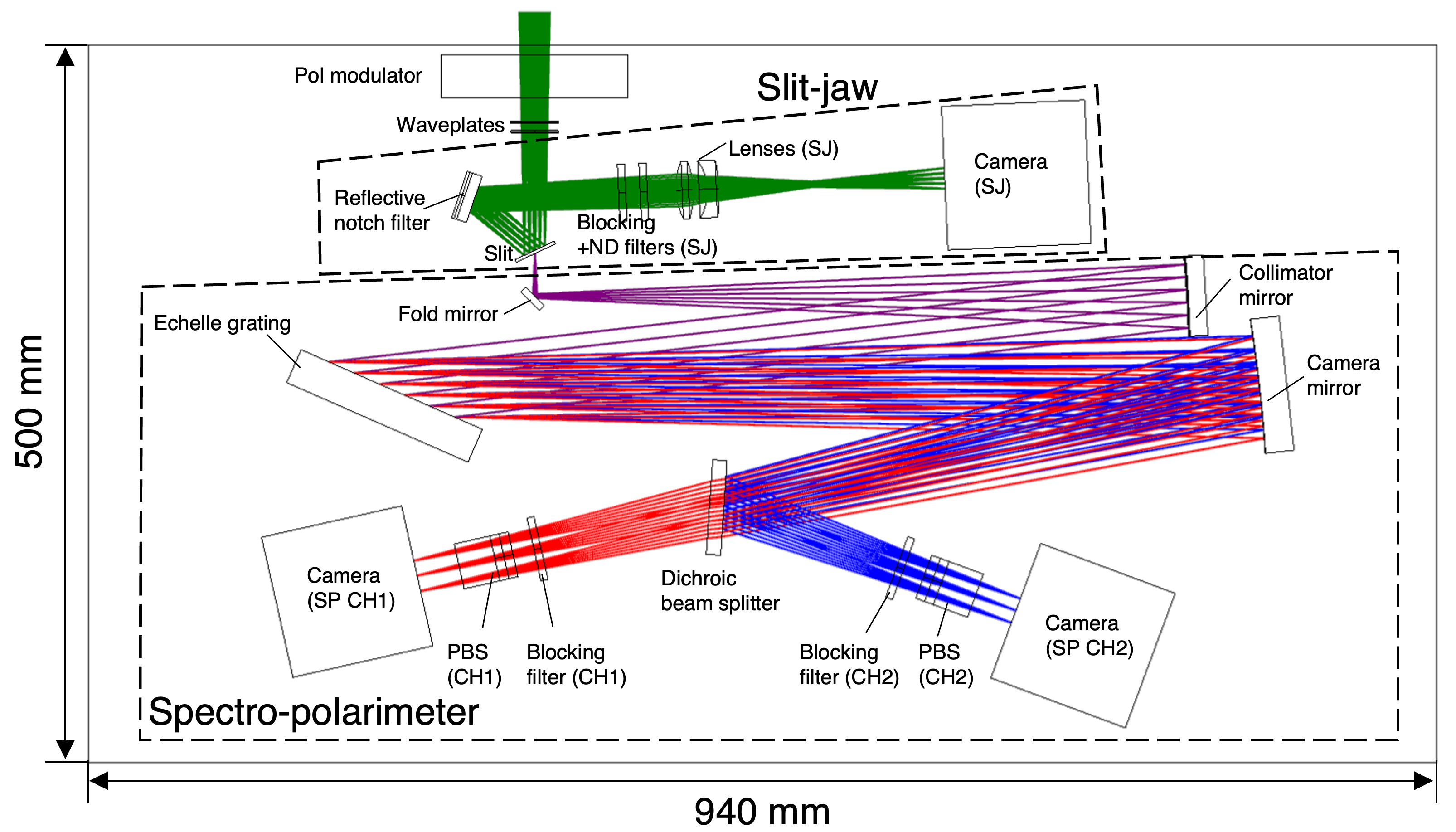}
\caption{Layout of the SCIP optics. The colors of the rays represent the following: (red) SP CH1, (blue) SP CH2, and (green) SJ. Modified from \cite{2020SPIE11447E..AJT}.}
\label{fig:optical}
\end{figure}

\begin{figure}[t]
\centering
\includegraphics[width=0.9\linewidth]{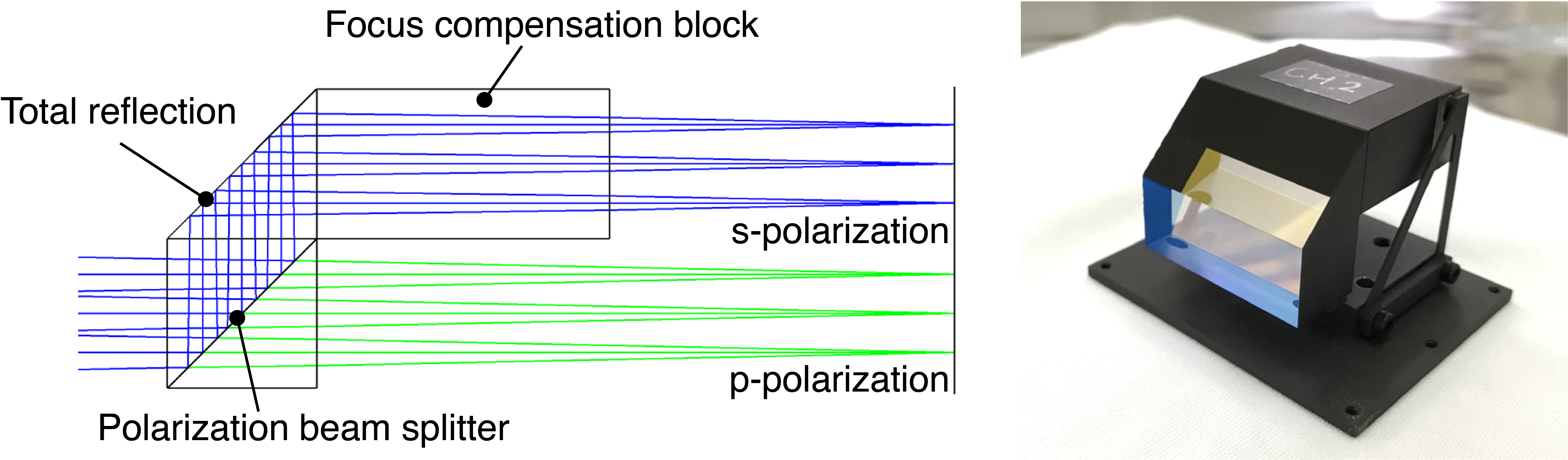}
\caption{(left) Optical design of the \acs{pbs}, using a cube beam splitter with a dielectric coating at the 45\degree\ surface, and with anti-reflection coatings on the entrance and exit surfaces. To adjust the focus position, a glass block is inserted in the optical path of the s-polarization that is reflected at the 45\degree\ surface. The thickness of the glass block is designed to make s- and p-polarizations in focus on the same plane. (right) Picture of the \acs{pbs} for the 770~nm wavelength band (CH2). Modified from \cite{2020SPIE11447E..AJT}.}
\label{fig:pbs}
\end{figure}

\textsc{Sunrise} employs a 1~m aperture telescope with a Gregorian design. It feeds light into \ac{islid}.
The \ac{islid} system of {\sc Sunrise i} and {\sc ii} \citep{2011SoPh..268...35G} \textbf{was} completely redesigned for \sunriset\ to accommodate the new instrument SCIP in addition to SUSI, TuMag, and CWS. The new \ac{islid}, described in \cite{2025SoPh..300...75K}, consists of two Offner relay optical systems \citep{smith2007modern} that \textbf{transfer} the Gregorian focus (F2) to the instruments. The first Offner system, consisting of \textbf{the} two spherical mirrors M5 and M6 feeds light into SUSI, TuMag, and CWS by dichroic beam splitters. M6 is used as a tip-tilt mirror for correlation tracking driven by CWS. The second Offner system, consisting of two spherical mirrors M7 and M8 with three folding mirrors (M9, M10, and M11), is used to feed the F/24.2 beam into SCIP. One of the spherical mirrors, M8, is mounted on a scan mirror mechanism (SMM) and is used by SCIP to scan a \ac{fov} perpendicular to the slit (see Section \ref{SS-smm}).

The SCIP optical layout is shown in Figure \ref{fig:optical} by the red and blue lines. The incident convergent light of F/24.2 first passes through the rotating waveplates and reaches the slit. The SP uses light passing through the slit. The light transmitted through the slit spreads out more in the direction perpendicular to the slit (i.e., dispersion direction) due to diffraction. Because the observing wavelength is in the near-infrared and longer than visible light, we need to take into account the effect of diffraction caused by the 11~$\mu$m-wide slit. In the optical design after the slit, we use the F/11 beam in the direction perpendicular to the slit, while the F/24.2 beam is used in the direction parallel to the slit. The energy contained in the F/11 beam is larger than 85\% at both 850~nm and 770~nm. 

The SP optics is based on the modified Czerny-Turner mounting comprising an echelle grating and two mirrors (collimator and camera mirrors) in a quasi-Littrow configuration. The light from the slit is collimated by the collimator mirror and \textbf{impinges on} the grating. The collimated light is diffracted by the echelle grating with a groove density of 110/mm, so that the two wavelength bands are observed simultaneously using different diffraction orders: the 19th order to observe the 850~nm wavelength band (CH1), and the 21st order to observe the 770~nm wavelength band (CH2). The grating has a blaze angle of 64\degree, but the angle of incidence to the grating was designed to be 59.6\degree\ to maintain the clearance between the collimator mirror and the camera mirror. The diffracted parallel light from the echelle grating becomes converging light by the camera mirror.  In order to achieve high imaging performance, the mirror surface shapes of the collimator and camera mirrors were determined by optimization. These surface shapes are similar to off-axis parabolas but include higher-order aspheric terms to sufficiently suppress the aberration \citep[see details in][]{2020SPIE11447E..AJT}. 

\begin{figure}[t]
\centering
\includegraphics[width=\linewidth]{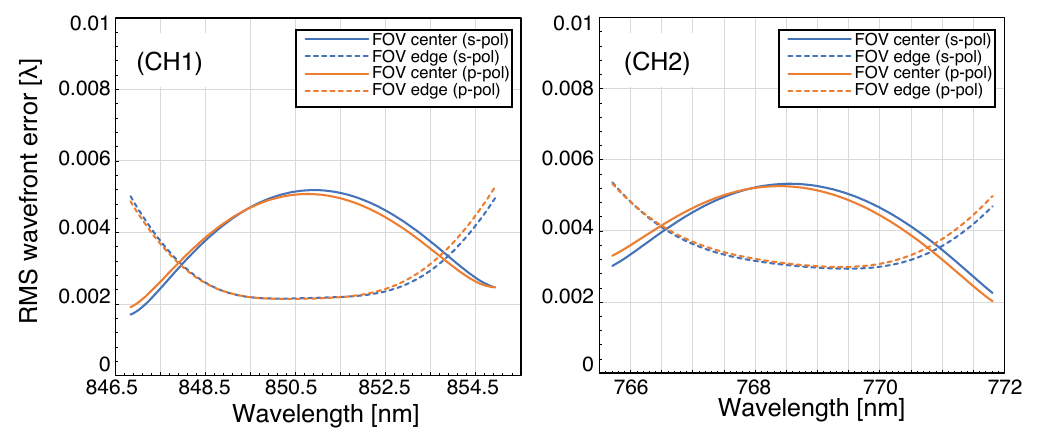}
\caption{RMS wavefront error of SP as a function of the wavelength for the 850~nm channel (CH1; left) and 770~nm channel (CH2; right). The plots show the wavefront errors at the center and edges of the \ac{fov} for both s- and p-polarizations. Modified from \cite{2020SPIE11447E..AJT}.}
\label{fig:waveerror}
\end{figure}

\begin{figure}[t]
\centering
\includegraphics[width=\linewidth]{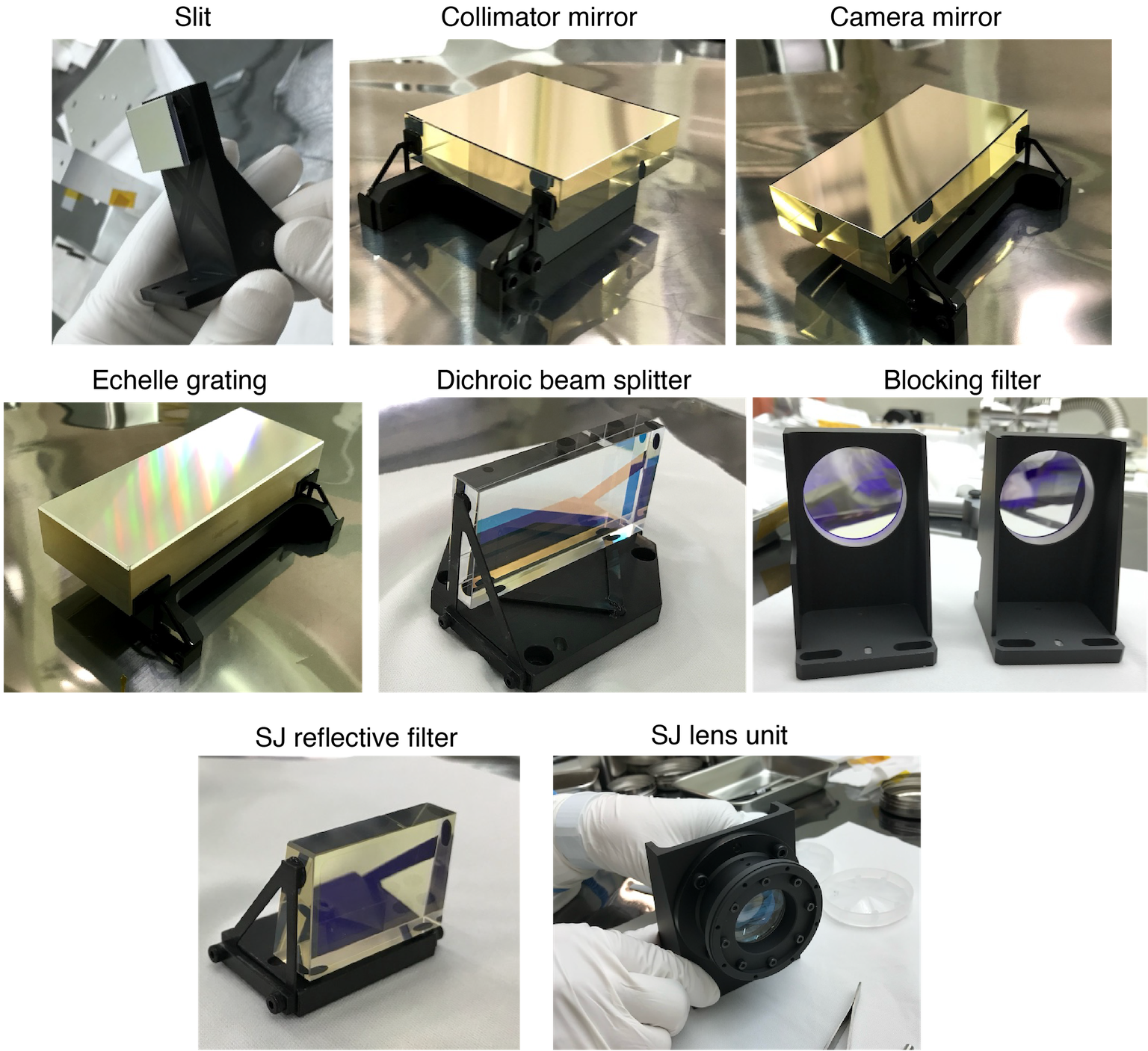}
\caption{Optical elements and their mounting structures in the SCIP O-unit.}
\label{fig:ocompo} 
\end{figure}

The two wavelength bands are split by a dichroic filter with a long-pass coating. In each channel, the required wavelength bands are extracted through a blocking filter. For simultaneous observation of two orthogonal linear polarizations, a \ac{pbs} is placed in front of each detector, which creates the orthogonal polarization images on each single detector (Figure \ref{fig:pbs}). The \ac{pbs} has an extinction ratio specification of more than 100, as verified by the vendor Nitto Optical. Figure \ref{fig:waveerror} shows RMS wavefront errors of the SP optics based on the optical design, which shows that the worst wavefront error is 4.6~nm rms as a design performance in the two channels. The total wavefront error, considering fabrication and alignment errors, is estimated \textbf{to be 48~nm rms from} a tolerance analysis \citep{2020SPIE11447E..AJT}. The optical parameters of the SP are summarized in Table \ref{tab:spoptics}. Pictures of the optical elements and their mounting structures are shown in Figure \ref{fig:ocompo}.

\begin{table}
\caption{SCIP SP optical parameters}
\label{tab:spoptics}
\begin{tabular}{ll}
\hline
\multicolumn{2}{l}{Basic features}\\
\hline
Spectrograph type&Modified Czerny-Turner in quasi-Littrow configuration\\
FOV & 11~$\mu$m (0.094", slit width)$\times$6.8~mm (58", slit length)\\
Wavelength range&(CH1): 846.830--854.930~nm (in vacuum, 19th order)\\
&(CH2): 765.708--771.808~nm (in vacuum, 21st order)\\
Plate scale (spatial)&0.0935"/pixel\\
Plate scale (dispersion)&(CH1) 42.3~m\AA/pixel at 846.6~nm (R=$2\times10^5$)\\
&(CH2) 38.5~m\AA/pixel at 765.5~nm\\
\hline
Opt. elements&Features and specifications\\
\hline
Waveplate\#1&Plane parallel plate (t=1.43~mm), quartz\\
Waveplate\#2&Plane parallel plate (t=1.64~mm),  sapphire\\
Slit&Width: 11~$\mu$m (corresponding to 0.0936")\\
     &Length: 6.8~mm (corresponding to 58")\\
     &Plane parallel plate (t=2~mm), fused silica\\
     &25\degree\ to the incident optical axis from \ac{islid}\\
Folding mirror&Planar, fused silica\\
Collimator mirror&Spherical plus XY polynomial, CLEARCERAM-Z\\
     &Curvature radius R=1060.6~mm\\
     &Protected silver coating\\
Grating&COTS grating by Richardson Gratings\textsuperscript{\texttrademark}, Newport\\
     &Planar constant-line-space grating, ZERODUR\\
     &Groove density 110/mm, blaze angle 64\degree\\
     &Angle of incidence 59.6\degree relative to the grating normal\\
     &Protected silver coating\\
     &Clear aperture: 46~mm$\times$130~mm\\
Camera mirror&Spherical plus XY polynomial, CLEARCERAM-Z\\
     &Curvature radius R=1060.8 mm\\
     &Protected silver coating\\
Dichroic beam splitter&Reflectivity$>98$\% in $\lambda=760 - 780$~nm \\
     &Transmission$>98.5$\% in $\lambda=840-860$~nm \\
     &Specified for an angle of incidence (AOI) of 12 -- 21$^{\circ}$ \\
     &\hspace{0.2cm}for both s- and p-polarizations.\\
     &Wedged plate (t=10~mm), fused silica\\
     &Wedge angle 0.176\degree\ relative to the front surface\\
Blocking filters&Bandpass:  10.9~nm around 850.75~nm (CH1), \\
    &\hspace{1.5cm}9.9~nm around 768.55~nm (CH2)\\
    &Plane parallel plate (t=5~mm), fused silica\\
PBSs&13~mm block prism, BSL7Y\\
     &Comp. block thickness 25.49~mm (CH1) and 25.41~mm (CH2)\\
Camera sensor&2048$\times$2048 pixels, 11 $\mu$m/pixel\\ 
\hline
\end{tabular}
\end{table}

\subsubsection{SP throughput for polarimetric measurements}

To achieve a high polarization sensitivity, we have to collect sufficient photons within the integration time. When considering (1) the reflectances and transmittances of all the optical components, including the telescope, \ac{islid}, and the \ac{scip} O-unit, and (2) the quantum efficiency of the image sensor, the total efficiency of the system is 2.4\% and 4.8\% in the 850 and 770~nm bands, respectively. The worse efficiency at 850~nm is because the mirrors in the telescope and the first half of \ac{islid} have an aluminum coating on the reflecting surfaces to observe the UV wavelengths \citep{2025SoPh..300...65F}, but this coating has lower reflectivity around 850~nm. The image sensor (see Section \ref{SS-cameras}) is also less sensitive to 850~nm. The number of detected electrons is estimated to be $1.1\times10^6$~$\rm{e}$/(pixel$\cdot$s) and $2.3\times10^6$~$\rm{e}$/(pixel$\cdot$s) at continuum wavelenghts around 850 and 770~nm, respectively, in a solar disk center observation. Because the full-well capacity of the image sensor is $\sim9\times10^4$~$\rm{e}$, each exposure should be shorter than 32~ms. To achieve a polarimetric sensitivity of 0.03\% (1$\sigma$) of the continuum intensity, a 10~sec image accumulation, combining s- and p-polarization beams, and \textbf{binning over two pixels} (0.19” sampling) along the slit are required in the 850~nm band, given the polarization modulation efficiency, whereas no summing is needed in the 770~nm band.

\subsubsection{SJ optics}
The SJ optical layout is shown in Figure \ref{fig:optical} with the green lines. The SJ observes light reflected by the slit plane, which first \textbf{reaches} a reflective filter \textbf{with} a notch filter coating to reflect the wavelength of 770.5$\pm$30~nm (FWHM) on the front side and transmits unwanted wavelengths. The transmitted unwanted light has a large energy compared to the reflected light and is attenuated by a black baffle around the filter. The rear surface of the reflective filter has a wedge to deflect the unwanted light. The light reflected from the reflective filter is then passed through an SJ blocking filter to extract only the required wavelength band of the continuum wavelength at 770.7$\pm$0.5~nm (FWHM). The light passing through the filter then enters the doublet lens, which forms a $1\times$ magnified image of the slit plane on the detector. Since the SJ has a sufficiently narrow wavelength band, the doublet lens is able to achieve adequate imaging performance with the same glass material, instead of multiple materials required for chromatic aberration correction. We selected BSL7Y (OHARA Inc.) as the lens material. The \ac{fov} is 64"$\times$64", with a pixel scale of 0.0936"/pixel. The optical parameters of the SJ are shown in Table \ref{tab:sjoptics}. The worst wavefront error is 3.5~nm rms as a design performance in the \ac{fov}. The total wavefront error, considering fabrication and alignment errors, is estimated to be 34~nm rms by the tolerance analysis \citep{2020SPIE11447E..AJT}. 

\begin{table}
\caption{SCIP SJ optical parameters}
\label{tab:sjoptics}
\begin{tabular}{ll}
\hline
\multicolumn{2}{l}{Basic features}\\
\hline
Imaging optics&Doublet (air-spaced two lenses)\\
FOV&7.6~mm$\times$7.6~mm (64"$\times$64")\\
Wavelength band&770.208--771.208~nm (in vacuum)\\
Plate scale&0.0936"/pixel\\
Magnification&1.0\\
\hline
Opt. elements&Features and parameters\\
\hline
SJ reflective filter& Bandpass: 770.5$\pm$30~nm (FWHM)\\
  &Wedged plate (t=9~mm), CLEARCERAM\\
  &9.0\degree\ wedge relative to the front surface\\
SJ blocking filter& Bandpass: 770.5$\pm$0.5~nm (FWHM)\\
  &Plane parallel plate (t=5~mm), BK7\\
SJ ND filter&Plane parallel plate (t=5~mm), BK7\\
SJ lens\#1&Bi-convex spherical lens (t=12~mm), BSL7Y\\
  &Curvature radii R1=48.646~mm, R2=62.758~mm\\
SJ lens\#2&Meniscus spherical lens (t=8~mm), BSL7Y\\
  &Curvature radii R1=36.294~mm, R2=112.642~mm\\
Camera sensor&Same as SP\\
\hline
\end{tabular}
\end{table}

\subsection{Opto-mechanical design}
\label{SS-optmecha}
Figure \ref{fig:cad} shows the opto-mechanical layout of the SCIP O-unit. All optical components are mounted on an optical bench consisting of a carbon fiber reinforced plastic (CFRP)-aluminum-honeycomb sandwich panel. To meet the strict wavefront error requirements for high spatial and spectral resolution, the optical elements are held by supporting mechanisms that minimize deformation of the optical surfaces. The CFRP sandwich panel offers significant advantages in low thermal deformation, weight and stiffness. The \ac{cte} of the CFRP sandwich panels was measured using a test piece of the panel, yielding $-0.15$~ppm/K and $0.4$~ppm/K in the two in-plane directions. Deformation due to moisture desorption was also measured and was found to be less than 5~ppm over 7~days. Note that the \ac{cte} in the out-of-plane direction does not need to be considered because it does not affect the relative position and orientation of the optics on the bench \citep{2020SPIE11447E..ABU}.

\begin{figure}
\centering
\includegraphics[width=\linewidth]{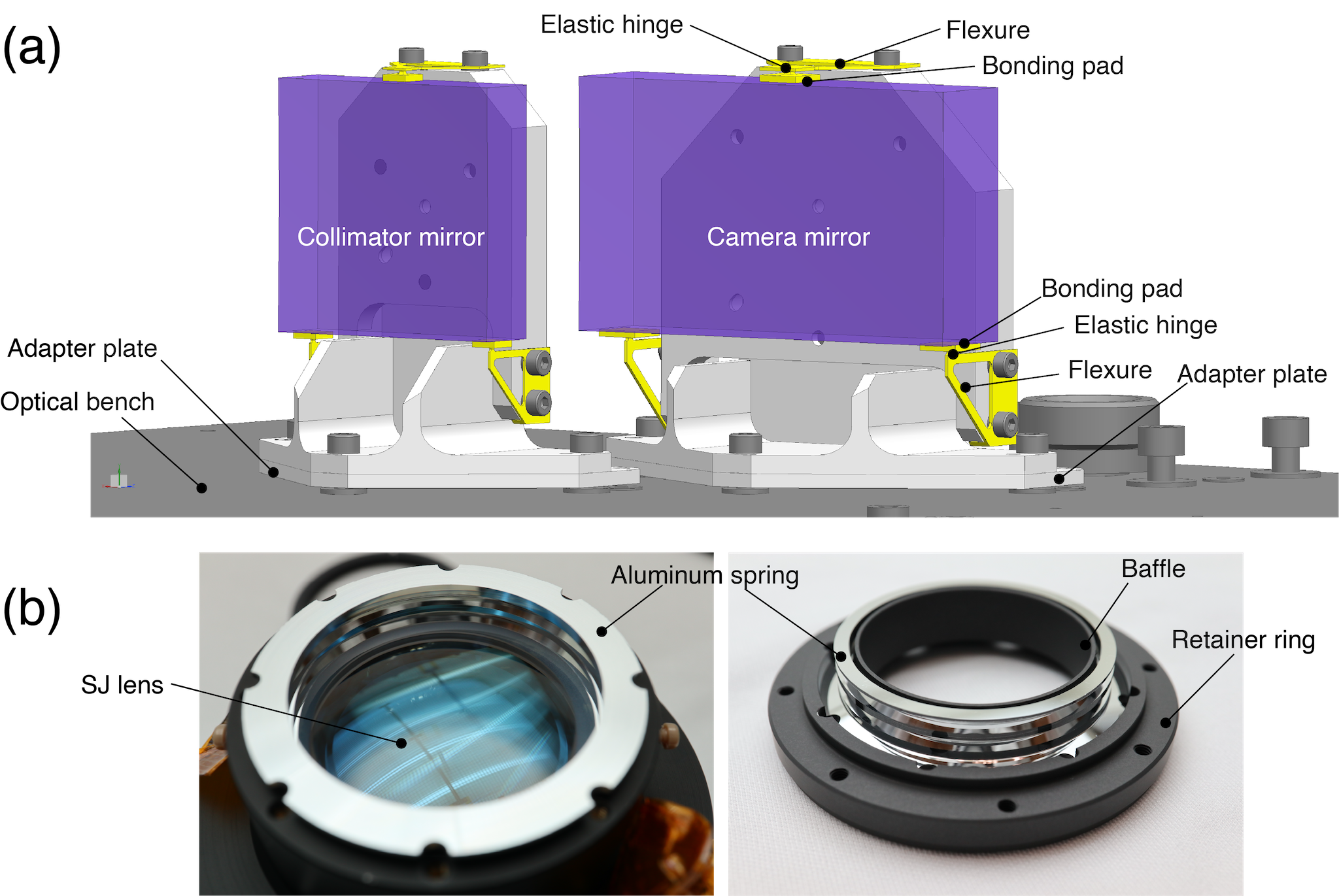}
\caption{(a) Illustration of the mirror support mechanisms for the collimator and camera mirrors. (b) Aluminum spring inserted into the SJ lens barrel (left) and the retainer ring (right). Modified from \cite{2020SPIE11447E..ABU}.}
\label{fig:mirrorlenssupport}
\end{figure}

The reflective and diffractive optical elements (collimator and camera mirrors, and grating) are supported by a quasi-kinematic mount consisting of flexures at its top and bottom surfaces, as shown in Figure \ref{fig:mirrorlenssupport}(a). The thickness of the flexure is set to 0.5 to 0.9~mm, i.e. as thin as possible to avoid inducing surface deformation while meeting the required mechanical tolerances. The flexure includes a monolithically machined bonding pad connected by elastic hinges with cross-sections ranging from $0.5\times0.5$~mm square to $1\times1$~mm square to provide the appropriate rotational degree of freedom. The optical element is bonded to the pad with an adhesive. The adhesive areas range from $6\times6$~mm square to $8\times20$~mm rectangular to adapt to the sizes and the masses of the substrates, and the adhesive layer thickness is commonly set to $0.15\pm0.05$~mm. The mirror substrate is also designed to minimize the surface deformation due to mounting. The thickness of the substrate is set to greater than $1/6$ of the diagonal dimension of the substrate to ensure a sufficient area on the side faces for bonding. In addition, 5~mm margins are provided on the adhesive side at the top and bottom surfaces to reduce the effect of adhesion on the clear aperture. The material of the flexure pad is selected to match the \ac{cte} of the substrate material to the extent possible. The super invar alloy is used for CLEARCERAM-Z, fused silica, and ZERODUR, and the titanium alloy is used for BSL7Y.

The refractive optical elements, such as waveplates, SP and SJ blocking filters, and SJ lenses, are held by a preloaded spring made of \ac{ptfe} or aluminum \textbf{(Figure \ref{fig:mirrorlenssupport} [b])}. The springs are not only to apply the preload to avoid separation at the expected accelerations but also to absorb the thermal strain difference between the mounting structure and the substrate. Figure \ref{fig:mirrorlenssupport}(b) shows a metal spring made of aluminum alloy 6061 used for the SJ lenses. This aluminum spring has a 0.15~mm thick bellows-like cross-section that provides good mechanical stability and thermal conductivity compared with the \ac{ptfe} springs. The mounting structures fabricated are shown in Figure \ref{fig:ocompo}.  

\begin{figure}[t]
\centering
\includegraphics[width=\linewidth]{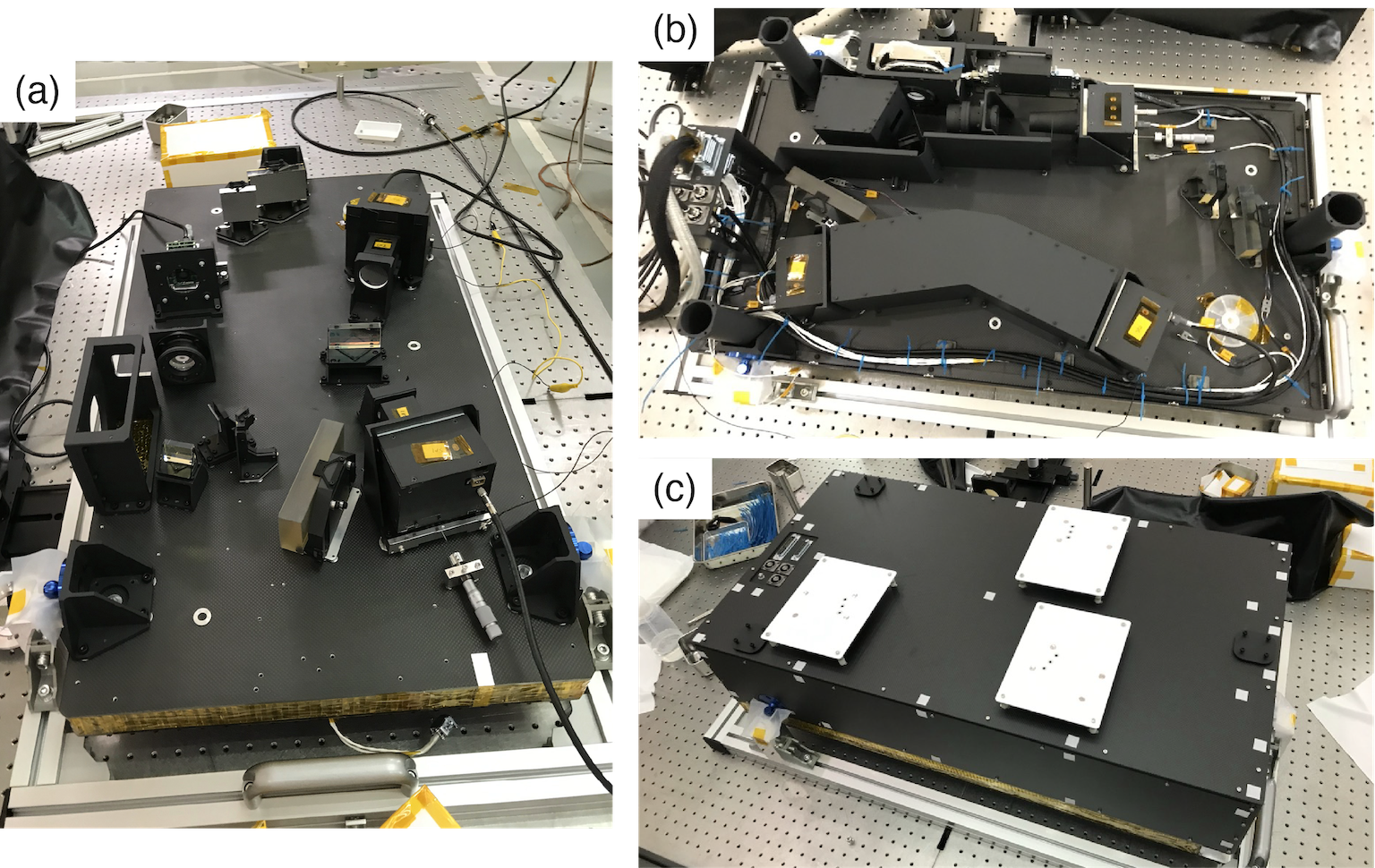}
\caption{Pictures of the SCIP O-unit (a) before installing the light trap structures, in which the SJ camera is not installed yet, (b) after installation of the light-trap structure around the SP cameras and SJ reflective filter as well as the SJ camera baffle, and (c) after installation of the CFRP enclosure and the camera radiators. }
\label{fig:ounit}
\end{figure}

Figure \ref{fig:ounit} shows pictures of the SCIP O-unit at various stages of the integration process. Some optical elements and paths are covered by the light-trap structures to suppress stray light in the O-unit as specified in \cite{2020SPIE11447E..AJT}. Blackening by plasma electrolytic oxidation (PEO) is applied to the aluminum light-trap structures. The O-unit is enclosed by a thin CFRP plate to minimize thermo-elastic deformation while maintaining sufficient stiffness and reducing weight. The SCIP O-unit has a mass of 27~kg in its flight configuration.

Integration of the SCIP O-unit into the \ac{pfi} is done through the three mount interface devices, which provide point fixation, line fixation, and flat fixation to constitute a kinematic mount that ensures the appropriate degrees of freedom to avoid over-constraining the O-unit. The point fixation device is located at the point closest to the optical interface, the line fixation device is aligned along the incident optical axis, and the flat fixation is applied to the remaining interface point.  

To demonstrate that the opto-mechanical design of the O-unit meets the required wavefront error budget, it is necessary to quantify the errors due to environmental effects such as gravity and temperature variation under the observation conditions. We conducted an integrated opto-mechanical analysis by (1) incorporating mechanical and thermal disturbances into a finite element model of the O-unit to obtain the nodal displacements of each optical element, (2) feeding them into the optical analysis in the form of rigid-body displacement and surface deformation fitted by polynomials, and (3) analyzing wavefront errors of the SP and SJ optical systems \citep{2020SPIE11447E..ABU, uraguchi2023jaxarr}. The analyses showed that mounting and environmental effects caused wavefront errors of less than 2~nm rms and 8~nm rms, respectively, for \textbf{the} SP, which \textbf{is} well within the wavefront error budget provided by \cite{2020SPIE11447E..AJT}. The opto-mechanical effects on wavefront errors were much smaller for \textbf{the} SJ than for \textbf{the} SP. The analysis also explained changes of the image position observed in the thermal vacuum optical test (see Section \ref{SSS-otv}).

\subsection{Polarization modulation}
\label{SS-polari}

\begin{figure}[t]
\centering
\includegraphics[width=\linewidth]{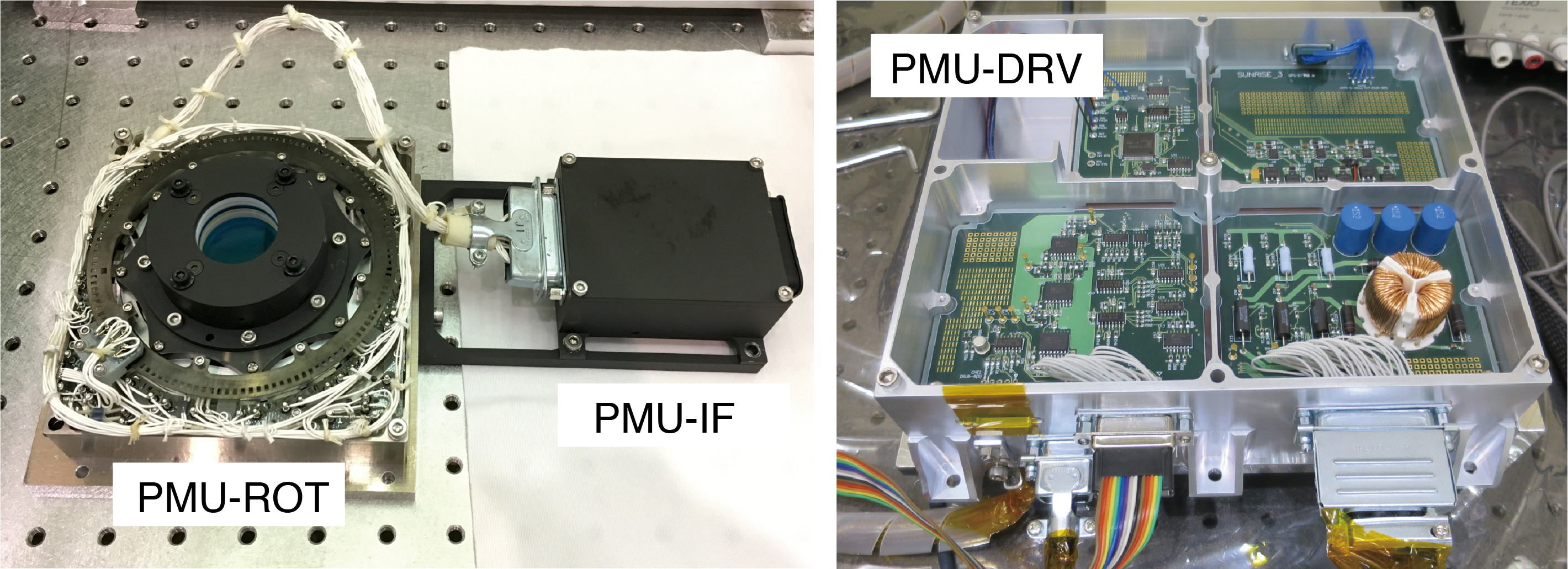}
\caption{(left) The rotating mechanism of the waveplate (\ac{pmu}-ROT) and its filter electronics (\ac{pmu}-IF) which are installed in the SCIP O-unit.  (right) The driver electronics (\ac{pmu}-DRV) for the \ac{pmu}-ROT, which is installed in the gondola E-rack. Modified from \citet{2020SPIE11447E..A3K}.}
\label{fig:pmu}
\end{figure}

The polarization measurements are done by a combination of a rotating waveplate, \acp{pbs} (Section \ref{SS-opt}), and cameras (Section \ref{SS-cameras}) in the SCIP O-unit. The SCIP O-unit uses a polarization modulation unit (PMU) to continuously rotate a waveplate at a constant speed. The \ac{pmu} consists of (1) a rotating mechanism (\ac{pmu}-ROT), (2) its driver electronics (\ac{pmu}-DRV), and (3) filter electronics (PMU-IF), as shown in Figure \ref{fig:pmu}. The \ac{pmu}-ROT and \ac{pmu}-IF are installed in the SCIP O-unit as one of the components. The PMU-DRV is installed in the gondola E-rack (Figure \ref{fig:diagram}). Table \ref{tab:pmu} summarizes the characteristics of the PMU. The rotating mechanism \ac{pmu}-ROT uses a DC brushless motor which was originally developed as a breadboard mechanism to evaluate its performance for future solar observations by a large-aperture optical telescope on a spacecraft \citep{2014SPIE.9151E..38S}. The mechanism provided a successful supreme performance exceeding 10 million times of continuous rotations in a lifetime testing under a high vacuum thermal environment over three years. After the lifetime test, it \textbf{was} adopted for the polarization modulation on the SCIP instrument \citep{shimizu2018jaxarr}. Thus, the mechanism may not be optimally designed, as the diameter of the hollow aperture (68~mm) is much larger than necessary for the beam size entering the waveplate. The \ac{pmu}-IF is added to reduce noise in the encoder signals for the rotational phase from the \ac{pmu}-ROT. Since the spatial resolution of SCIP is approximately 10 times higher than that of the CLASP rocket experiment, which used the same type of rotating mechanism \citep{2015SoPh..290.3081I}, the rotation speed of the \ac{pmu}-ROT must be approximately 10 times faster than that of CLASP (4.8 sec/rotation) to suppress polarization crosstalk due to platform jitter when observing the fine-scale dynamics in the solar atmosphere. The driver electronics PMU-DRV \textbf{was} newly developed to make it optimized for the rotation speed of 0.512 sec/rotation. Another set of the same rotating mechanism and its driver electronics \textbf{was} manufactured and installed in the \ac{susi} for UV spectropolarimetry \citep{2025SoPh..300...65F}.

\begin{table}[t]
\caption{Basic features of \acs{pmu} mechanism and the waveplate}
\label{tab:pmu}
\begin{tabular}{ll}
\hline
Mechanism\\
\hline
Rotating mechanism&DC brushless motor\\
Waveplate rotation speed&0.512 sec/rotation\\
Number of exposures&16 exposures/rotation\\
Angle error&$\pm1$\degree\\
\hline
Waveplate\\
\hline
Material&quartz and sapphire\\
Wavelength&762--775~nm, 842--859~nm\\
Retardation&127\degree$\pm3$\degree\\
Diameter&$\phi$30~mm\\
Clear aperture&$\phi$27~mm\\
Parallelism&2” for each plate\\
Transmission wavefront error&$\lambda/15$ PV for each plate\\
AR Coating&Reflectivity $<1$\% on both sides\\
\hline
\end{tabular}
\end{table}

\begin{figure}
\centering
\includegraphics[width=\linewidth]{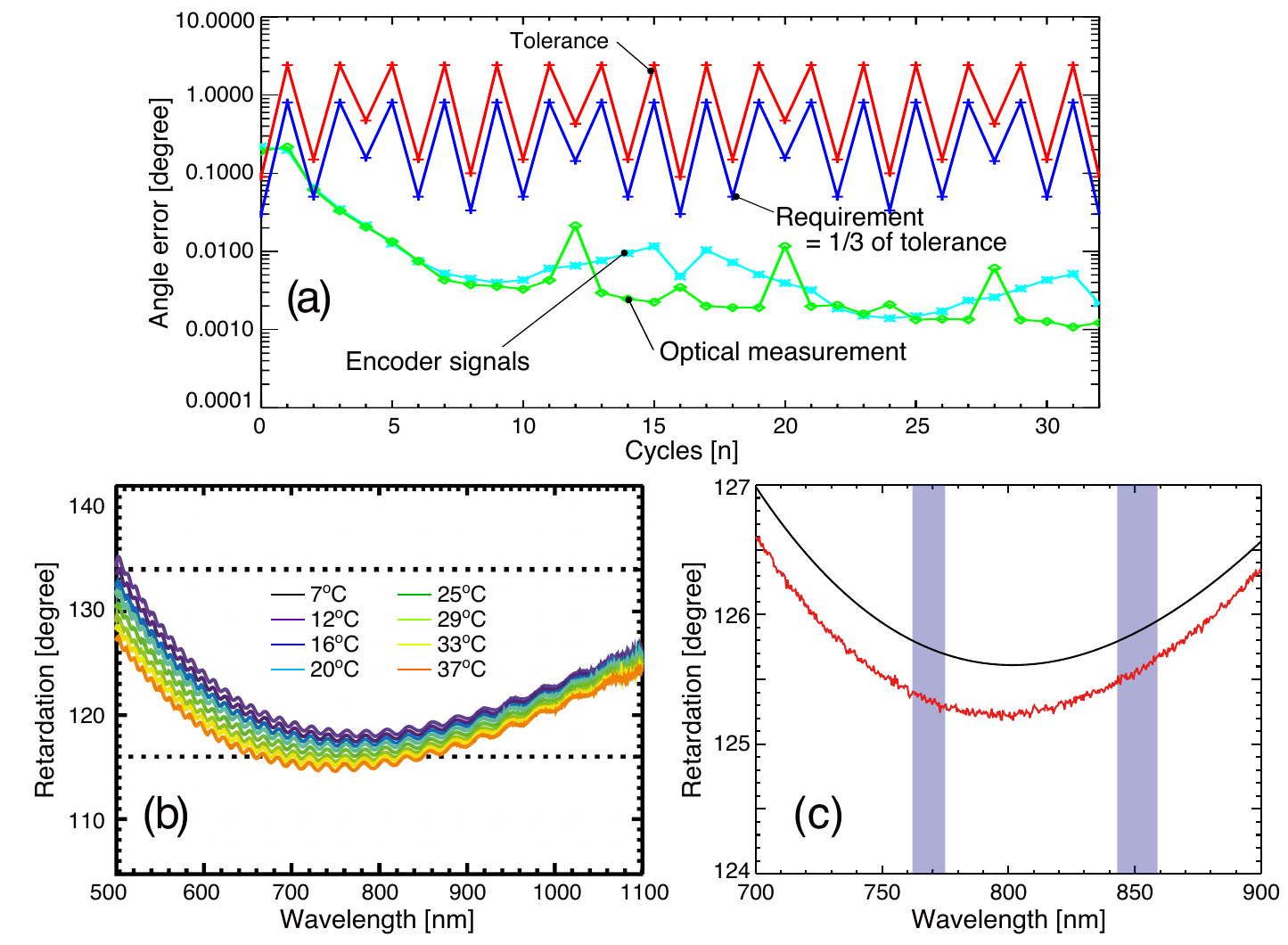}
\caption{(a)Rotation angle error as a function of the cycle in one rotation, in which the red and blue lines show tolerance coming from the required polarization accuracy and requirements allocated to each cycle component defined by $1/3$ of the tolerance \citep{2020SPIE11447E..A3K}. The green and light blue lines show optical and encoder measurements of the angle errors, respectively. (b) Temperature dependence of the retardation angle for a prototype waveplate consisting of quartz and sapphire plates, with temperatures indicated by different colors. (c) Wavelength dependence of the retardation for the flight waveplate, in which the black and red lines show designed and measured (at 23\degree C) retardation, respectively. Modified from \citet{2020SPIE11447E..A3K}.}
\label{fig:pmuerr-retardation}
\end{figure}

The camera operations are synchronized with phase signals sent from the \ac{pmu}-DRV, performing an exposure every 22.5\degree\ in modulator angle and having 16 images \textbf{during} one rotation of the waveplate. The full Stokes vector is obtained by onboard polarization demodulation by the SCIP E-unit, which \textbf{uses} an integration of the acquired images multiplied by coefficients according to the rotational phase of the waveplate \citep{2023JATIS...9c4003K}. If the waveplate does not rotate uniformly or the camera exposure is out of synchronization, polarization errors are induced by the difference in the rotational phase between the polarization modulation and polarization demodulation. We developed the control logic in the PMU-DRV to achieve the required rotation uniformity with a speed of 0.512 sec/rotation. \textbf{The} polarization modulation was optically measured by mounting a non-flight half-waveplate simulating the same mass and inertia, demonstrating that the rotation angle errors derived were within $\pm1$\degree\ by the encoder signals and the optical measurement. Figure \ref{fig:pmuerr-retardation} (a) shows the frequency dependence of the rotation angle errors averaged over 500 rotations to verify that the rotation angle error at each cycle is within the requirement determined by the polarization accuracy. The angle errors are below the requirement except for $n=0$ and 2 cycles. The rotation angle error at $n=2$ cycles exceeds the requirement but is within the tolerance. The polarization error due to $n=0$ cycles appears as a crosstalk between Stokes Q and U and can be corrected using housekeeping encoder data even after the onboard demodulation. Details of the tests and their results are provided in \cite{2020SPIE11447E..A3K} and \cite{kubo2021jaxarr}.

As stated in Section \ref{SS-opt}, the waveplate used in SCIP consists of a quartz (\ce{SiO2}) and sapphire (\ce{Al2O3}) plates. The combination \textbf{was} selected because it can cancel the wavelength dependence and temperature dependence of the retardation \textbf{over} a wide wavelength range. The thickness of the quartz and sapphire plate is designed so that the retardation is close to 127\degree\ to have good polarization modulation efficiency for both the circular and linear polarization. The two plates are independently mounted in the waveplate holder by \ac{ptfe} springs, considering different \ac{cte}. Anti-reflective coatings are applied on both sides to avoid multiple reflections between the two plates. Table \ref{tab:pmu} summarizes the specifications of the waveplate. The temperature dependence of the retardation was measured using a prototype of the waveplate consisting of the same material \citep{2020SPIE11447E..A3K}. Figure \ref{fig:pmuerr-retardation} (b) shows the measured retardation as a function of the wavelength at various temperatures from 7\degree C to 37\degree C, demonstrating that the change in the retardation is as small as $\pm1$\degree\ when the temperature is maintained at $20\pm5$\degree C (see Section \ref{SS-oth} for details on O-unit thermal control). In the flight waveplate, the optical axes of the quartz and sapphire plates \textbf{were} carefully aligned by rotating one of the plates to suppress ripples seen in the prototype waveplate. After the final adjustment, the retardation of the flight waveplate is almost constant at 125.5\degree\ in the two wavelength bands as shown in Figure \ref{fig:pmuerr-retardation} (c). The beam wobbling by the waveplate rotation \textbf{was} minimized by adjusting the tilt of the waveplate relative to the rotating axis of the mechanism using shims and is smaller than 0.01\arcsec~\ac{rms}.

\subsection{Scan mirror}
\label{SS-smm}

\begin{figure}[t]
\centering
\includegraphics[width=\linewidth]{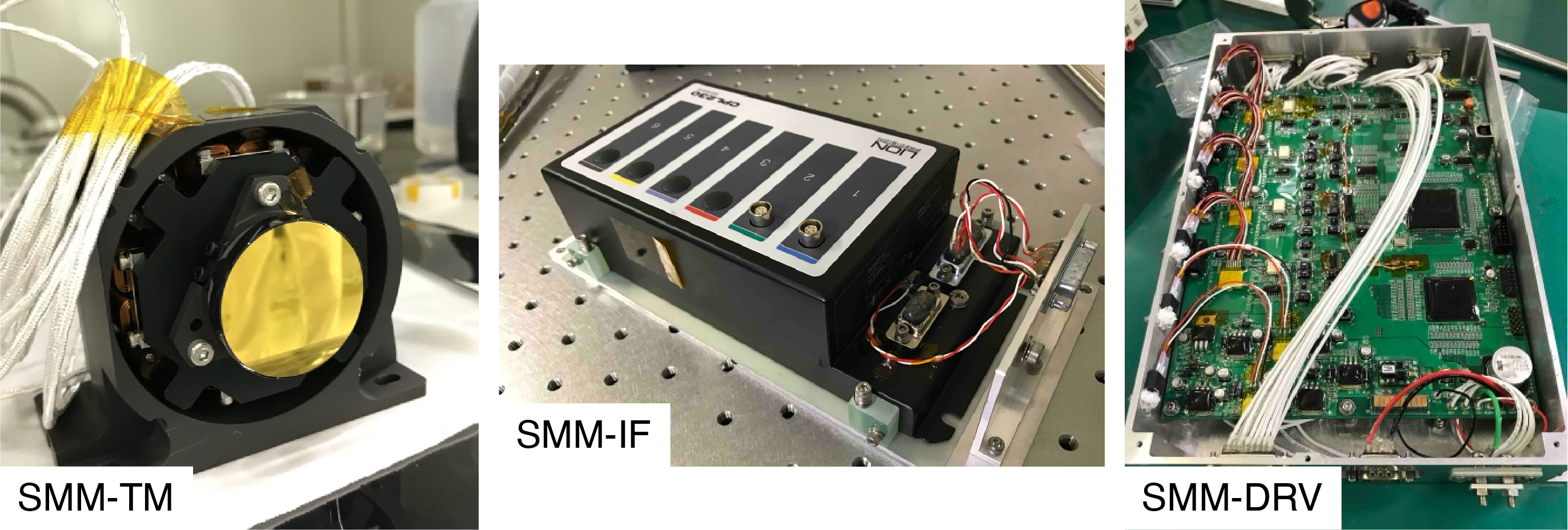}
\caption{Tip-tilt mirror mechanism (\ac{smm}-TM) attached with the flight mirror with gold coating, electronics unit precisely reading gap-based capacitance sensors (\ac{smm}-IF), and its driver electronics (\ac{smm}-DRV). Modified from \cite{2022SoPh..297..114O}.}
\label{fig:smm}
\end{figure}

\begin{table}[t]
\caption{Basic features of SMM}
\label{tab:smm}
\begin{tabular}{ll}
\hline
Mechanism\\
\hline
Tilt sensor&Gap-based capacitance sensor (2 sensors)\\
Actuator&Electromagnetic actuator (4 actuators)\\
Control frequency&Closed-loop control with $>90$~Hz\\
Step resolution&0.0936" sky angle (2.857" mechanical angle)\\
Scanning range&$\pm32.9"$ sky angle ($\pm1005$" mechanical angle)\\
Stability ($3\sigma$) &0.033" sky angle (1.0"  mechanical angle)\\
Linearity&$<0.15$\%\\
Step speed&$<31.3$~ms, nominal 20~ms\\
\hline
Mirror\\
\hline
Material&fused silica\\
Diameter&$\phi$35~mm\\
Clear aperture &$\phi$25~mm\\
Shape&convex sphere, curvature radius 425.4~mm\\
Coating&protected Ag coating\\
\hline
\end{tabular}
\end{table}

Because the \ac{scip} O-unit provides a single-slit spectrograph, a scan mirror mechanism (SMM) is required to map a two-dimensional \ac{fov} by tilting a mirror. The \ac{smm} consists of three units: a tip-tilt mirror mechanism (\ac{smm}-TM), electronics that control a mirror tilt (\ac{smm}-DRV), and an electrical unit reading the output signals from gap-based capacitance sensors installed in the tip-tilt mirror (\ac{smm}-IF), as shown in Figure \ref{fig:smm} \citep{2020SPIE11445E..4FO,2022SoPh..297..114O}. The SCIP E-unit transmits a tilt command to the \ac{smm}-DRV that drives a tip-tilt motion of the \ac{smm}-TM. The \ac{smm} operations are synchronized by the E-unit using the PMU phase signals, as well as the camera operations (Section~\ref{SS-polari}). The E-unit triggers a camera exposure 32~ms later than the timing of the tip-tilt command to avoid an exposure during mirror motion. This means the mirror tilt has to be moved and stabilized within 32 ms after the tilt command. Table \ref{tab:smm} summarizes the basic features of the \ac{smm}. The \ac{smm}-TM and \ac{smm}-IF are installed not in the SCIP O-unit but in the \ac{islid} optics because the \ac{smm}-TM uses a convex spherical mirror located at a pupil position in the Offner relay-optical system feeding and focusing a solar image onto the SCIP slit \citep{2025SoPh..300...75K}. The \ac{smm}-IF has to be located close to the \ac{smm}-TM. The \ac{smm}-DRV is mounted on the gondola E-rack as with the \ac{pmu}-DRV. 

In the \ac{smm}-TM, a mirror is mounted on a stage supported by two sets of two flexural pivots, \textbf{allowing for a} two-dimensional tilt. The scan motion is performed only in one direction perpendicular to the slit in a standard observation, but the other direction can be used for coalignment purposes between \ac{islid} and \ac{scip} and with the other on-board instruments. To precisely control the tilt of the mirror, a closed-loop control is implemented, which requires feedback to measure the tilt angles. Gap-based capacitance sensors (CPL230, Lion Precision) are installed behind the mirror of the SMM-TM. The output of the sensors is processed and immediately fed back to drive the electro-magnetic actuators with a control frequency larger than 90~Hz (crossover frequency).

\begin{figure}[t]
\centering
\includegraphics[width=\linewidth]{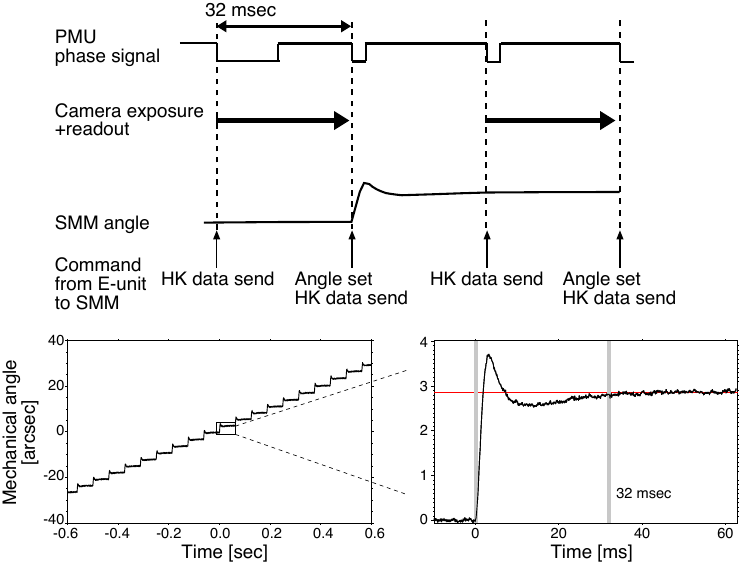}
\caption{(top) Schematic illustration showing timing relationship among the \ac{pmu} phase signals, camera exposures plus readouts, commands from the E-unit to SMM, and the SMM tilt angle. Note that this is the case for the fastest scan mode, where the camera exposure and readout are completed in 32 ms. (bottom) Test result of the scan motion in the fastest scan mode (64 ms step interval), measured by a position sensitive detector (PSD). A camera exposure starts from the gray line marked with 32 msec. Modified from \cite{2022SoPh..297..114O}.}
\label{fig:smm-motion}
\end{figure}

Figure \ref{fig:smm-motion} (top) illustrates the time sequence among the SMM tilt angle, camera exposures plus readouts, and the \ac{pmu} phase signals for the non-polarimetric, rapid mode (see Section~\ref{SS-sync_test} for details). As mentioned above, the \ac{smm} tilts the mirror toward the next position when receiving a tip-tilt command from the E-unit within 32~ms during non-exposure periods. The performance of the \ac{smm} was optically evaluated by installing a non-flight flat mirror of the same size and weight as the flight spherical mirror. The flat mirror allowed measuring tilt angles using an auto-collimator and a position sensitive detector (PSD). The optical measurements confirmed that the SMM fulfilled (1) the scan-step uniformity (linearity of 0.08\%) across the wide scan range ($\pm1005$" in the mechanical angle), (2) high stability of 0.1" ($3\sigma$) in the mechanical angle, and (3) fast stepping speed shorter than 26~ms, as shown in Figure \ref{fig:smm-motion} (bottom). Details of the SMM test configuration and results are shown in \cite{2022SoPh..297..114O}. After verification of the SMM performance, the non-flight flat mirror was replaced with the flight spherical one.

\subsection{Cameras}
\label{SS-cameras}

\begin{figure}[t]
\centering
\includegraphics[width=0.8\linewidth]{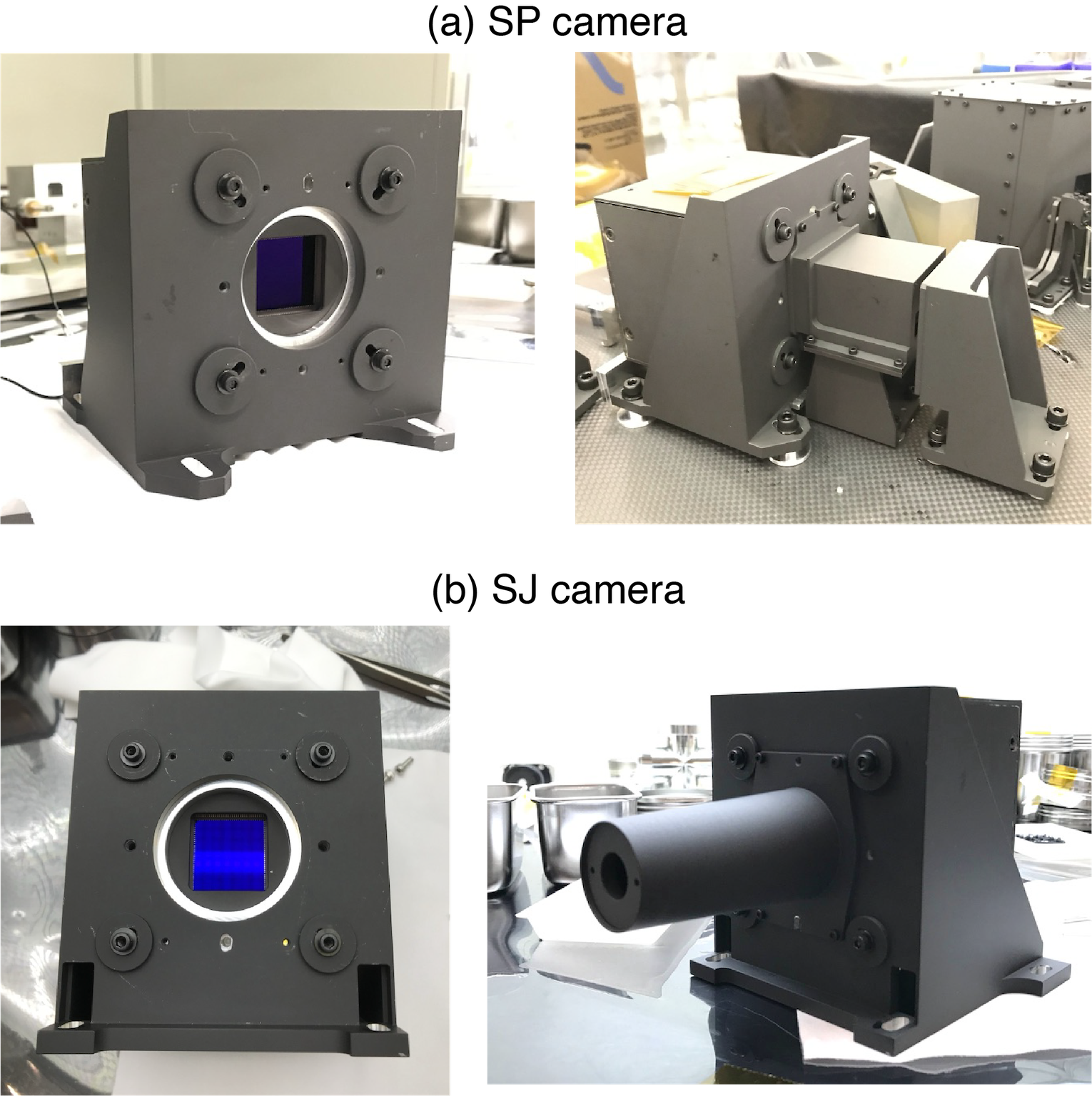}
\caption{(a) One of the two SP cameras mounted on the camera bracket. A camera baffle structure was installed to enclose the optical path between \ac{pbs} and the camera. (b) SJ camera mounted on the camera bracket. The SJ camera also has a camera baffle to suppress unwanted stray light.}
\label{fig:cameras}
\end{figure}

Specific, custom cameras were designed and manufactured for both the TuMag and SCIP instruments aboard \sunriset\ \citep{2023FrASS..1067540O}, based on the back-thinned CMOS sensor GSENSE400BSI (Gpixel Inc.). In both instrument cameras, the detector has $2048\times2048$ pixels with an 11 $\mu$m pixel size, which, in the case of SCIP SP, images the $\sim8.5$~nm (CH1) and $\sim7.5$~nm (CH2) wide spectrum of a one-dimensional, 58\arcsec-long region of the Sun with 0.0935"/pixel spatial sampling for both orthogonal polarization states. The SJ camera has a wider \ac{fov} coverage than 60\arcsec$\times$60\arcsec. The exposure time is adjustable up to 32~ms and depends on the SCIP observing modes (Section~\ref{SS-sync_test}). The quantum efficiency of the sensor is as high as 60\% and 40\% at 770~nm and 850~nm, respectively. The sensor works in the rolling shutter configuration. The rolling shutter direction is selected to achieve a synchronous exposure in orthogonal polarization states at a specific wavelength in the SP observation, while an exposure is not exactly simultaneous in the wavelength direction. The scientific and system-engineering requirements for the SCIP cameras are summarized in Table~\ref{tab:tablecam}. Figure \ref{fig:cameras} shows pictures of the SP and SJ cameras. The SP and SJ cameras have \textbf{adopted horizontal and vertical rectangular shapes}, respectively, as a result of a 90\degree\ rotation in the rolling shutter direction.

   \begin{table}
      \caption[]{Main camera scientific and system-engineering requirements}
         \label{tab:tablecam}
      \begin{minipage}{\textwidth}
         \begin{tabular}{llll}
            \hline
            {\rm Name}      &  {\rm Symbol}   &   {\rm Value}   &  {\rm Units (Type)}\footnote{Units are per pixel where necessary.} \\
            \hline
            Sensor's quantum efficiency   &   $Q$   &   $\geq 0.8$ \\
            & &\multicolumn{2}{l}{(peak in the visible light)} \\
            & &$\geq 0.4$ \\
            & &\multicolumn{2}{l}{(SCIP wavelength ranges)}\\
            {\rm Pixel full well}   &  & $\geq 6 \cdot 10^{4}$   &  e$^-$  \\
            Photon flux budget  & $F$  & $\geq 3 \cdot 10^{4}$   &   e$^-$   \\
            Photon response non-uniformity  & $f_{\rm prnu}$  & $< 0.01\, F$  &   \\
            Digital depth  &   &  12 & bit \\
            Pixel area  &    &   $11 \times 11$ & $\mu$m$^2$\\ 
            Effective collecting area   &      &   $2048 \times 2048$   & pixels   \\
            Configurable readout speed  &  & $\leq 48$ & frames$\,{\rm s}^{-1}$  \\
            Readout noise   & $N_{\rm r}$  & $<10$   &  e$^{-}$  \\
            Fixed pattern noise\footnote{For the specified photon flux.}   &   $N_{\rm dc,fpn}$   &   $<100$   & e$^{-}$ \\
            Dark current noise at 20$^{\circ}$C   & $N_{\rm dc}$ & $< 50$   &   e$^{-}$  \\
            Maximum camera envelope   &     &   $100 \times 80 \times 73$ & mm$^{3}$  \\
            Optical entrance window   &    &  No  &    \\
            Housing material   &    &  PCO black  &    \\
            Thermal interface   &     &   Yes   &   Cold finger   \\
            CoaXPress coaxial connector & & Yes & DIN 1.0/2.3 jack \\
            \multicolumn{2}{l}{Sensor's default operational temp.} & 20.0 $\pm$\, 0.5 & $^{\circ}$C \\
            Power consumption & & $< 6.9$ & W  \\ 
            \hline 
         \end{tabular}
      \end{minipage}
   \end{table}

\subsection{Thermal design}
\label{SS-oth}
The thermal design of the SCIP units is driven by the Earth's stratospheric thermal environment, where \sunriset\ operates. For a flight altitude of 37 km, the air pressure is around 500~Pa, and rarefied gas heat transfer phenomena must be considered. The primary thermal interaction between the instrument and its environment is through radiation; therefore, a similar approach was followed to that taken towards the thermal design of \ac{leo} satellites, and hot and cold cases were studied. However, in contrast to \ac{leo} satellites, a stratospheric balloon stays above the same region of Earth for a long period of time, and steady-state extreme case analyses can be used for dimensioning the thermal control subsystem. The hot and cold float cases are determined by the worst-case possible combination of the direct sunlight, albedo, Earth’s \ac{olr}, and the \ac{sza}.  However, using the extreme values for the previously mentioned loads directly would lead to an oversized system, as the solar elevation angle, the albedo coefficient, and the \ac{olr} are partially correlated, and extreme values for all the loads do not occur at the same time. In the case of \sunriset, a new approach was followed to determine the albedo and \ac{olr} loads using measured data from NASA's Clouds and Earth Radiant Energy System (CERES) for the summer season and the region of interest \citep{2018AcAau.148..276G}. The obtained worst-case environmental conditions are summarized in Table \ref{tab:tworstcase}.
  
\begin{table}
\caption{Worst-case thermal environmental conditions for SCIP.}
\label{tab:tworstcase}
\begin{tabular}{ccc}
\hline
Parameter & Hot operational case & Cold operational case \\
\hline
Albedo & 0.64 & 0.77 \\
OLR (W/m$^2$) & 237.5 & 176.5 \\
SZA (\degree) & 45.5 & 87.7 \\
\hline
\end{tabular}
\end{table}

\begin{figure}[t]
\centering
\includegraphics[width=\linewidth]{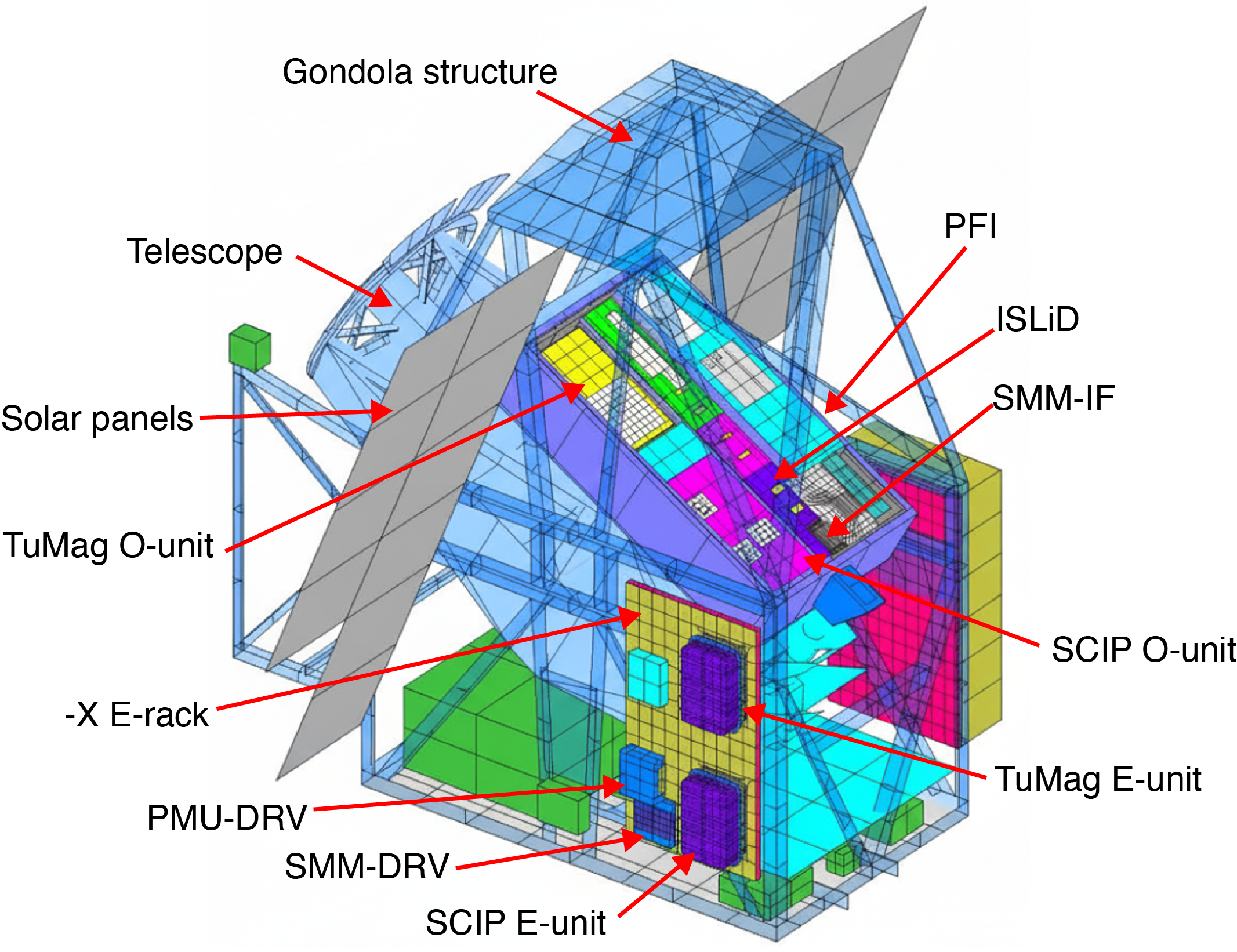}
\caption{Geometrical Mathematical Model (GMM) for thermal analysis of \sunriset\ implemented in ESATAN-TMS, showing the mounting locations of the SCIP components. Modified from \cite{piqueras2025jaxarr}.}
\label{fig:sunrisethermal}
\end{figure}

During the ascent phase, due to the low temperatures and the relative air speed, it is possible to achieve the lowest temperatures. In the case of \sunriset, the SCIP O-unit is inside the \ac{pfi}, protected against external winds by the so-called windshields. Additionally, the \ac{pfi} receives direct sunlight that contributes to heating up the units because the gondola is allowed to spin during the first phase of the ascent. After studying worst-case hot and cold launch scenarios, it was concluded that the float conditions shown in the previous paragraph and Table \ref{tab:tworstcase} were more extreme and therefore the ones used for sizing the thermal hardware \citep{fernandezsoler2021}.

\begin{figure}[t]
\centering
\includegraphics[width=\linewidth]{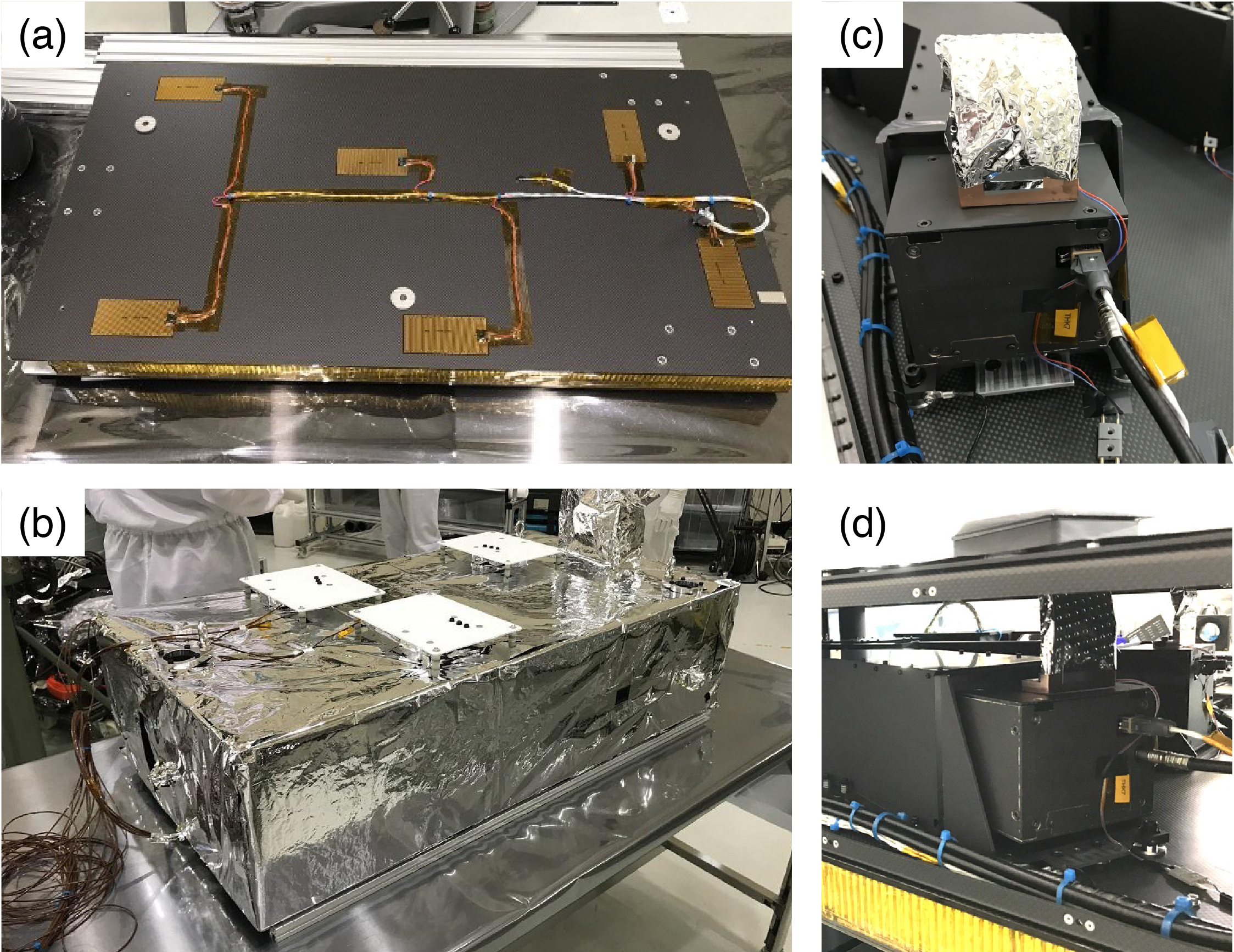}
\caption{(a) Operational heaters attached on the bottom surface of the optical bench. (b) The SCIP O-unit covered by \acs{sli} in which the three radiators on the top face are connected to the two SP cameras and the SJ camera. (c) A flexible copper strap attached to the SP2 camera in which the strap is wrapped by SLI. (d) Connection of the SP2 camera and its radiator by the copper strap. Modified from \citet{2022SPIE12184E..2BK}.}
\label{fig:othermal}
\end{figure}

The SCIP O-unit is located in one of the corners of the \ac{pfi}, as shown in Figure \ref{fig:sunrisethermal}, along with the other units. All these units share a similar environment. However, the thermal requirements and operating conditions are different for each one, and so is their thermal design. The O-unit is intended to operate at near room temperature ($\sim20$\degree\ C), but the PFI, in general, is colder. Therefore, in order to reduce heat losses to the \ac{pfi}, and to decouple the behavior of the O-unit from the other PFI units and the PFI itself, the unit has been insulated conductively and radiatively. The O-unit is mechanically attached to the PFI by means of three titanium mounting points. These interfaces provide a high thermal resistance path, conductively decoupling the \textbf{O-unit} from the PFI. Additionally, to insulate radiatively, the \textbf{O-unit} is wrapped in \ac{sli} made of aluminized polyimide films on both sides (Figure \ref{fig:othermal} [b]). We choose \ac{sli} over \ac{mli} because residual air makes gas conduction the dominant heat transfer mechanism, reducing the effectiveness of \ac{mli} and making the added complexity unnecessary.

Three cameras (SJ, SP1, and SP2) are located inside the O-unit. Each camera dissipates around 3.6~W during nominal operation, distributed as $\sim0.6$~W within the CMOS sensor and $\sim3$~W within the \ac{pcb} with the control electronics (\ac{pcb}--FPGA). The camera thermal requirements are the main drivers of the O-unit thermal design. The CMOS sensor of each camera needs to be maintained between 15 and 30\degree C with a temperature stability of $\pm 0.1$\degree C. The CMOS sensor is in direct contact with a copper cold finger, which is mechanically attached to the top part of the camera. The top part of each of the cameras is refrigerated by three dedicated white-painted (Aeroglaze A276) radiators placed on top of the instrument (Figure \ref{fig:othermal} [b]). The camera and the radiators are connected by a flexible copper strap (TAI Corp. P5-503, Figure \ref{fig:othermal} [c][d]). A thermal filler (Gap Pad GPHC5.0) is inserted between the strap and the camera, as well as between the radiator and the strap to have good conductive coupling. The CMOS sensor operating temperature and thermal stability are achieved by a dedicated heater located on the cold finger and capable of dissipating up to 6~W.

The top part of the unit has direct view to the cold sky, but is also subjected to solar, albedo, and OLR loads. The camera radiators have been sized to maintain the CMOS sensor at 15\degree C for the worst hot operating conditions, so after adding temperature uncertainties, there is enough margin for the heater to stabilize the sensor below 30\degree C. The top part of the instrument, not used as the radiators, is insulated with \ac{eps} foam, wrapped in aluminized Mylar with the Mylar facing outside. Although this outer finishing does not provide the best radiative insulation, it \textbf{was} decided to use it to avoid overheating the top part during the ascent phase, when the telescope is fixed in horizontal position and there is direct sunlight, or in case of malfunction with the telescope pointing during float operation. In order to control the temperature of the rest of the components of the O-unit, operational heaters are placed on the optical bench. The heaters are controlled by the E-unit and have a maximum dissipation of approximately 25~W. Figure \ref{fig:othermal} (a) shows the operational heaters (6 pieces of Minco HK6913) attached to the bottom surface of the optical bench. For the ascent phase, and in case the E-unit is turned off, there are also non-operational heaters to maintain the optical bench above 0\degree C, in which Minco HK5171 together with CT198 are used for thermostat control.

\begin{figure}[t]
\centering
\includegraphics[width=\linewidth]{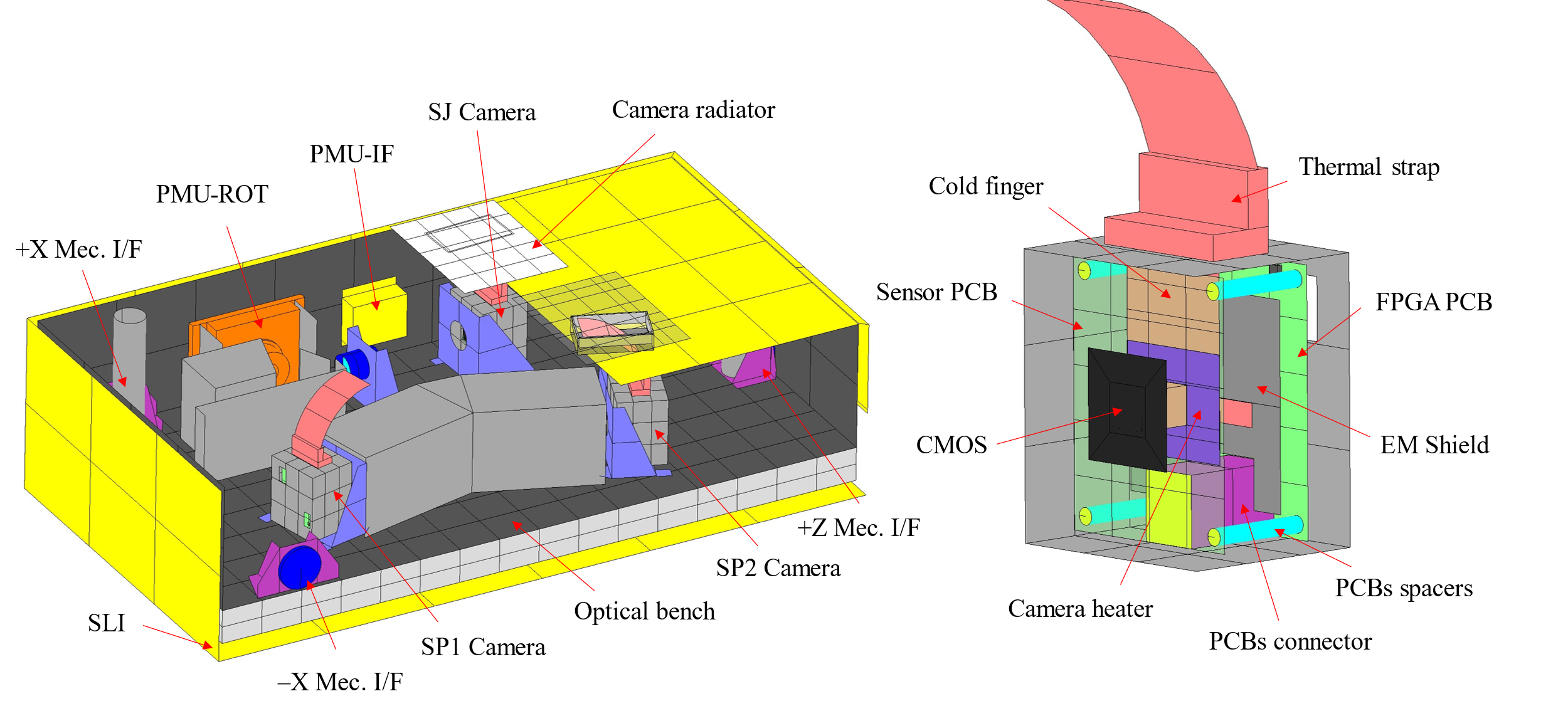}
\caption{ESATAN-TMS GMM of the SCIP O-unit (left) and one of its cameras (right). Reproduced with permission from \citet{piqueras2025jaxarr}.}
\label{fig:otmodel}
\end{figure}

The analysis and thermal design of the O-unit \textbf{was} performed with the help of an ESATAN-TMS, a widely used software package supported by the European Space Agency for conducting in-orbit thermal analysis based on the lumped parameter method, with a model including approximately 1000 nodes. The model implements all the main elements of the O-unit, as shown in Figure \ref{fig:otmodel}, and is integrated into the complete \sunriset\ thermal model shown in Figure \ref{fig:sunrisethermal}. The \sunriset\ thermal model contains detailed models of most of the instruments, but due to its complexity, each analysis can last for several hours. In order to speed up calculations for the different units, equivalent boundary conditions are calculated for each one using the \sunriset\ thermal model. Then, an equivalent standalone analysis can be carried out independently for each unit. After a major update in the unit’s model, the unit is integrated again in the \sunriset\ model, and new boundary conditions are calculated. Details on the thermal analysis of the \ac{scip} O-unit are described in \cite{piqueras2025jaxarr}. 

\section{SCIP E-unit}
\label{S-eunit}

The SCIP E-unit houses electronic components on three specifically tailored \acp{pcb}: (1) the \ac{dpu}, (2) the \ac{amhd} and (3) the \ac{pcm}, as shown in the block diagram (Figure~\ref{fig:diagram}). To meet the harsh environment of a stratospheric flight, along with the limited budget compared to space missions, we designed the E-unit based on \ac{cots} components. Consequently, the electronic components and boards of the E-unit have to be enclosed in a pressurized vessel under temperature control, so that we \textbf{can} use commercial devices without any concern about their near-vacuum behavior. The development time is another important factor to take into account: the use of general-purpose, modern components ---not qualified for space, together with state-of-the-art software tools, sharply reduces the engineering efforts. A further driver of this development \textbf{was} the maximization of commonalities between the electronics units of TuMag and SCIP, which \textbf{were} developed by the same group at the same time. The same three \acp{pcb} are used in TuMag, whereas a fourth board used in TuMag, the high-voltage power supply board, is not needed for SCIP. \textbf{The electrical harness connections between the SCIP components are summarized in Appendix \ref{SS-compression}.}

\subsection{The electronic boards inside the E-unit}
\subsubsection{The DPU board}
\label{SS-dpu}
\begin{figure}[t]
\centering
 \includegraphics[width=\linewidth]{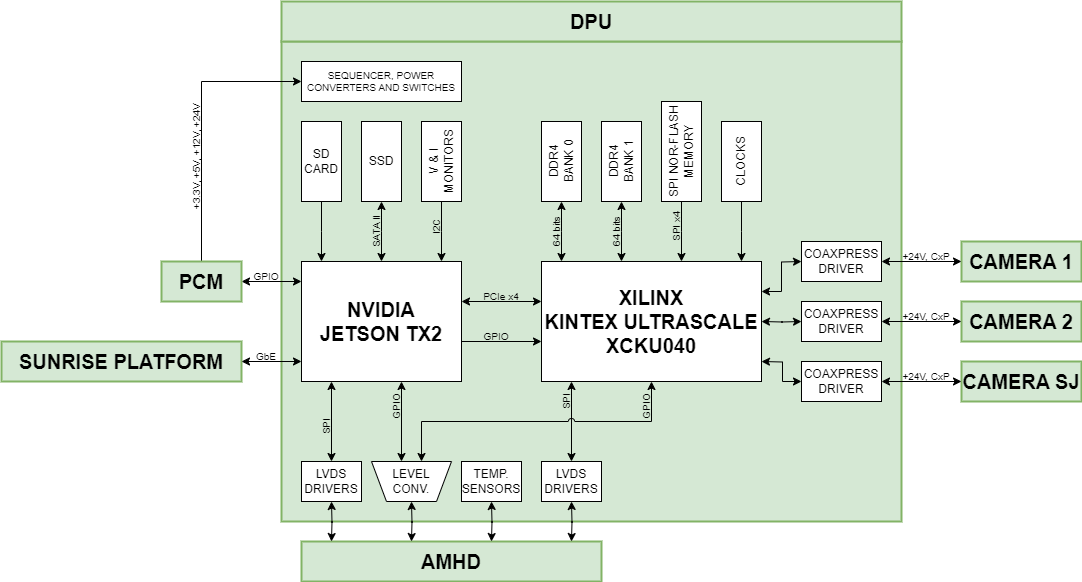}
 \caption{Functional block diagram of the DPU board. The central blocks represent the system controller (SyC; NVIDIA\textsuperscript{TM} Jetson TX2) and the frame grabber (FG; Xilinx\textsuperscript{TM} Kintex Ultraescale).}
\label{fig:DPU} 
 \end{figure}

The DPU board includes two main blocks: a \ac{syc} and a \ac{fg}, which also acts as an image processor, as shown in Figure \ref{fig:DPU}. 
The \ac{syc} performs tasks such as image storage management, housekeeping data acquisition, instrument state monitoring, and communications with the balloon platform and the ground support software. The FG has to guarantee synchronization between subsystems and to perform intensive image processing tasks such as acquisition, polarization demodulation, accumulation, and bit compression of the images (Appendix~\ref{S-firmware-implementation}). Both devices are the same as those used in TuMag, and their detailed description can be found in Section~6.1 of \citet{2025SoPh..300..148D}.

One difference from the TuMag's DPU is that the \ac{fg}'s \ac{fpga} carries out the image acquisition and processing from the three scientific cameras (SP1, SP2, and SJ), instead of two cameras in the TuMag case. The processing pipeline for SP1 and SP2 is nearly identical and includes image demodulation-accumulation, bit-compression, and standard image compression. By contrast, the SJ camera is processed in the SP1's pipeline using a time-multiplexed scheme. In order to keep synchronization with the PMU phase, the timing of the camera triggers and the SMM commanding are controlled by the \ac{fg}. For this purpose, it includes a custom core called SCIP Control Unit, which decodes the \ac{pmu} phase signal and generates the sequence of \ac{cxp} camera triggers and SMM-movement interrupt signals. The latter signals are received by the Microblaze processor, which executes a specific interrupt routine to command the \ac{smm} by the \ac{spi}. These \ac{spi} commands are eventually interpreted by the AMHD board, which implements the physical interface with the SMM-DRV.

 \subsubsection{The AMHD board}
 \label{SS-amhd}

   \begin{figure}
   \centering
   \includegraphics[width=\linewidth]{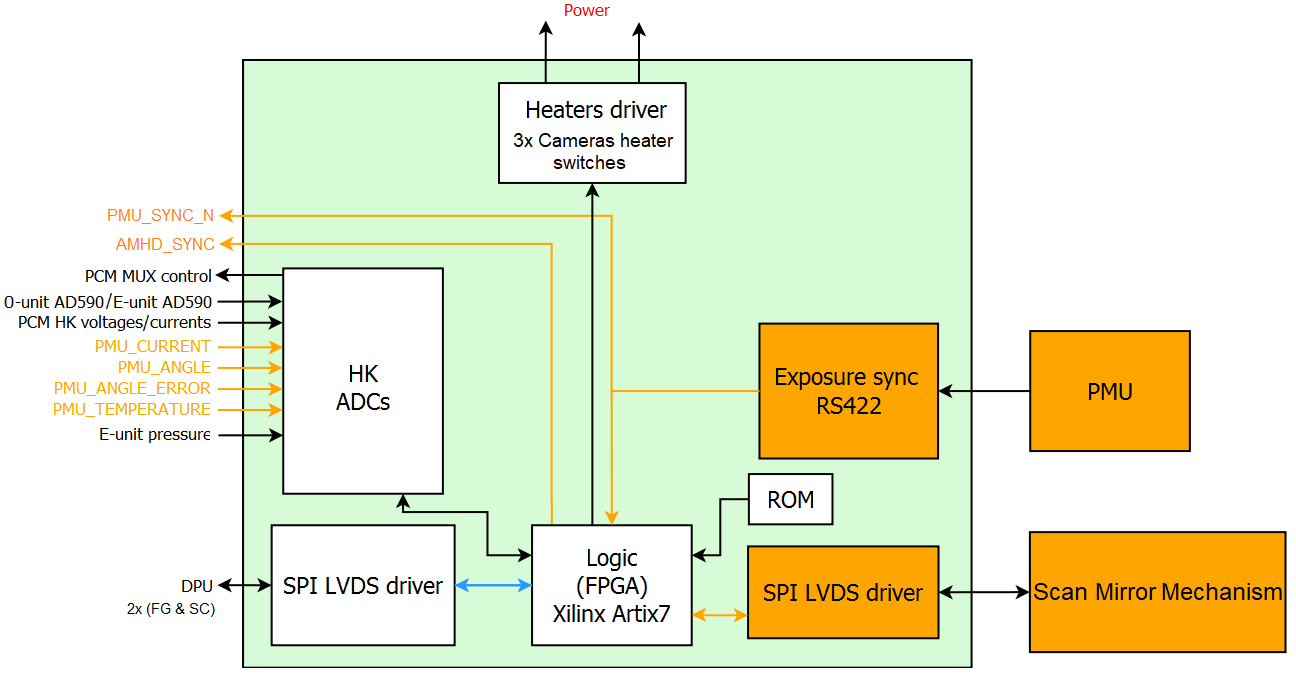}
   \caption{Functional block diagram of the \ac{amhd} board.}
   \label{fig:AMHD}%
   \end{figure}
 
The \ac{amhd} interfaces the O-unit as well as the DPU and PCM boards in the E-unit to acquire the instrument status telemetry, drive the mechanisms, and stabilize the instrument temperatures by controlling the heaters. An \ac{amhd} functional diagram is shown in Figure~\ref{fig:AMHD}. 
The blocks of the \ac{hk} \acp{adc}, the driver controlling the interfaces with the \ac{dpu}, the heater driver, the logic ``brain'' of the board, and  a read-only memory (ROM) are the same as those in the TuMag AMHD \citep[see Section 6.2 in][]{2025SoPh..300..148D}. The following blocks are specific to SCIP:

    \begin{enumerate}    
        \item An RS-422 receiver to obtain the PMU synchronization signal. This logic has two main features. First, it routes the PMU synchronization signal to the \ac{fg}. Second, it ensures signal integrity at the FG even if communication with the PMU is lost.

        \item A dedicated \ac{spi} with the SMM to control it and acquire its \ac{hk} data. The frequency for regular \ac{hk} is configurable, but the \ac{pmu} and \ac{smm} subsystems' \ac{hk} is acquired every 32~ms at the configured frequency.
    \end{enumerate}

\subsubsection{The PCM board}
 \label{SS-pcm}

The \ac{pcm} is the electronic subsystem that powers the SCIP E-unit and provides it with the required voltages. 
Five secondary output lines of the \ac{pcm} and their approximate maximum average power consumption in the so-called operational mode of the instrument are summarized in Table~\ref{tab:PCM-OutputPower}. In addition, an unregulated $\sim+25.6$~V (average of $V_{\rm in}$) output is also delivered from the \ac{pcm} primary side, after a common-mode filter and an inrush-current limiter circuit, through the \ac{amhd} board. The design of the \ac{pcm} is almost the same as that in TuMag \citep[Section~6.4 in][]{2025SoPh..300..148D}. One difference from the TuMag \ac{pcm} is that the number of dedicated \ac{cots} $+24$~V DC/DC converters is reduced from two to one because the high-voltage power-supply board is not needed for SCIP. Another difference is that the unregulated power ($V_{\rm in}$) is delivered to the PMU and SMM, as well as the various heaters in the instrument.

        \begin{table}
           \caption{{SCIP operational-mode maximum average power of each output.}}
           \label{tab:PCM-OutputPower}
           \begin{tabular}{r r l}
        	  \hline
        	  {Voltage} & Power & Subsystem \\
        	  \hline
        	  +\,3.3~V & 5.0~W  & DPU, AMHD \\
        	  +\,5.0~V & 4.0~W & DPU, AMHD\\
        	  +\,12.0~V & 36.0~W & DPU, AMHD, Mechanisms \\
        	  $-$\,12.0~V & 0.5~W & AMHD \\
        	  +\,24.0~V & 10.0~W & Cameras\\
        	  $V_{\rm in}$ & 45.0~W & Heaters \\
                        & 31.0~W & PMU, SMM \\
        	  \hline
        	  Total approx. & 131.5~W & \\
        	  \hline
          \end{tabular}
        \end{table}

\subsection{The pressurized electronics box}
\label{SS-ebox}

\begin{figure}[t]
\centering
\includegraphics[width=0.8\linewidth]{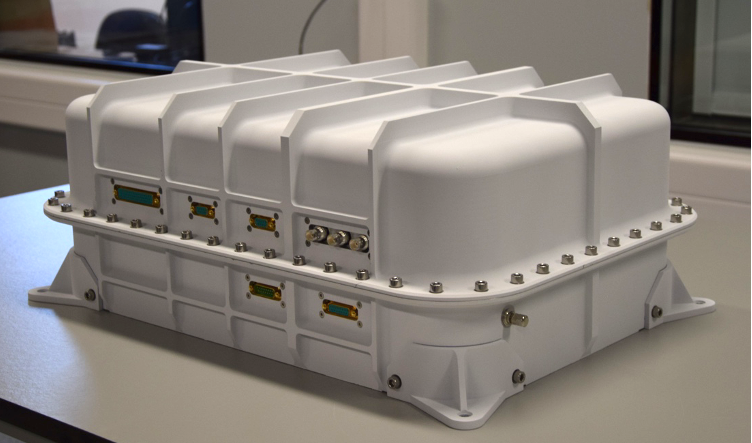}
\caption{Enclosure box of the SCIP E-unit. Connectors can be distinguished.}
\label{fig:EBox}
\end{figure}
 
The pressurized enclosure of the SCIP E-unit (Figure~\ref{fig:EBox}) houses the three electronic boards (\ac{pcm}, \ac{dpu}, and \ac{amhd}) in charge of the necessary functions for the instrument. The use of \ac{cots} electronic components, instead of military- or space-grade ones, is enabled by a pressurized box capable of maintaining an internal pressure slightly above 1~atm to guarantee the proper functioning of the circuits. The box is filled with dry nitrogen, preventing water condensation at low temperatures. As with TuMag, a ``book-like'' design has been devised so that the box is opened in two halves during \ac{aiv} stages in order to ease the assembly and verification of the electronic components and their interconnections. Once all the components are attached, the two halves are closed and screwed together through a flange with an O-ring seal to prevent pressure losses. Stiffeners are included in the design to guarantee the required rigidity of the structure while maintaining a low wall thickness to reduce mass. The box envelope is 584~mm~$\times$~404~mm~$\times$~226~mm. It hosts electrical connectors and provides the interface attachments with the \sunriset\ electronic rack. Connectors are also included for data transmission and for power supplies. The box surfaces are treated with a chromium-process coating in order to obtain good thermal and electrical conductivity. The outer ones are white painted (using SG121-FD from MAP company) for radiative cooling.\\

\subsection{Thermal design}
\label{SS-eth}

\begin{figure}[t]
\centering
\includegraphics[width=0.8\linewidth]{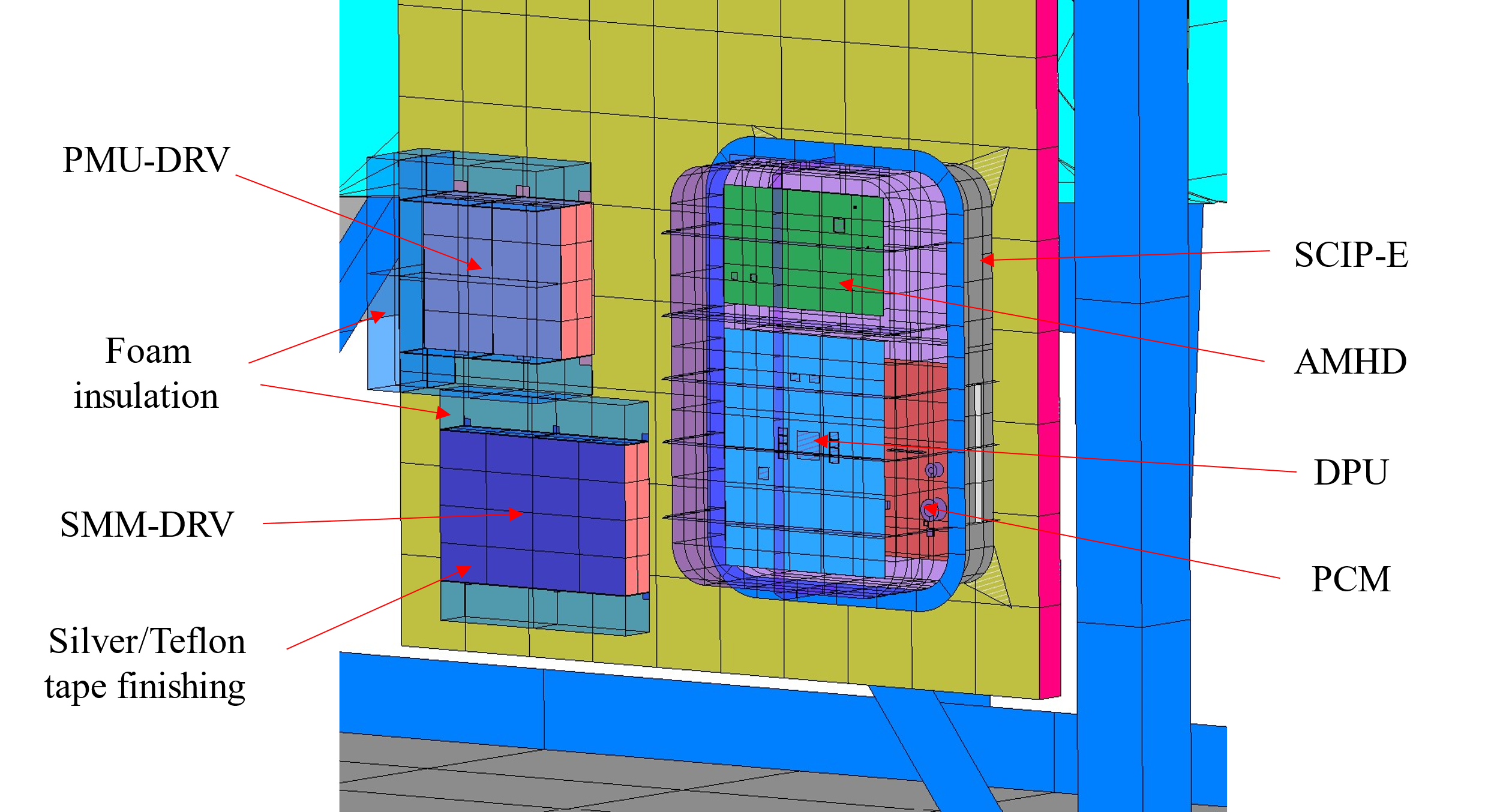}
\caption{ESATAN-TMS GMM of the SCIP E-unit, PMU-DRV, and SMM-DRV mounted on the E-rack. Reproduced with permission from \citet{piqueras2025jaxarr}.}
 \label{fig:GMM}
\end{figure}

The SCIP E-unit, PMU-DRV, and SMM-DRV are mounted on one of the E-racks of \sunriset\ (Figure~\ref{fig:sunrisethermal}). The E-racks are shielded from external winds by transparent windshields allowing solar and infrared radiation to pass through. The thermal-design concept for the SCIP E-unit is identical to that of the TuMag E-unit because the two E-units are located on the same E-rack and experience the same thermal environment \citep[see][for details]{2025SoPh..300..148D}. Thermal models for the E-unit, together with PMU-DRV and SMM-DRV, \textbf{was} developed using ESATAN-TMS. While the stratospheric thermal environment resembles the space environment in terms of heat transfer and environmental loads, certain situations still require consideration of convection interactions. Because the E-unit is pressurized, additional couplings among electronic components, the housing, and the nitrogen inside have been accounted for. A simplified thermal model (Figure~\ref{fig:GMM}) is integrated into the complete \sunriset\ model to obtain thermal boundary conditions for each electronic unit. A detailed thermal model \textbf{was} then analyzed separately for each unit, following the methodology outlined in \cite{2017EUCASS...G}. Detailed thermal analyses of the E-unit, PMU-DRV, and SMM-DRV are summarized in  \cite{piqueras2025jaxarr}. 

 \section{On-board software, firmware, and data flow}
 \label{S-software}
 
The \ac{scip} on-board software is divided into two main blocks: the \ac{csw} and the \ac{fgsw}. The \ac{csw}, which runs in the \ac{syc}, is mainly devoted to telecommand (TC) and telemetry (TM) management,\footnote{Telecommands are orders to the instrument through the \sunriset\ \ac{ics}. Telemetries are data transferred from SCIP to the \ac{ics}.} transferring images from the \ac{fg} to the \sunrise\ \ac{ics}, generating thumbnails from selected images, housekeeping acquisition, and contingency management. The \ac{fgsw}, which runs within the \ac{fg}'s MicroBlaze processor, commands the different subsystems during the observation modes, handles the commands from the SyC, and manages the image transfer from the \ac{fg} to the \ac{syc}. 

The basic design of the software (\ac{csw} and \ac{fgsw}) and the data flow are identical to those of TuMag. A graphical user interface called GUISCIP runs on ground support equipment. It provides the same functionality as TuMag's interface but is customized for SCIP operations. Details of the software and the ground support equipment are described in Sections~8 and~9 of \citet{2025SoPh..300..148D}, respectively. In contrast, the firmware implementation is tailored to the on‑board processing required by the SCIP observing modes, and its details are provided in Appendix~\ref{S-firmware-implementation}.

\section{Assembly and test of SCIP}
\label{S-aitv}

\subsection{Optical alignment}
\label{SS-oalign}
\begin{figure}[t]
\centering
\includegraphics[width=\linewidth]{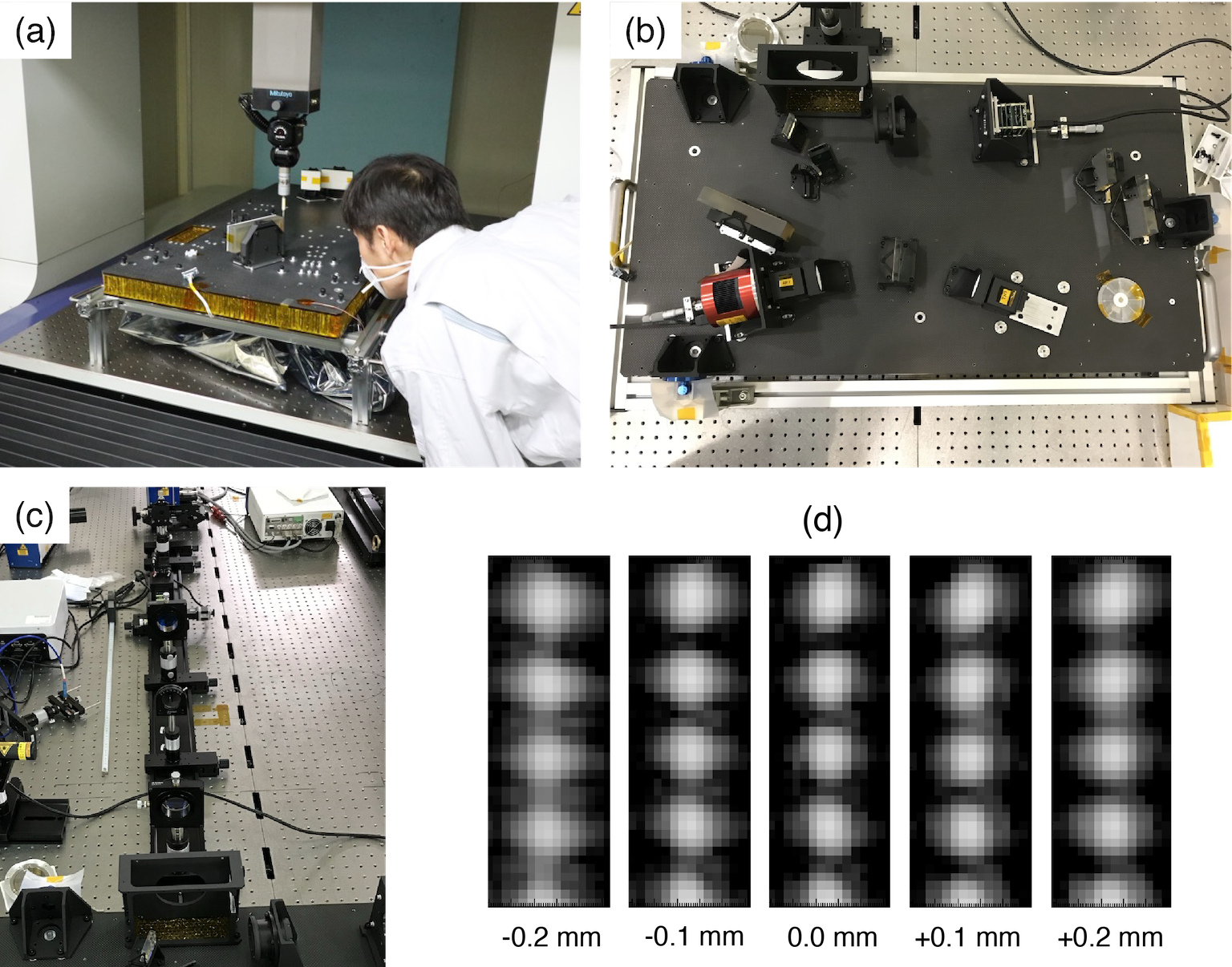}
\caption{(a) Mechanical alignment measurement with CMM for the SCIP spectrograph optical components. (b) The SCIP O-unit configured for the optical alignment measurement using a non-flight camera mounted on the camera bracket. (c) Feed optical system for the optical alignment measurement. (d) An example of a focus scan result measured with the non-flight camera. Modified from \citet{2022SPIE12184E..27K}.}
\label{fig:oalign}
\end{figure}

\begin{figure}[t]
\centering
\includegraphics[width=0.8\linewidth]{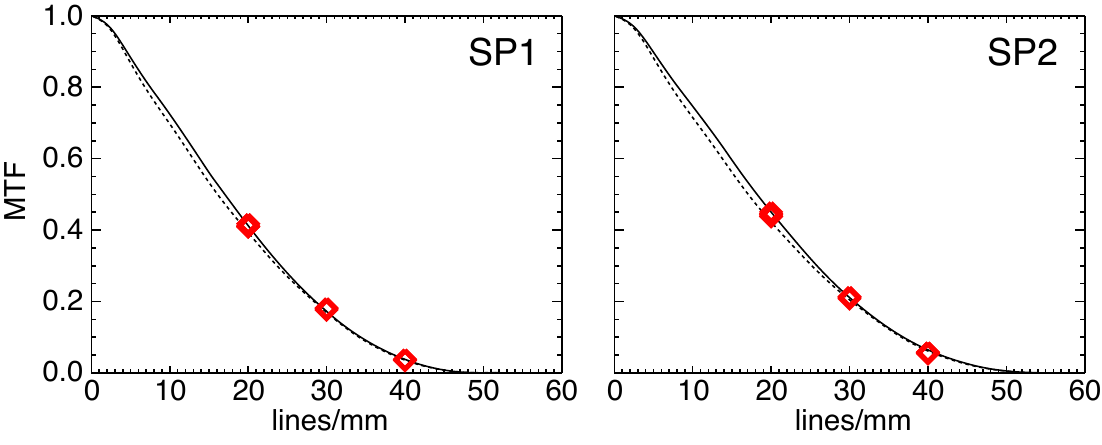}
\caption{\acs{mtf} of SP1 (left) and SP2 (right). The solid and dashed lines show an ideal \acs{mtf} with no wavefront error (WFE) and an \acs{mtf} with WFE of 40~nm rms, respectively. The red diamonds represent the Ronchi ruling measurements at 20, 30, and 40~lines/mm. Modified from \citet{2022SPIE12184E..27K}.}
\label{fig:mtf}
\end{figure}

Because the SCIP O-unit does not have any adjustable mechanisms except the rotating waveplate and the scan mirror mechanism, careful optical alignment is necessary to achieve the spatial and spectral resolution described in Section \ref{SS-opt}. It is especially important to coalign positions and tilts of the collimator and camera mirrors as well as the tilt of the echelle grating with respect to the slit in the spectrograph \citep{2022SPIE12184E..27K}. These optical elements were first installed on the optical bench, and their rear and side surfaces were measured using a coordinate measuring machine (CMM, Mitsutoyo LEGEX 910) as shown in Figure \ref{fig:oalign} (a). These elements were mechanically aligned by fabricating the adapter plates (Figure \ref{fig:mirrorlenssupport} [a]), which were inserted between the optical bench and the mirror support mechanism. 

After the mechanical alignment, the optical elements were finely coaligned by shimming based on optical measurements. A telecentric feed optical system was constructed to feed the F/24.2 beam, same as \ac{islid}, into the O-unit (Figure \ref{fig:oalign} [c]). The coalignment measurement was done by feeding light from tunable lasers for 850~nm and 770~nm bands (Sacher Lasertechnik TEC-520). A non-flight CMOS camera (ASI1600MMPro) was mounted on the camera bracket for the optical alignment test because it had a smaller pixel size of 3.8~$\mu$m than the flight camera (Figure \ref{fig:oalign}[b]). 

We measured \textbf{the} astigmatism aberration by manually defocusing the camera (Figure \ref{fig:oalign}[d]) and subsequently co-aligned the collimator and camera mirrors to minimize astigmatism aberration. After the coalignment of the mirrors, we mounted the flight camera and made a focus scan to find the best focus position. We evaluated the spatial \ac{mtf} of the spectrograph by feeding light from a white light lamp (NIHON PI PIS-UHX-NIR) through Ronchi ruling targets of 20, 30, and 40 lines/mm which were arranged perpendicular to the slit direction in the feed optical system. We demonstrated that the spectrograph met the wavefront error required by the optical design (Figure \ref{fig:mtf}). 

\begin{figure}[t]
\centering
\includegraphics[width=0.9\linewidth]{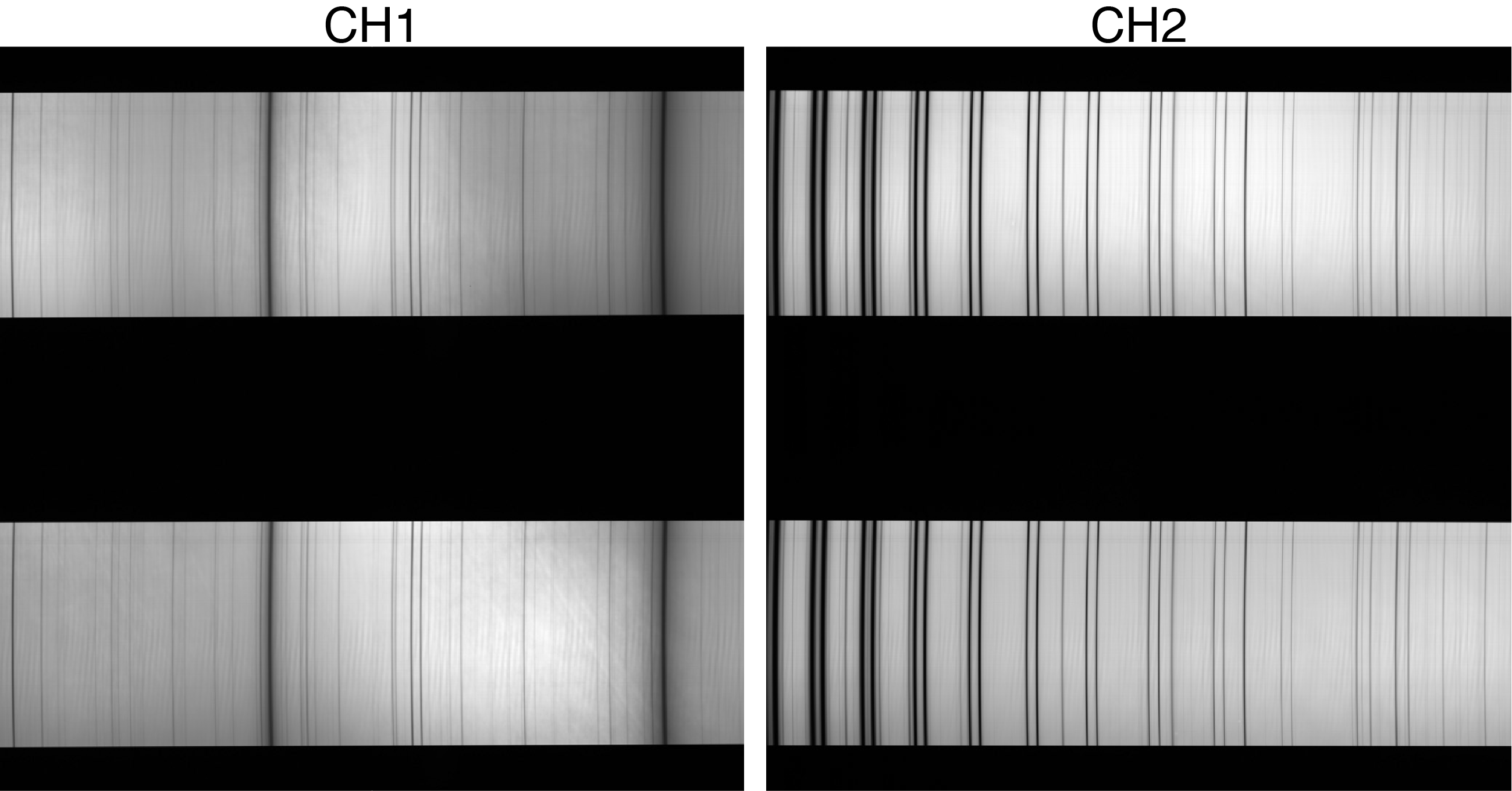}
\caption{SP1 (left) and SP2 (right) data taken by feeding the natural sunlight. The horizontal and vertical directions correspond to the wavelength and spatial directions, respectively. The s- and p-polarization spectra are vertically separated by the \ac{pbs} (Figure \ref{fig:pbs}) and focused onto the same sensor. Modified from \citet{2022SPIE12184E..27K}.}
\label{fig:suntest}
\end{figure}

The spectrum and spatial positions measured on the SP cameras were verified by feeding natural sunlight to measure the solar spectral lines using a heliostat implemented at the NAOJ \ac{atc} cleanroom facility, as shown in Figure \ref{fig:suntest}.

The SJ optical system \textbf{is} less sensitive to the coalignment, but we checked the tilt of the SJ reflective filter and the SJ lens unit by measuring a star target image created by the feed optical system. The focus position of the SJ camera was adjusted by looking at the sharpness of the slit. Because the SJ optics \textbf{use} a lens system that causes a non-negligible focus offset of 0.33~mm between air and vacuum, the offset was applied in the SJ camera position after finding the best position in air. The focus offset was verified in the thermal vacuum test described in Section \ref{SSS-otv}.

\subsection{Synchronization test}
\label{SS-sync_test}
\begin{figure}[t]
\centering
\includegraphics[width=0.8\linewidth]{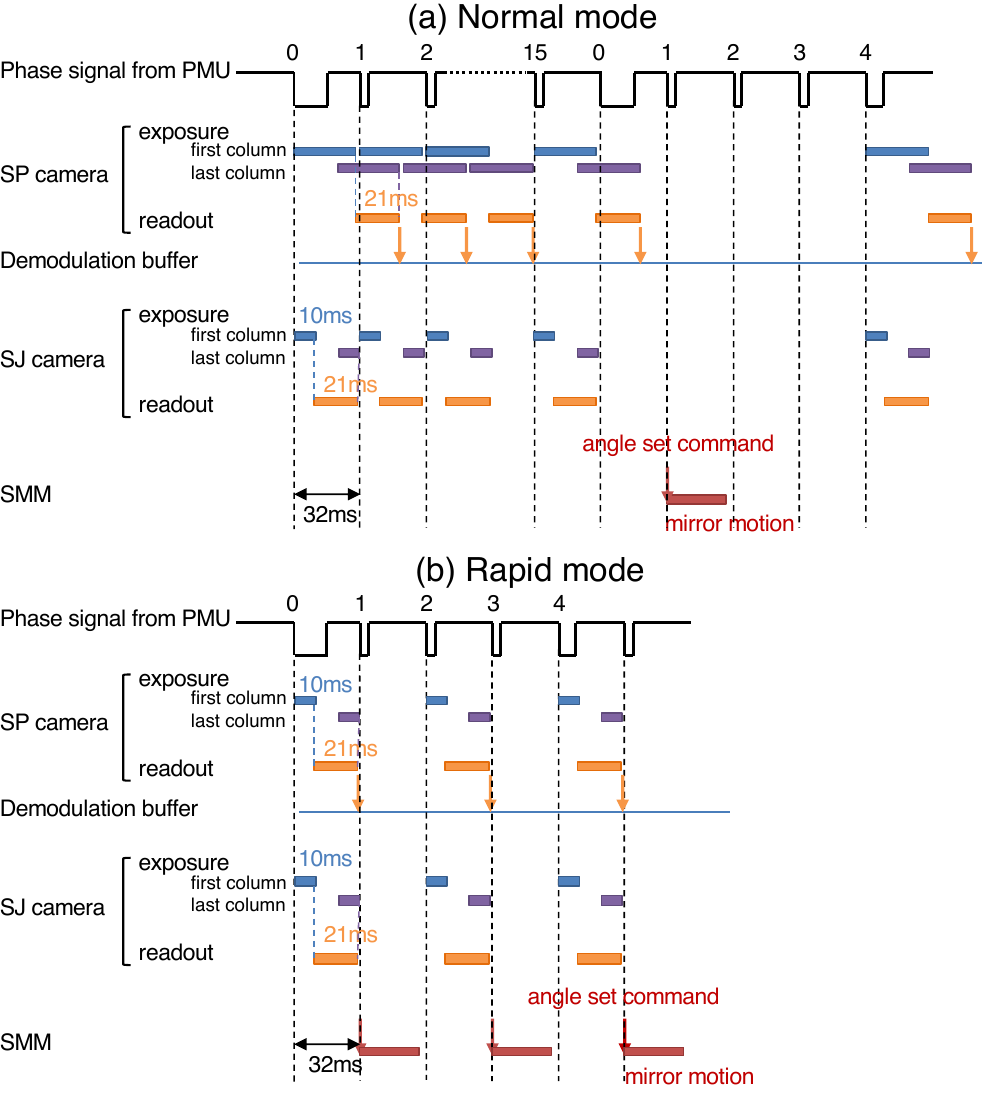}
\caption{Timing diagram of the \ac{pmu} phase signal, camera exposures and readouts by the SP and SJ cameras, onboard demodulation, and SMM angle set commands and associated scan motion in the (a) normal and (b) rapid modes. Modified from \citet{2023JATIS...9c4003K}.}
\label{fig:sync}
\end{figure}

The \ac{scip} observation requires synchronous control among the rotating phases of \ac{pmu}, the scan motion of \ac{smm}, and the exposures by the three cameras. The SCIP E-unit manages synchronization control of the components according to the phase signal sent from the \ac{pmu}. Figure~\ref{fig:sync} shows the timing diagram of the \ac{pmu} phase signal, camera exposures and their readout for the SP and SJ cameras, and SMM angle set commands and associated motions. The length of the pulse in the PMU phase signal represents the rotation phase of the waveplate: the longest, mid-long, and nominal pulses arrive every 360\degree, 90\degree, and 22.5\degree, respectively. The SP cameras start an exposure of the first column in its sensor at the leading edge of the pulses and then send images to the demodulation buffer in the \ac{dpu}. The exposure start times of the first and last columns differ by 21~ms because of the rolling shutter of the sensor. To avoid mirror motion during exposure, the angle set command is sent to \textbf{the} SMM after the exposure of the last column is completed in both normal and rapid modes. In the normal mode, SP images sent to the \ac{dpu} are demodulated and accumulated to produce a set of images corresponding to Stokes $I_d$, $Q_d$, $U_d$, $V_d$, and $R_d$ parameters ({\b Appendix~\ref{SS-demodulation-and-accumulation}}), while SJ images are transferred without onboard processing. Note that $I_d$, $Q_d$, $U_d$, $V_d$, and $R_d$ are results of the demodulation operation and therefore differ from the true Stokes parameters. In particular, the new parameter $R_d$ is used to introduce a 45\degree phase shift with respect to Stokes $V$ and to compensate for the rolling shutter effect of the SP cameras \citep{2022ApOpt..61.9716K}. In the rapid mode, both SP and SJ cameras obtain eight images per one \ac{pmu} rotation, i.e., every 64~ms, and no onboard image processing is applied in the \ac{dpu}.

After the component-level testing, the PMU, SMM, cameras, and E-unit were integrated to demonstrate the synchronization performance with the required accuracy. The timing relationship between image acquisition by the camera and the SMM mirror motion was investigated by \cite{2022SoPh..297..114O}, who demonstrated that the scan mirror motion and an exposure did not overlap with the 10~msec exposure duration used in the rapid mode observation. To verify the synchronization between the PMU and the SP cameras before installing them into the O-unit, using a similar configuration as used for measuring the rotation angle errors (Section~\ref{SS-polari}), we measured intensity modulation with one of the SP cameras and verified that the exposure timing was consistent with the design. After installing the \ac{pmu} and cameras into the O-unit, the synchronization between the \ac{pmu} and the SP cameras was finally verified by the instrument-level polarization calibration described in Section~\ref{SS-polcal}.

\begin{figure}[t]
\centering
\includegraphics[width=\linewidth]{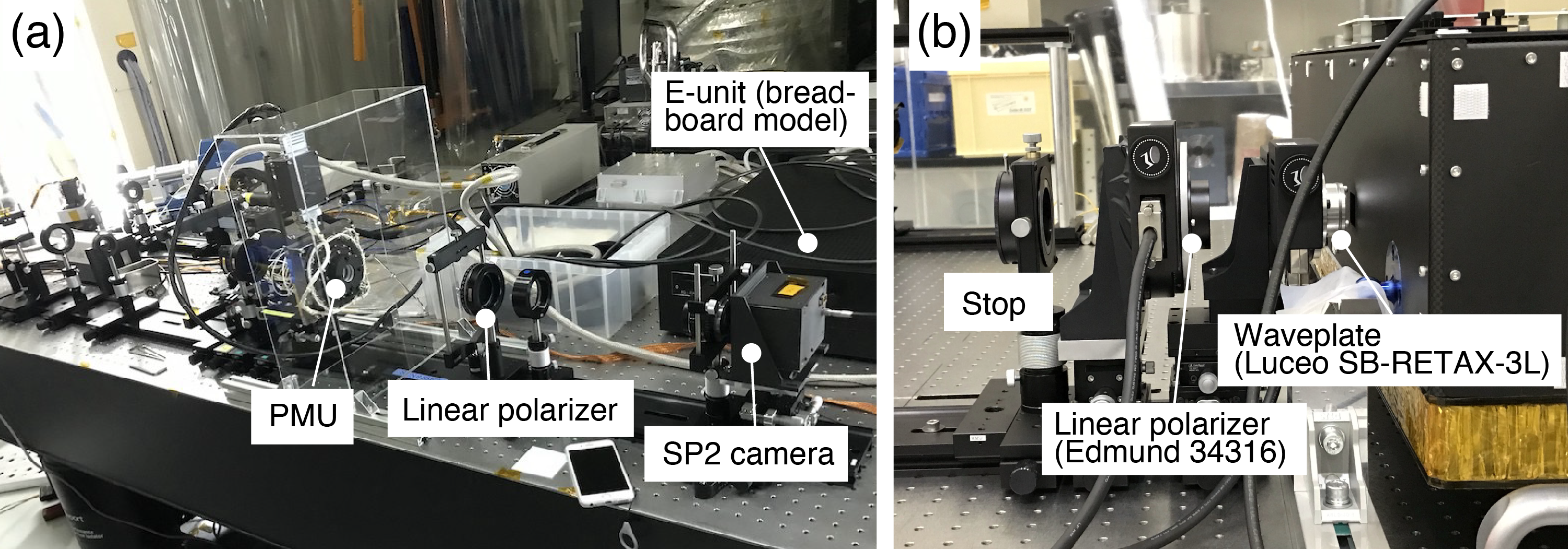}
\caption{(a) Test configuration to verify the synchronization between the \ac{pmu} and the SP camera. (b) Configuration of the instrument-level polarization calibration, in which a linear polarizer and an achromatic waveplate are located in front of the O-unit optical entrance. Modified from \citet{2022ApOpt..61.9716K}.}
\label{fig:polcal}
\end{figure}

\subsection{Polarization calibration}
\label{SS-polcal}
The onboard demodulation operation produces Stokes $I_d$, $Q_d$, $U_d$, and $V_d$ (and $R_d$) data from images taken with the SP cameras using the fixed arithmetic scheme described in Appendix~\ref{SS-demodulation-and-accumulation}. The incident Stokes parameters and the demodulation output can be related by a response matrix $\bm{X}_0$ in the following equation \citep{1990NCATN.355.....E,2008SoPh..249..233I},

\begin{equation}
\begin{pmatrix}
 I_d \\ Q_d \\ U_d \\ V_d 
\end{pmatrix}
= \bm{X}_0
 \begin{pmatrix}
I\\ Q\\ U\\ V
\end{pmatrix}
\end{equation}
where $\left(I_d, Q_d,U_d,V_d\right)^{\rm T}$ is the output of the demodulation operation and $\left(I,Q,U,\right.$ $\left.V\right)^{\rm T}$ is the incident Stokes vector. Because the CMOS sensors used in the SCIP cameras work in rolling shutter configuration, the exposure start time linearly varies along the rolling shutter direction, which causes polarization crosstalk between $Q_d$ and $U_d$, and between $V_d$ and $R_d$. They are easily corrected by applying a rotation matrix corresponding to the phase shift due to the rolling shutter to $(Q_d,U_d)^{\rm T}$ and $(V_d,R_d)^{\rm T}$, respectively, to obtain $(Q_d',U_d',V_d')^{\rm T}$. Then the response matrix $\bm{X}$ is obtained to relate $(I_d, Q_d',U_d',V_d')^{\rm T}$, demodulation output after the rolling shutter correction, with $\left(I,Q,U,V\right)^{\rm T}$, incident Stokes vector, i.e., 
\begin{equation}
\begin{pmatrix}
 I_d \\ Q_d' \\ U_d' \\ V_d'
\end{pmatrix}
= \bm{X}
 \begin{pmatrix}
I\\ Q\\ U\\ V
\end{pmatrix}.
\end{equation}

Extensive instrument-level polarization calibration was employed to get the response matrix $\bm{X}$ considering spatial and wavelength dependence for both SP channels and both s- and p-polarizations \citep{2022ApOpt..61.9716K}. In the feed optics used in the optical alignment (Section \ref{SS-oalign}), a wire-grid linear polarizer and an achromatic waveplate (Luceo SB-RETAX-3L) were inserted in front of the O-unit optical entrance (Figure~\ref{fig:polcal} [b]). They were rotated to specific angles by motorized rotation stages to feed specific polarization states from the halogen lamp. In total, 68 sets were taken at different angles of the linear polarizer and the waveplate. Along with the elements of the response matrix at each pixel, the origin of the waveplate angle and the waveplate retardation were simultaneously derived by a least-squares fitting.

Because \ac{scip} is designed to have the polarization modulator at the optical entrance and to minimize internal instrumental polarization crosstalk, the response matrix is expected to be close to an ideal one. When the retardation of the rotating waveplate is 125\degree, the ideal response matrix of \ac{scip} is given as
\begin{equation*}
\bm{X}_{idel}=
\begin{pmatrix}
1 & 0.213 & 0 & 0 \\
0 & 0.502 & 0 & 0 \\
0 & 0 & 0.502 & 0 \\
0 & 0 & 0 & 0.502 
\end{pmatrix}.
\end{equation*}
The measured response matrices, averaged over spatial and wavelength directions, yielded  
\begin{gather*}
\bm{X}_{ch1}=
\begin{pmatrix}
1.000 & 0.207 & 0.003 & -0.001 \\
0.000 & 0.500 & 0.049 & 0.000 \\
0.000 & -0.049 & 0.500 & 0.000 \\
0.000 & -0.001 & 0.000 & 0.435 
\end{pmatrix}\\
\bm{X}_{ch2}=
\begin{pmatrix}
1.000 & 0.208 & -0.002 & -0.004 \\
0.000 & 0.497 & 0.054 & 0.000 \\
0.000 & -0.054 & 0.497 & 0.000 \\
0.000 & -0.004 & 0.000 & 0.434 
\end{pmatrix}
\end{gather*}
for each channel. Detailed results of the instrument-level polarization calibration including the stability of the above response matrices against temperature variations are summarized in \cite{2022ApOpt..61.9716K}.

After \ac{scip} was mounted into the \ac{pfi} and the telescope, the polarization response matrices were updated by the polarization calibration measurements from the primary focus (F1), accounting for the polarization effect of the telescope except for the primary mirror. We mounted the same achromatic wave plate and a Glan-Taylor linear polarizer at F1 and produced various polarization states from a LED lamp, which were specific for the \ac{scip} wavelength range. The response matrices by the F1 polarization calibration are used to calibrate the flight polarization data.

\subsection{Thermal-vacuum tests}
\subsubsection{O-unit TV test}
\label{SSS-otv}
\begin{figure}[t]
\centering
\includegraphics[width=\linewidth]{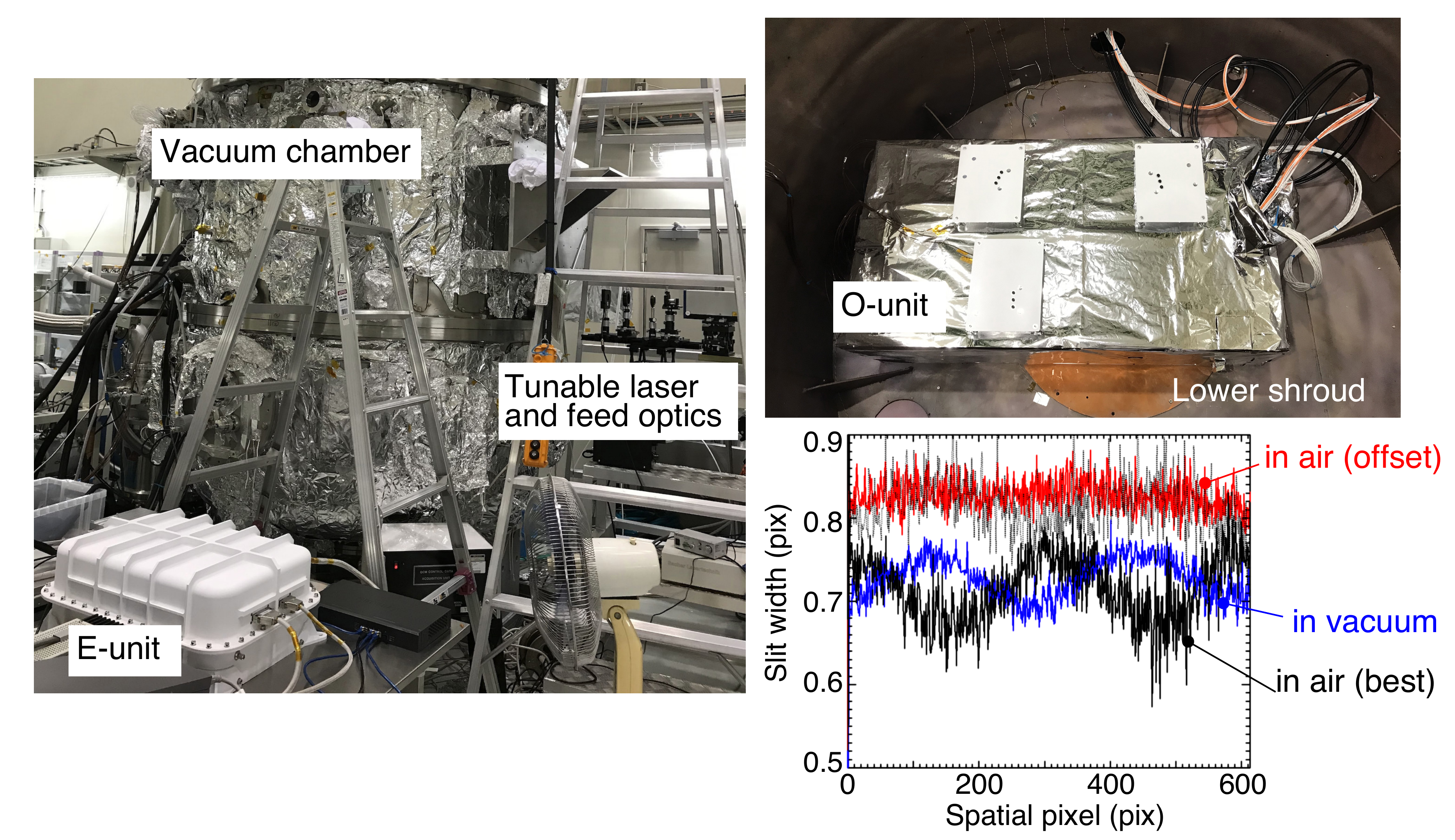}
\caption{Outlook of the vacuum chamber for the O-unit \acs{tvt} (left) showing the layout of the feed optics for the optical test and the SCIP E-unit. The SCIP O-unit was positioned on the lower shroud of the vacuum chamber (top right). The slit width measured with SJ in the \acs{tvt} (bottom right) shows the best focus in air (black), offset position in air (red), and offset position in vacuum (blue). Modified from \citet{2022SPIE12184E..2BK}.}
\label{fig:otv}
\end{figure}

\begin{figure}[t]
\centering
\includegraphics[width=\linewidth]{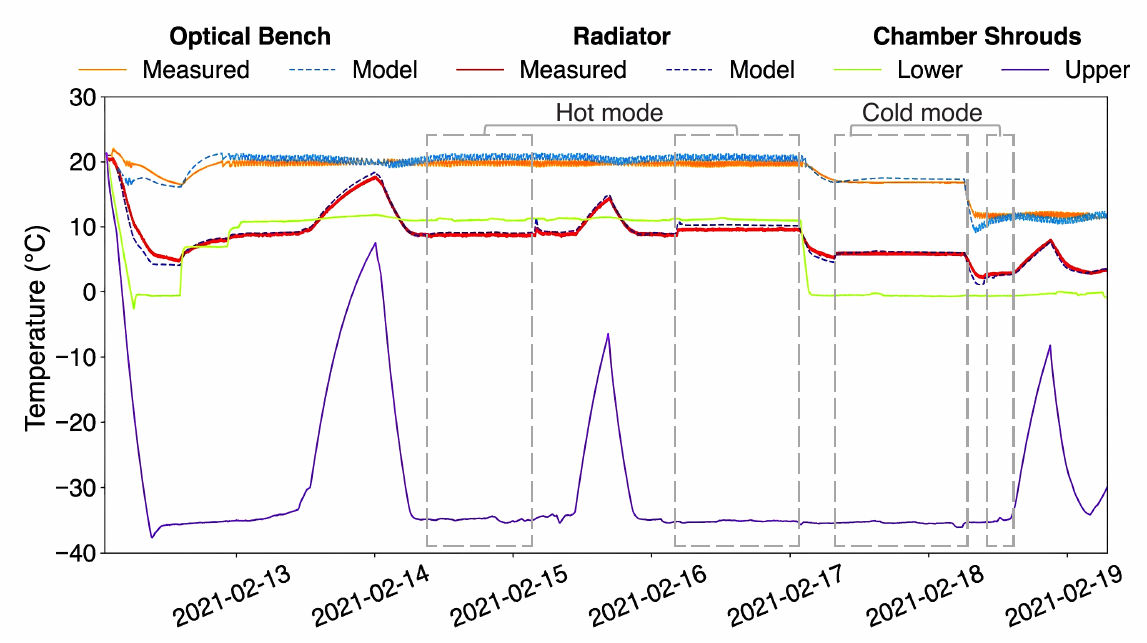}
\caption{Temperature plots of the O-unit \acs{tvt}. The orange and red lines show measured temperatures of the optical bench and camera radiator, respectively. The light-blue and blue dashed lines show their model temperatures. The light-green and purple lines show the chamber shroud temperatures. The gray dashed boxes indicate the period during which the measured and model temperatures were correlated in \textbf{the} hot and cold modes. Modified from \citet{2022SPIE12184E..2BK}.}
\label{fig:otv-t}
\end{figure}

The SCIP O-unit was subjected to a \ac{tvt} in which the O-unit was installed in a vacuum chamber and was exposed to the thermal vacuum environment expected \textbf{during} the balloon flight. The test was aimed (1) to correlate \textbf{predictions from the O-unit thermal model with the measured} temperature distribution in the O-unit (Section \ref{SS-oth}) and (2) to verify the expected spatial and spectral resolutions in the thermal vacuum environment. The \ac{tvt} was carried out using a vacuum chamber at NAOJ \ac{atc}. As shown in Figure~\ref{fig:otv} (top right), the \ac{scip} O-unit was placed in the lower shroud of the chamber, which simulated the \ac{pfi} environment, with insulation by glass-epoxy spacers, while the upper shroud, not shown in the picture, simulated the radiator environment by maintaining a lower temperature than the lower shroud. The E-unit was located outside the vacuum chamber and was connected to the O-unit through a vacuum feedthrough in this \ac{tvt} (Figure \ref{fig:otv} left). This is because the E-unit was tested in a separate \ac{tvt} in Spain to simulate the different thermal environments between the O-unit and the E-unit (Section \ref{SSS-etv}). The detailed test configuration is shown in \cite{2022SPIE12184E..2BK}. Correlating the thermal model \textbf{predictions with the temperature distribution measured through} the \ac{tvt} was especially important for verifying the thermal design, because the flight conditions could not be reproduced completely on ground. Therefore, the in-flight behavior needed to be predicted by the thermal model. The agreement between the thermal model and the measured data is also shown for the optical bench (orange and light-blue lines) and the camera radiator (red and blue lines) in Figure \ref{fig:otv-t} during some phases of the \ac{tvt}. Detailed model correlation results are provided in \cite{piqueras2025jaxarr}.
  
Because some of the optical properties in the O-unit were designed to behave differently in air and in vacuum, we constructed the test configuration to verify those properties by injecting laser and white light through a vacuum window (Figure \ref{fig:otv} left). The air-to-vacuum shift of spectral positions was measured by feeding tunable laser light into the spectrograph for both the 850~nm and 770~nm bands. Spectral profiles of the laser verified the spectral resolution in different temperature conditions. We found that the spatial position of the spectrum data varied associated with the temperature ripple caused by switching the heater ON/OFF of the optical bench. This was caused by a slant of the collimator and camera mirrors, and the grating, due to a bi-metallic effect in their support mechanisms \citep{uraguchi2023jaxarr}. The temporal variation of spatial and spectral positions of the spectrum data could be corrected by post-processing during the flight. The air-to-vacuum focus shift of the SJ was verified by measuring the width of a slit image illuminated with white-light. In air, the SJ camera focus was shifted by a known amount of 0.33~mm from the best position determined by the narrowest slit width in a focus scan (Section \ref{SS-oalign}). In the \ac{tvt} we confirmed that the slit width measured by the SJ recovered the narrowest width at the best focus position in air (Figure \ref{fig:otv} bottom right).

\begin{figure}[t]
\centering
\includegraphics[width=0.9\linewidth]{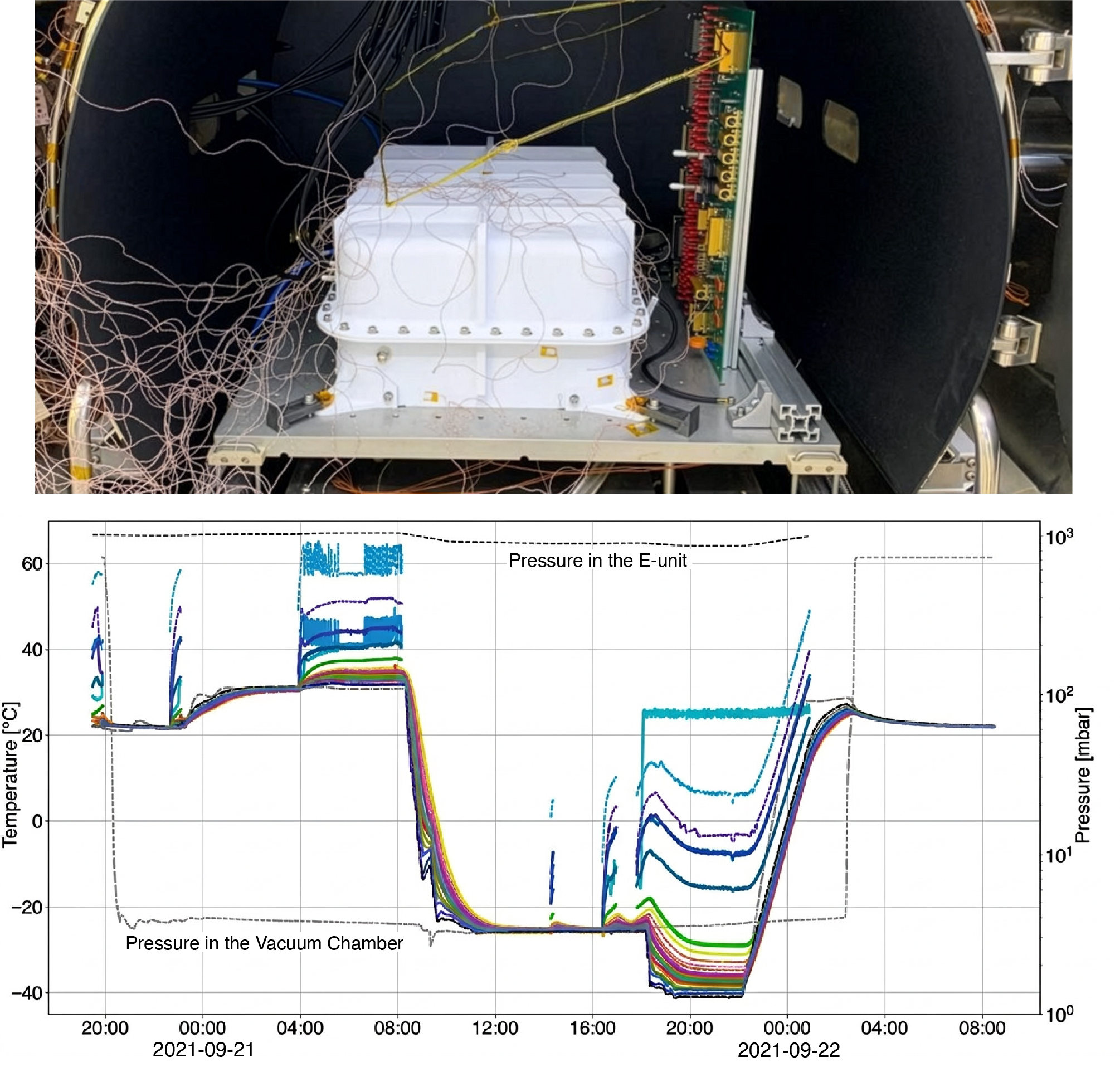}
\caption{SCIP E-unit during the TVT with the electrical ground support equipment on the right (top). TVT temperatures and pressure plot (bottom). The dashed lines show pressures inside the vacuum chamber and the E-unit, while the other solid lines show temperatures of components in the E-unit.}
\label{fig:etv}
\end{figure}
 
\subsubsection{E-unit TV test}
\label{SSS-etv}
The E-unit was tested under low vacuum conditions at IDR/UPM facilities (see Figure~\ref{fig:etv}). Testing was conducted in a nitrogen atmosphere at flight pressure to verify proper operation and to ensure that no leakage occurred. Because reproducing the precise stratospheric environment was not feasible, the chamber's conductive and radiative interfaces were configured to achieve the worst calculated temperatures at the unit reference point with qualification margins of $\pm 15$\degree C. The design process was completed by correlating the detailed model of the E-unit (Section \ref{SS-eth}) with the test data. After this correlation, the model predictions showed a maximum discrepancy of less than 3\degree C compared with the test data.

\section{Summary}
\label{S-Summary}
\ac{scip} is a new spectropolarimeter instrument that was developed for the third flight of the \sunrise\ balloon-borne solar telescope. It \textbf{simultaneously} measures full Stokes profiles of multiple spectral lines in near-infrared wavelength ranges around 770~nm and 850~nm. The wavelength ranges include photospheric \fei\ 846.8~nm, upper photospheric \ki\ 766.5~nm and 769.9~nm, and \caii\ 849.8~nm and 854.2~nm, which provides good sensitivity to the Zeeman effect so that we can measure magnetic fields from the photosphere to the chromosphere. Owing to the 1~m aperture telescope and the greatly reduced influence of atmospheric seeing at the balloon altitude, \ac{scip} can measure these spectral lines with good spatial and spectral resolutions of 0.21" and $1\times10^5$, respectively, enabled by corresponding samplings of 0.094” and $2\times10^5$. Considering the telescope and instrument throughput, the instrument achieves a polarization sensitivity of 0.03\% (1$\sigma$) with a 10~sec integration time and 2 pixel binning along the slit in the 850~nm band, whereas no binning is needed in the 770~nm band. The \ac{scip} O-unit is designed to achieve these performances by a combination of spectrograph and slit jaw optics, opto-mechanics and thermal control, the \ac{pmu} utilizing the rotating waveplate, and the \ac{smm} for scanning the \ac{fov}. The E-unit is designed to control synchronization among the PMU, SMM, and cameras, process the acquired data, output it to the \ac{ics}, and perform housekeeping for the entire \ac{scip} system. After verifying the performance of \ac{scip} through ground testing, it was integrated into the \sunriset\ \ac{pfi} and the telescope. During the flight of \sunriset\ in July 2024, \ac{scip} successfully operated and obtained spectropolarimetric data for various solar targets, including the quiet Sun, sunspots, plages, etc. By combining the data from \ac{susi} and TuMag, which are also onboard \sunriset, \ac{scip} is expected to provide unique data to understand energy transfer and dynamical processes taking place through the photosphere--chromosphere.

\begin{acks}
We would thank the significant technical support given by the Advanced Technology Center (ATC), NAOJ. Engineering support from Mitsubishi Precision Co., Technocraft, Mitsubishi Electric Co., and Syouei System Co. is gratefully acknowledged for the development of the PMU and SMM mechanisms. High-precision optics mounted on the SCIP O-unit were fabricated by Natsume Optical Co., Koshin Kogagu Co., Keihin Coat Co., Geomatec Co., Technical Co., Nitto Optical Co., and Kogakugiken Co. The development of the structural and thermal systems of the O-unit was supported by MRJ Co. and Orbital Engineering Inc. 
\end{acks}

\begin{fundinginformation}
The \sunriset\ project is funded in Japan by the ISAS/JAXA Small Mission-of-Opportunity program for novel solar observations, JSPS KAKENHI Grant Number 18H05234 and 23K25916 (PI: Y. Katsukawa), and the grant of Joint Research by the National Institutes of Natural Sciences (NINS) (NINS program No. OML032402). The Spanish contribution to \sunriset\ \ac{scip} has been supported by the Spanish MCIN/AEI under projects RTI2018-096886-B-C5, and PID2021-125325OB-C5; and from the ``Center of Excellence Severo Ochoa'' awards to IAA-CSIC (SEV-2017-0709, CEX2021-001131-S), all co-funded by European REDEF funds, ``A way of making Europe''. D.O.S. acknowledges financial support from a {\em Ram\'on y Cajal} fellowship. C. Q. N. acknowledges support from the Projects PID2022-136563NB-I00/10.13039/501100011033 and ICTS2022-007828, funded by MICIN and the European Union NextGenerationEU/RTRP. {\sc Sunrise III} is funded by the Max Planck Foundation, the Strategic Innovations Fund of the President of the Max Planck Society (MPG), the Deutsches Zentrum f\"ur Luft und Raumfahrt (DLR), and private donations by supporting members of the Max Planck Society, which is gratefully acknowledged.  This project has received funding from the European Research Council (ERC) under the European Union's Horizon 2020 research and innovation programme (grant agreement No. 101097844 — project WINSUN). 
\end{fundinginformation}

\appendix
\section{Firmware implementation of SCIP observing modes}
 \label{S-firmware-implementation}

\begin{figure}
\centering
\includegraphics[width=\textwidth]{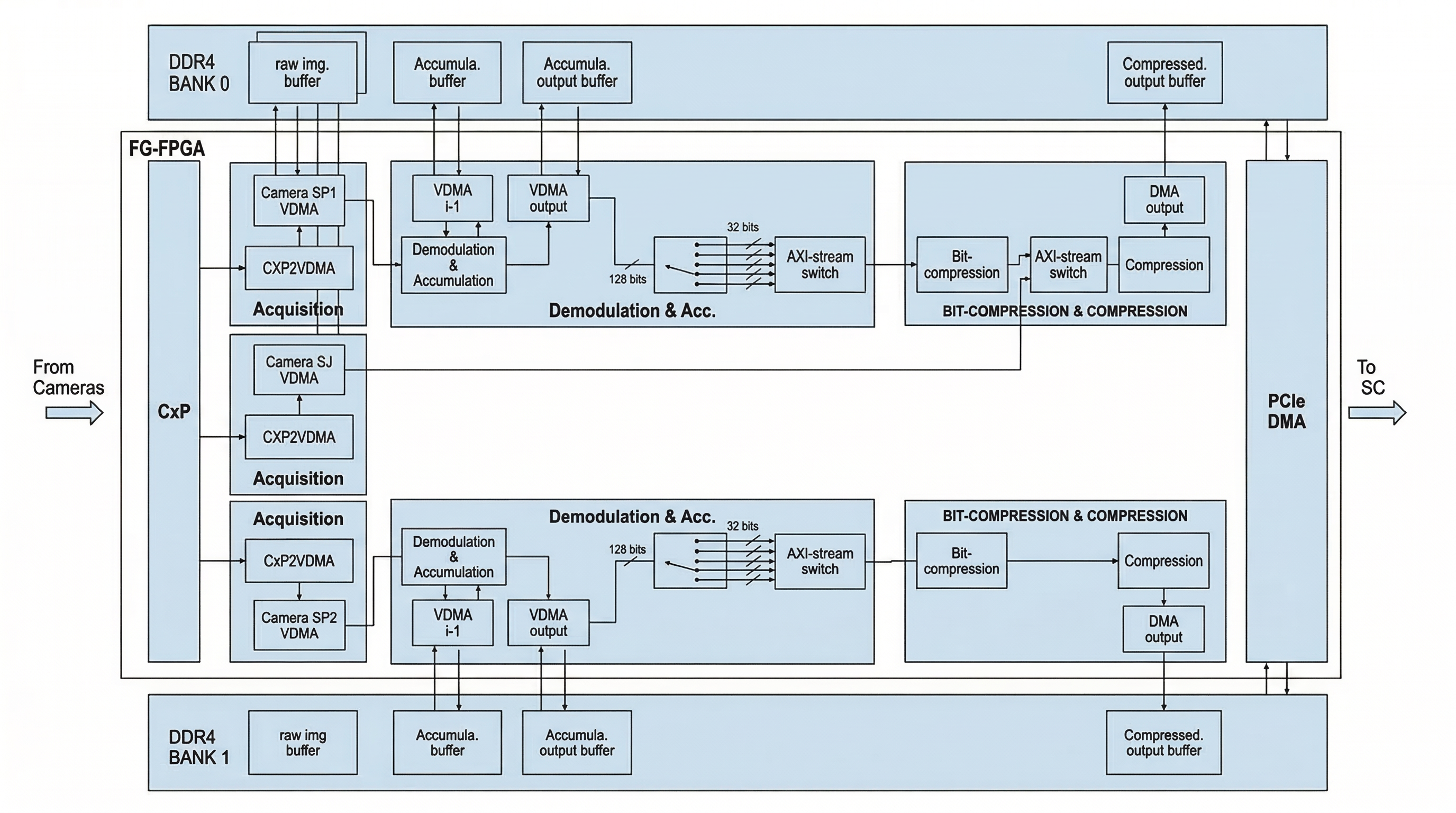}
\caption{FG processing block diagram.}
\label{fig:FGblockdiag}
\end{figure}

Figure~\ref{fig:FGblockdiag} shows a simplified block diagram of the image processing blocks inside the architecture of the \ac{fg}'s \ac{fpga} (some elements, such as the \ac{ddr} memory controllers, bus controllers, etc., are not depicted for simplicity). It consists of two almost identical pipelines that process images from the cameras and transfer the reduced scientific data to the \ac{syc} through the \ac{pcie} interface.

The pipeline at the bottom of the diagram processes images from the SP2 camera and uses the external DDR bank 1 for buffering intermediate results. The other pipeline, \textbf{at the} top of the image, processes images from the SP1 and SJ cameras and uses the DDR bank 0 in the same way. Since SJ's stream is not demodulated, its acquisition block is directly connected to the compression block of SP1 through an \ac{axi}-Stream switch.

Several Video Direct Memory Access (VDMA) Intellectual Property cores (IPs) are instantiated in the design. These blocks are provided by Xilinx\textsuperscript{TM} as part of its \acs{ip} repository and are intended for video frame handling. In the design, each of them is responsible for creating a circular buffer in external memory and performing read/write transfers from/to it with minimum microprocessor intervention. After being configured, a synchronization pulse is enough to trigger a whole image transfer from the buffer to the next processing block. All circular buffers are accessible from the \ac{syc} through the \ac{pcie} interface, implemented with a Xilinx IP XDMA. During nominal observations, the SyC reads final results from the compression output buffer. However, for testing purposes, it can also read intermediate results from the ``raw img.'' or ``Accumula. output'' buffers.

All processing blocks are configurable so that they can be enabled, disabled, or bypassed. In such a way, they can accomplish SCIP observation requirements. In the normal mode, all processing blocks are enabled. However, in the rapid mode, the demodulation and accumulation cores as well as the bit-compression core are bypassed. For testing purposes, different configurations can be set to obtain, for example, demodulated images without compression from the Accumula. output buffer.

\subsection{Acquisition block}
\label{SS-acquisition-block}

\begin{figure}
\centering
\includegraphics[width=\textwidth]{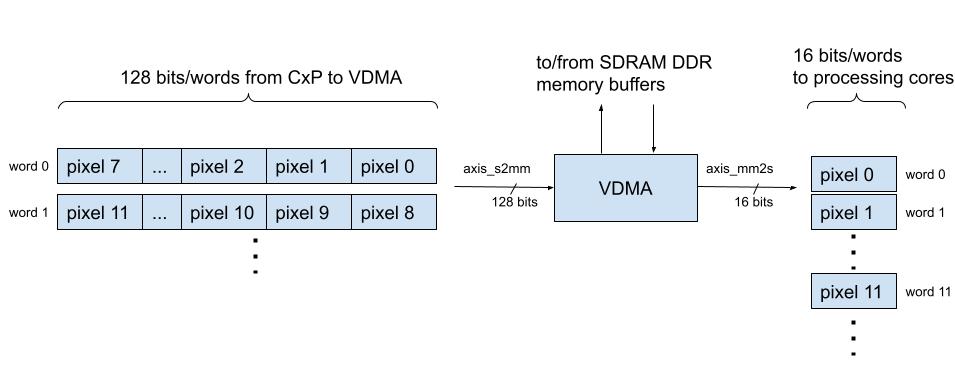}
\caption{Graphical illustration of the CoaXPress-to-VDMA pixel ordering.}
\label{fig:VDMA}
\end{figure}

The CoaXPress-to-VDMA block (CxP2VDMA) is the core of the acquisition stage. This block implements a bus translation between the \ac{axi}-Stream output bus of the \ac{cxp} Host core and the \ac{axi}-Stream input bus of the VDMA core. This adaptation pursues two main aims: preventing CxP metadata from reaching VDMA, hence avoiding metadata to be mistaken with pixel data, and translating 12 \ac{bpp} format to 16 \ac{bpp} format. The latter is needed since images from the cameras arrive at the \ac{fg} in 12 \ac{bpp} format packed into 32-bit words. This pixel-packed format contains a non-integer number of pixels per word (i.e., two pixels + 8 bits of the next pixel), which is undesirable for the rest of the pipeline operations, which are executed in one-pixel per clock cycle flavor. In this context, the CxP2VDMA block unpacks the received pixels, extends its depth to 16 bits and finally pack them into 128-bit words. The Camera VDMA's \ac{axi}-Stream ports are asymmetrically configured to match such a bus width and supply the demodulator core with 16 \ac{bpp} at one pixel per clock cycle as described in Figure~\ref{fig:VDMA}. For testing purposes, the \ac{syc} can read raw images with a pixel de-interleaving format from the raw image buffer.

 \subsection{Demodulation and accumulation}
 \label{SS-demodulation-and-accumulation}

The demodulation operation carried out during the normal mode consists of multiplying the demodulation matrix by the modulated images pixel by pixel. The demodulation matrix is a 5 $\times$ 5 matrix of elements {[}-1, -0.5, 0, 0.5, 1{]} whose order (on each row) depends on the PMU rotation phase. As a result, five images corresponding to the four Stokes parameters $I_d$, $Q_d$, $U_d$, $V_d$, and the pseudo-Stokes parameter $R_d$ are obtained \citep{2022ApOpt..61.9716K,2023JATIS...9c4003K}. 
Such a set of images are accumulated along a configurable number of PMU cycles ranging from ${\rm n_{PMU}} = 2$ (1.024~s, the shortest integration) to ${\rm n_{PMU}} = 4$ (2.048~s). The longest integration time of 10.24~s is achieved by acquiring five datasets with ${\rm n_{PMU}} = 4$ at each slit position. After demodulation and accumulation, the data format changes from unsigned 12~\ac{bpp} to unsigned 19~\ac{bpp} for $I_d$ and signed 18~\ac{bpp} for $Q_d$, $U_d$, $V_d$, and $R_d$. Note that $I_d$, $Q_d$, $U_d$, $V_d$, and $R_d$ are results of the demodulation operation and are different from the true Stokes parameters. The true Stokes parameters are retrieved by the polarization calibration described in Section \ref{SS-polcal}.

The image demodulation and accumulation block has been designed to perform both operations in a pipeline. In this configuration, demodulation and accumulation are carried out simultaneously (after a small initial latency). The data flow through the core for a demodulation of \(N_a=16\, {\rm n_{PMU}} \) images consists of the following steps:

\begin{enumerate}
\def\labelenumi{\arabic{enumi}.}
\item The first image is received, demodulated, and buffered in the accumulation buffer. The PMU phase of the first image is latched and is incremented with every new image received. The output of the demodulation is buffered in a ping-pong buffer, which allows reading from one frame position and writing the result in another frame position without read/write conflicts. Notice that the result of demodulating one single spatial pixel is a set of five interleaved pixel values ($I_d$, $Q_d$, $U_d$, $V_d$, and $R_d$), 24-bit wide each, arranged in words of 128-bit each (this includes five padding bits to complete a power-of-two bit width).
\item The following images, up to \(N_a\), are demodulated and accumulated with the previous demodulated image. To do that, the previous demodulated image is read from the accumulation buffer as soon as a new image is received and demodulated. In such a way, synchronization is ensured, and demodulated images can be accumulated pixel by pixel. The resulting image is buffered in the accumulation buffer, except for the last one which is routed to the demodulation output buffer.
\item The demodulation output buffer stores the final demodulation results of $I_d$, $Q_d$, $U_d$, $V_d$, and $R_d$ in the pixel-interleaving format described above. These images are de-interleaved in order to supply the bit-compression and compression stages with a single image stream. Such a de-interleaving operation is done by means of a pixel selector and an AXI-Stream switch. For each Stokes parameter (including $R$), the whole buffer is read, but only the bits containing that parameter are selected and packed into 32-bit words.
\end{enumerate}

\subsection{Bit-compression}
\label{SS-bit-compression}

Demodulated and accumulated images, coded as unsigned 19~\ac{bpp} (for Stokes $I_d$) and signed 18~\ac{bpp} (for Stokes $Q_d$, $U_d$, $V_d$, and $R_d$ ) values, are pre-processed before applying the standard image compression algorithm. Such a process, named bit-compression, transforms the unsigned 19~\ac{bpp} and signed 18~\ac{bpp} data to unsigned 16~\ac{bpp} data. To ensure precision in the small value range, a threshold $N_{\rm c}$ is established, below which no compression is carried out. $N_{\rm c} = 5000$ DN by default, but it can be parameterized after laboratory tests. Bit-compression is given by 
\begin{equation}
\label{eq:DN}
X = {\tt round}(a + \sqrt{bN + c}),
\end{equation} 
where $N$ stands for the input pixel value ($N > N_{\rm c}$) and $X$ stands for the bit-compressed pixel value. Table~\ref{tab:bitcom} shows parameters $a$, $b$, and $c$ for the different types of compression: no bit-compression, unsigned 19~\ac{bpp} to unsigned 16~\ac{bpp}, and signed 18~\ac{bpp} to unsigned 16~\ac{bpp}. Signal losses induced by the bit-compression step are more than five times smaller than the expected photon noise and can therefore be considered negligible \citep{2023JATIS...9c4003K}. The bit-compression parameters listed in Table~\ref{tab:bitcom} differ from those described in \citep{2023JATIS...9c4003K} because they were updated to better align with the planned flight operations in 2024.

\begin{center}
\begin{table}
\caption{Bit compression parameters}
\begin{tabular}{lllll}
\hline
Compression & $a$ & $b$ & $c$ & Note \\
\hline
(bpp) & (DN) & (DN) & (DN$^2$) \\
\hline
 No & $-$ & $-$ & $-$ & For TEST and RAPID mode \\
19U to 16U & 1006.027 & 7987.946 & -23987910 & For Stokes $I_d$ \\
18S to 16U & 1078.459 & 7843.082 & -23836926 & For Stokes $Q_d$, $U_d$, $V_d$, and $R_d$ \\
\hline
\end{tabular}
\label{tab:bitcom}
\end{table}
\end{center}
A custom bit-compression, FPGA core is designed to accomplish the SCIP's data reduction needs. It uses a dedicated fixed-point architecture together with a Coordinate Rotation Digital Computer (CORDIC) Xilinx IP core to calculate the square root. The core is implemented as a hardware block that works in a pipeline manner after an initial latency. That is, it accepts one input value per clock cycle and produces one result per clock cycle. All figures for $a$, $b$, and $c$ in Table~\ref{tab:bitcom} are significant to provide the required precision, once they are represented as fixed-point values (after multiplication by an integer power of 2), each with a given bit width. Note that, when working in fixed-point precision, the threshold application to the input data can simply be done by bit selection, discarding the fractional part.\footnote{The threshold application also ensures the radicand in Eq.~(\ref{eq:DN}) is positive.} Additionally, divisions by a power of 2 can be implemented through a bit shift or directly discarding the corresponding less-significant bits.

\subsection{Compression}
\label{SS-compression}

The last stage in the processing pipeline is an image compression block. The core of this block is the IP Sunrise Lossless Compression (SLOC) which implements the standard CCSDS-LDC 121.0-B2, a low-complexity solution to compress any kind of data including 2D images. This core performs one pixel per clock cycle at a frequency of 150 MHz to achieve the real-time processing required in SCIP and TuMag instruments \citep{2022_IEE...S}. Note that the image compression was not used during the flight in July 2024.

\section{SCIP harness}
\label{S-harness}

\begin{figure}[t]
\centering
\includegraphics[width=\linewidth]{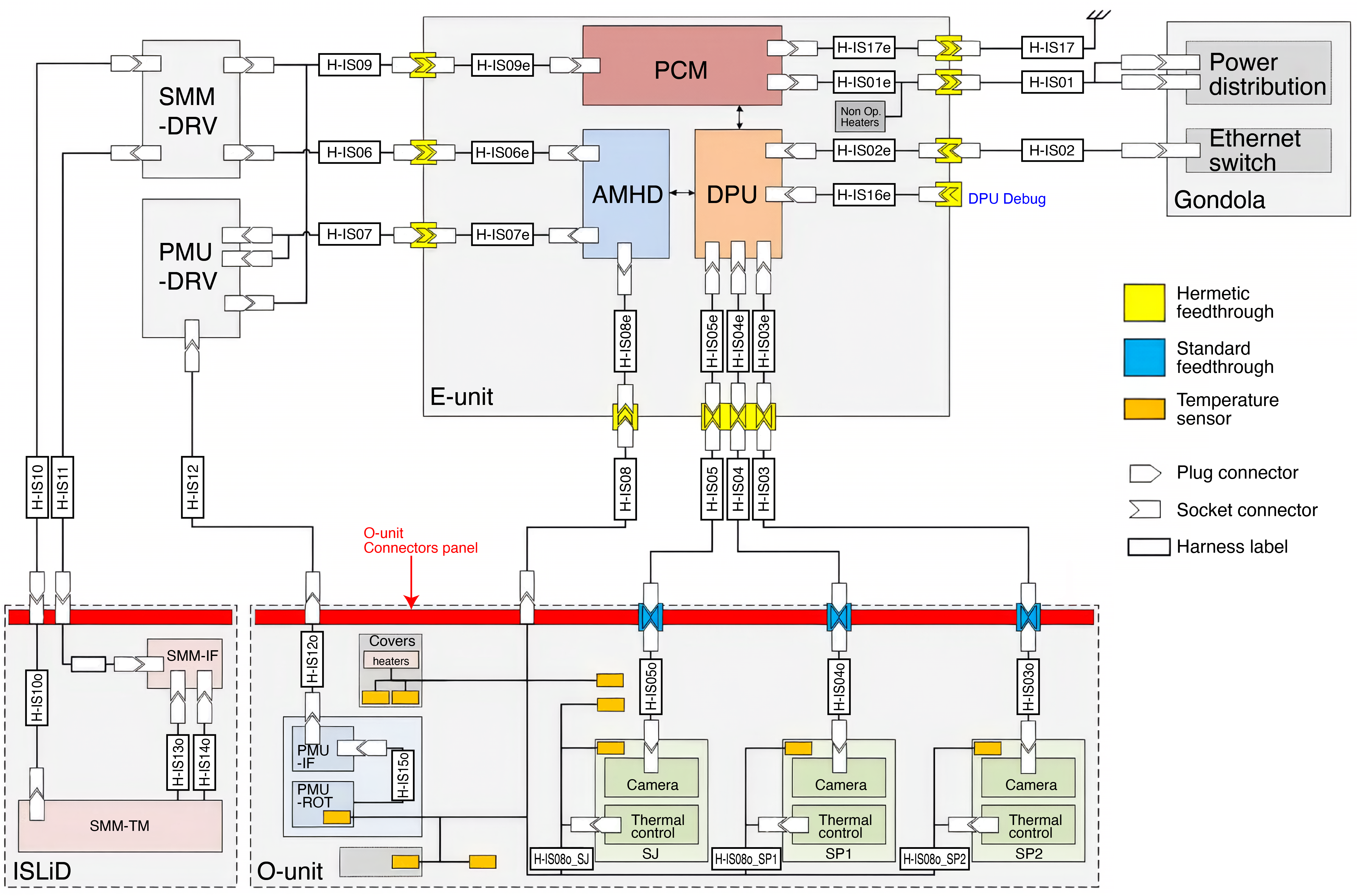}
\caption{Detailed harness block diagram of SCIP.}
\label{fig:harness}
\end{figure}
   
The SCIP harness consists of 13 cable bundles external to the E-unit plus some other bundles which are internal to the various units. A block diagram of the harness can be seen in Figure~\ref{fig:harness}. The external bundles interconnect 1) subsystems in the E-unit with those in the O-unit, in the gondola, and the SMM-DRV and PMU-DRV; 2) the SMM-DRV with the SMM-IF, located in the \ac{islid}; 3) the PMU-DRV with the PMU-IF, located in the O-unit; and 4) the SCIP ground with the system common ground. They are named H-ISXX, where XX is a consecutive number. Internal bundles to the units have a similar definition and materials to the external ones. They are labeled H-ISXXe/o, where ``e'' and ``o'' refer to the E-unit and O-unit, respectively. D-sub connectors have been used for the external harness and a combination of them and Samtec, high-density connectors for the internal bundles. The external bundles receive power from the gondola power distribution, provide communication between the SCIP E-unit and \ac{ics}, and carry the signals needed for global control of the instrument. Three of them (H-IS03, 04, and 05) are \ac{cxp} cables that provide power, control signals for the three cameras, and image transfer from the cameras to the E-unit.

Several considerations have been taken into account in the SCIP harness design. Special care has been taken with cable widths and selection of materials in order to avoid small voltage drops in the long harnesses connecting the E-unit and O-unit and to minimize the outgassing impact in the quasi-vacuum flight conditions. Differential techniques with twisted pairs and external shielding connected to the structure have been adopted for best common-mode noise rejection and ground loop prevention, in order to minimize electromagnetic interference of \ac{scip} with itself and with the other instruments.

\bibliographystyle{spr-mp-sola}
\bibliography{scip}  

\end{article}
\end{document}